\documentclass[twocolumn,superscriptaddress,longbibliography,aps,prx,floatfix,
preprintnumbers]{revtex4-2}

\usepackage[normalem]{ulem}
\usepackage{graphicx}
\usepackage{bm}
\usepackage{color}
\usepackage{epstopdf}
\usepackage{amsmath}
\usepackage{amssymb}
\usepackage{epstopdf}
\usepackage{lipsum}

\usepackage{gensymb}

\usepackage{multirow}

\usepackage{scrextend}

\usepackage[urlcolor=blue,colorlinks=true,citecolor=blue,linkcolor=blue,pdfstartview={FitH},bookmarks=false]{hyperref}

\usepackage[dvipsnames]{xcolor}

\graphicspath{{fig/}{./fig/}{.}}

\begin{document}

\title{
Lattice dynamics study of electron-correlation-induced charge density wave \\ in antiferromagnetic kagome metal FeGe
}

\author{Andrzej~Ptok}
\email[e-mail: ]{aptok@mmj.pl}
\affiliation{\mbox{Institute of Nuclear Physics, Polish Academy of Sciences, W. E. Radzikowskiego 152, PL-31342 Krak\'{o}w, Poland}}

\author{Surajit Basak}
%\email[e-mail: ]{surajit.basak@ifj.edu.pl}
\affiliation{\mbox{Institute of Nuclear Physics, Polish Academy of Sciences, W. E. Radzikowskiego 152, PL-31342 Krak\'{o}w, Poland}}

\author{Aksel~Kobia\l{}ka}
%\email[e-mail: ]{akob@kft.umcs.lublin.pl}
\affiliation{Department of Physics and Astronomy, Uppsala University, Uppsala SE-75120, Sweden}

\author{Ma\l{}gorzata~Sternik}
%\email[e-mail: ]{sternik@wolf.ifj.edu.pl}
\affiliation{\mbox{Institute of Nuclear Physics, Polish Academy of Sciences, W. E. Radzikowskiego 152, PL-31342 Krak\'{o}w, Poland}}

\author{Jan~\L{}a\.{z}ewski}
%\email[e-mail: ]{lazewski@wolf.ifj.edu.pl}
\affiliation{\mbox{Institute of Nuclear Physics, Polish Academy of Sciences, W. E. Radzikowskiego 152, PL-31342 Krak\'{o}w, Poland}}

\author{Pawe\l{}~T.~Jochym}
%\email[e-mail: ]{jochym@wolf.ifj.edu.pl}
\affiliation{\mbox{Institute of Nuclear Physics, Polish Academy of Sciences, W. E. Radzikowskiego 152, PL-31342 Krak\'{o}w, Poland}}

\author{Andrzej~M.~Ole\'{s}}
%\email[e-mail: ]{a.m.oles@fkf.mpg.de}
\affiliation{\mbox{Institute of Theoretical Physics, Jagiellonian University,
Prof. Stanis\l{}awa \L{}ojasiewicza 11, PL-30348 Krak\'{o}w, Poland}}

\author{Przemys\l{}aw~Piekarz}
%\email[e-mail: ]{piekarz@wolf.ifj.edu.pl}
\affiliation{\mbox{Institute of Nuclear Physics, Polish Academy of Sciences, W. E. Radzikowskiego 152, PL-31342 Krak\'{o}w, Poland}}

\date{\today}

\begin{abstract}
Electron-correlation-driven phonon soft modes have been recently reported in the antiferromagnetic kagome FeGe compound and associated with the observed charge density wave (CDW).
In this paper, we present a systematic investigation of the CDW origin in the context of the {\it ab initio} lattice dynamics study.
Performing the group theory analysis of the mentioned soft mode, we found that the stable structure has the Immm symmetry and can be achieved by shifts of Ge atoms.
Additionally, we show that the final structure realizes a distorted honeycomb Ge lattice as well as a non-flat kagome-like Fe net.
For completeness, we present the electronic properties calculations.
From the theoretical STM topography simulation, we indicate that the observed CDW occurs in the deformed honeycomb Ge sublattice.
\end{abstract}

\maketitle

\section{Introduction}
\label{sec.intro}

Kagome lattice materials attract significant attention due to their unique electronic properties~\cite{yin.lian.22}, such as flat bands, van Hove singularities (VHSs) at the M point, and Dirac cone dispersion at the K point.
From this, a wide range of exotic properties and behaviors emerge in the kagome lattice due to different degrees of electron filling~\cite{guterding.jeschke.16,basak.ptok.22,wang.wu.23}.
For example, a strong correlation can induce magnetic order~\cite{tanaka.ueda.03,pollmann.fulde.08}, while a VHS close to the Fermi level can lead to the lattice instability and charge density wave (CDW)~\cite{kiesel.platt.13,wang.li.13}.
The kagome lattices can also exhibit topological properties~\cite{guo.franz.09,ortiz.teicher.20,hu.wu.22,yin.lian.22}, not only limited to the electronic structure~\cite{chisnell.helton.15,ni.gorlach.17,chen.nassar.18,xue.yang.19,xing.chen.22,he.gao.23}.
Therefore, the kagome lattice materials provide an excellent platform to study new physical phenomena.

Recently, we have learned about several kagome lattice systems, which exhibit interesting properties.
Here, we can mention Weyl semimetal Co$_{3}$Sn$_{2}$S$_{2}$, with ferromagnetic Co kagome net~\cite{liu.sun.18,liu.liang.19,morali.batabyal.19,yin.zhang.19,xu.zhao.20,kanagaraj.ning.22}.
Topological properties and the break of the time reversal symmetry lead to the intrinsic giant anomalous Hall effect~\cite{liu.sun.18,wang.xu.18,yang.noky.20}, while hosting of the exotic Weyl fermions~\cite{liu.liang.19} induces the appearance of the Fermi arc~\cite{morali.batabyal.19}.
Another example, $A$V$_{3}$Sb$_{5}$ ($A=$K, Rb, and Cs) with a vanadium kagome net displays the CDW with the Star of David (SoD) pattern below $\sim 90$~K~\cite{ortiz.gomes.19,li.zhabg.21,liang.hou.21}.
Moreover, below $\sim 2$~K the coexistence of CDW and superconducting state is observed~\cite{ortiz.teicher.20,ortiz.sarite.21,gupta.das.22}.
Such compounds can also exhibit an unconventional anomalous Hall effect~\cite{yang.wang.20,yu.wu.21}.
Finally, also other well-studied kagome systems, such as FeSn~\cite{kang.ye.20,han.inoue.21,zhang.oli.23,meier.du.20}, CoSn~\cite{meier.du.20,kang.fang.20,liu.li.20}, Fe$_{3}$Sn$_{2}$~\cite{ye.kang.18,lin.choi.18,yin.zhang.18}, $RE$Ti$_{3}$Bi$_{4}$ ($RE =$ Yb, Pr, and Nd)~\cite{chen.zhou.23,sakhya.ortiz.23,mondal.sakhya.23}, $M$Mn$_{6}$Sn$_{6}$ ($M=$Y, Er, Tb)~\cite{yin.ma.20,ghimire.dally.20,li.wang.21,fruhling.streeter.21,ma.xu.21,zhang.koo.22,mielke.ma.22,riberolles.slade.22} and ScV$_{6}$Sn$_{6}$~\cite{suriya.meier.22,cao.xu.23}, or $A$V$_{6}$Sb$_{6}$ ($A=$K, Rb, Cs, or Gd)~\cite{yang.fan.21,yin.tu.21,shi.yu.22,mantravadi.gvozdetskyi.23,hu.wu.22} can be mentioned.

\begin{figure}[!t]
\centering
\includegraphics[width=\columnwidth]{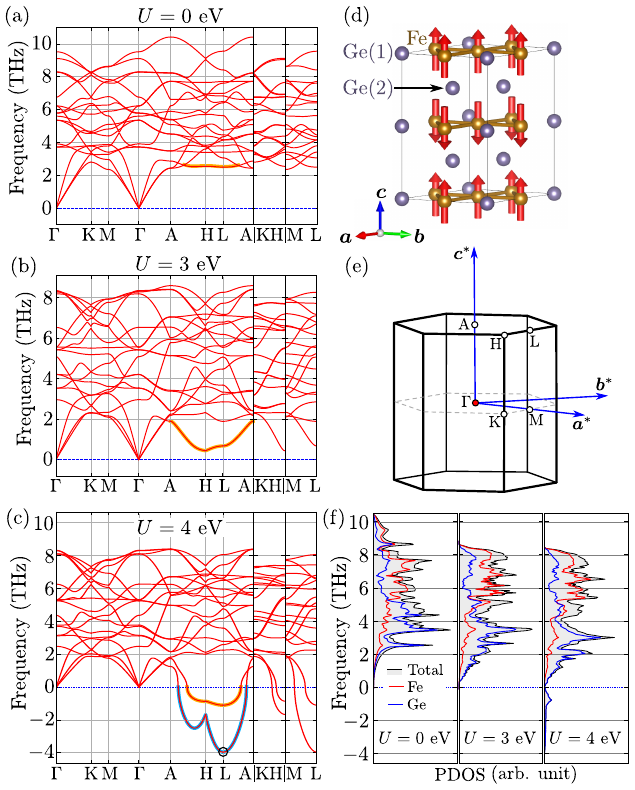}
\caption{
(a)-(c) The effect of the local Coulomb interaction on the phonon dispersion curves along high symmetry directions for FeGe with the P6/mmm symmetry.
Results for a different Hubbard $U$ parameter (as labeled).
(d) Magnetic unit cell of FeGe with the P6/mmm symmetry in the presence of the A-AFM order, and its Brillouin zone (e).
(f) The phonon DOS as a function of the local Coulomb interaction value.
\label{fig.191}
}
\end{figure}

In this paper, we focus on FeGe, in which CDW was recently discovered~\cite{teng.chen.22,yin.jiang.22}.
FeGe exhibits an antiferromagnetic (AFM) order below $410$~K~\cite{watanabe.kunitomi.66,tomiyoshi.yamamoto.66,beckma.carrander.72,haggstrom.ericsson.75,forsyth.wilkinson.78,bernhard.lebech.84,bernhard.lebech.88}.
At room temperature, the Fe atoms are ferromagnetically coupled with the kagome sublattice plane and antiferromagnetically coupled between kagome sublattices, i.e. along $c$ [so-called A-AFM order, see Fig.~\ref{fig.191}(d)].
At lower temperatures, a tilt of the Fe magnetic moments from the $c$ axis was reported.
The CDW phase is reported below $100$~K~\cite{teng.chen.22}.
The coexistence of these two ordered phases gives a great opportunity to study the interplay between them~\cite{teng.chen.22,teng.oh.23}.
The scanning tunneling microscopy (STM) of the surface charge distribution uncovers the $2 \times 2$ pattern~\cite{teng.chen.22,shao.yin.23,chen.wu.23}.

In this context, it is important to recognize correctly the origin of the CDW.
For example, in the case of vanadium kagome net systems ($A$V$_{3}$Sb$_{5}$) the CDW is associated with imaginary soft modes at the M and L points~\cite{tan.liu.21,ptok.kobialka.22,subedi.22,subires.korshunov.23}, which induce the structural phase transition~\cite{ptok.kobialka.22,gutierrez.dangic.23}.
These soft modes lead to a stable structure with the C2/m~\cite{ptok.kobialka.22} or Fmmm~\cite{subedi.22} symmetry.
The phonon spectrum of FeGe, calculated without electron correlations, is similar to the other CoSn-like compounds~\cite{ptok.kobialka.21}, and no imaginary soft modes are observed.
However, the introduction of the correlations leads to the softening of some optical phonons along the L--H direction~\cite{miao.zhang.23,chen.wu.23,ma.yin.23}.
Such soft modes generate the Ge-dimerization and were recognized as a source of CDW~\cite{wang.23}.
Nevertheless, a more thorough dynamical study of the CDW formation mechanism has not been performed so far.
Here, using the {\it ab initio} techniques, we derive a critical value of the on-site Coulomb interaction, which induces imaginary soft modes in the phonon spectrum.
Detailed analysis of the symmetry of these soft modes allow us to find a stable low-symmetry structure of FeGe.

The paper is organized as follows.
Details of the numerical calculations can be found in Sec.~\ref{sec.comp}.
Our results are presented and discussed in Sec.~\ref{sec.res}.
We start with a description of the dynamical properties of the FeGe system (Sec.~\ref{sec.dyn}).
Using the electron-correlation-driven phonon soft mode analysis, we show that the stable structure has the Immm symmetry.
Next, for this stable structure, we discuss the electronic properties (Sec.~\ref{sec.ele}).
Finally, we summarize and conclude our findings in Sec.~\ref{sec.sum}.

\section{Computational techniques}
\label{sec.comp}

The first-principles density functional theory (DFT) calculations were performed using the projector augmented-wave (PAW) potentials~\cite{blochl.94} implemented in 
Vienna Ab initio Simulation Package ({\sc Vasp})~\cite{kresse.hafner.94,kresse.furthmuller.96,kresse.joubert.99}.
For the exchange-correlation energy, the generalized gradient approximation (GGA) in the Perdew--Burke--Ernzerhof (PBE) parametrization was used~\cite{perdew.burke.96}.
The energy cutoff for the plane-wave expansion was set to $350$~eV.
We introduced the correlation effect on Fe $3d$ orbitals using the DFT+U scheme proposed by Dudarev {\it et al.}~\cite{dudarev.botton.98}.
It is worth mentioning that the use of Liechtenstein {\it et al.} approach~\cite{liechtenstein.anisimov.95} changed results only quantitatively [see Fig.~\ref{fig.compare} and Fig.~\ref{fig.compare_ph} in the Supplemental Material (SM)~\footnote{The Supplemental Material at [URL will be inserted by publisher] for additional theoretical results. We present the atom displacement induced by the soft modes, electronic band structure, and role of the correlation effects withing DFT+U.}].

The optimization of the lattice constants and atomic positions in the presence of the spin--orbit coupling (SOC) was performed for magnetic unit cells (with the collinear A-AFM order).
The structures were optimized with different {\bf k}-grids generated using the Monkhorst--Pack scheme~\cite{monkhorst.pack.76} depending on the investigated symmetry.
The initial structure with the P6/mmm symmetry was optimized using $8 \times 8 \times 5$ {\bf k}--point grid.
Structures with the Ibam and Immm symmetries were optimized with $15 \times 8 \times 8$ {\bf k}--point grids.
As a convergence criterion of the optimization loop, we took the energy change below $10^{-6}$~eV and $10^{-8}$~eV for ionic and electronic degrees of freedom, respectively.
The symmetry of the structures after optimization was analyzed with {\sc FindSym}~\cite{stokes.hatch.05} and {\sc Spglib}~\cite{togo.tanaka.18}, while momentum space analysis was performed within {\sc SeeK-path}~\cite{hinuma.pizzi.17}.

Dynamic properties were calculated using the direct {\it Parlinski--Li--Kawazoe} method~\cite{parlinski.li.97}, implemented in the {\sc Phonopy} package~\cite{togo.chaput.23,togo.23}. 
Within this method, the interatomic force constants (IFC) are calculated from the Hellmann-Feynman (HF) forces acting on the atoms after displacements of individual atoms inside the supercell.
We performed these calculations using the supercell with the shape corresponding to $2 \times 1 \times 2$ magnetic unit cells with the Ibam and Immm symmetries (approximately cubic shape, containing 24 formula units).
During these calculations, reduced $5 \times 5 \times 5$ {\bf k}-grid was used.

For the final crystal structure, the STM topographies were simulated within the Tersoff--Hamann approach~\cite{tersoff.hamann.85} for the slab containing four FeGe layers (with fixed atomic positions) separated by $\sim10$~\AA\ of vacuum.

\section{Results and discussion}
\label{sec.res}

\subsection{Dynamical properties and system stability}
\label{sec.dyn}

Let us start with a discussion of the lattice dynamics for the initial structure with the P6/mmm symmetry, in the presence of different values of Coulomb interaction on Fe $d$ orbitals (Fig.~\ref{fig.191}).
In the absence of the correlations ($U=0$~eV), none of the phonon dispersion curves show any noticeable softening nor imaginary values [see Fig.~\ref{fig.191}(a)]~\cite{ptok.kobialka.21}.
The introduction of Hubbard $U$ causes the softening of the lowest phonon mode along the H--L direction~\cite{miao.zhang.23,chen.wu.23,ma.yin.23}.
This is clearly visible when we compare that phonon branch for $U = 0$~eV and $U  = 3$~eV [cf.~Fig.~\ref{fig.191}(a) and~\ref{fig.191}(b), where the soft mode is marked with an orange background].
However, further increase of $U$ (e.g. to $4$~eV) leads to the emergence of imaginary soft modes [presented as negative frequencies in Fig.~\ref{fig.191}(c)].
As we can see, there is some critical value of the Hubbard parameter ($U_{c}$) for which the soft modes become imaginary.
At this higher $U$ value, there are two imaginary soft branches [marked with orange and blue in Fig.~\ref{fig.191}(c)].
The branch marked with an orange background corresponds to the soft mode visible for $U < U_{c}$ in Fig.~\ref{fig.191}(b).

In Fig.~\ref{fig.191}(f) the phonon density of states spectra were compared for various values of the Hubbard $U$ parameter. 
Increasing the local Coulomb potential (from $0$~eV to $4$~eV) causes an enhancement of both the magnetic moments of the iron atoms (from $1.53$~$\mu_B$ to $2.96$~$\mu_B$) and the crystal cell volume (by about 12\%). 
However, it is worth emphasizing that phonon softening is not connected with volume extension but 
only with correlation effects, which we verified by changing volume and $U$ separately.
What is more surprising, the increase in value of $U$ on Fe results in a softening of phonon modes associated mainly with the vibrations of Ge atoms: Ge(1) perpendicular and Ge(2) parallel to the Fe kagome layers (marked with orange and blue in Fig.~\ref{fig.191}(c), respectively).

\begin{figure}[!t]
\centering
\includegraphics[width=\columnwidth]{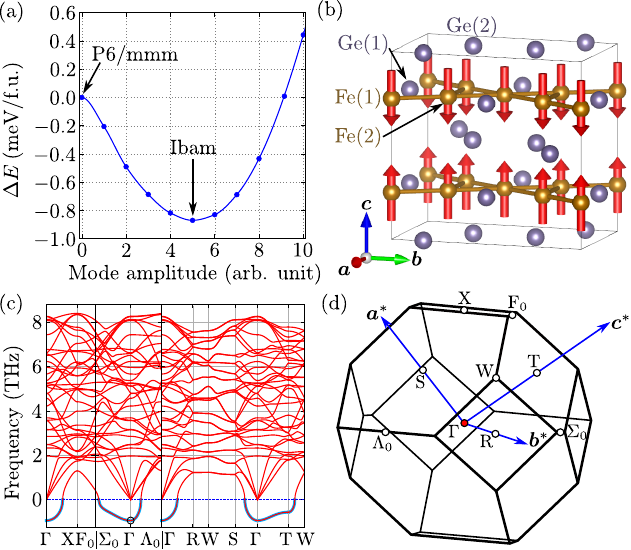}
\caption{
(a) The soft mode amplitude dependence of the system's energy for the distorted structure induced by the soft mode at the L point of the P6/mmm structure.
The zero of energy scale is set at the energy of the initial P6/mmm structure.
The optimized system with the lowest energy has the Ibam symmetry and magnetic unit cell presented in panel (b). 
In (c), the phonon dispersion curves for FeGe with the Ibam symmetry along high symmetry directions of the Brillouin zone are presented following the scheme displayed in (d).
\label{fig.72}
}
\end{figure}

\begin{figure*}
\centering
\includegraphics[width=\textwidth]{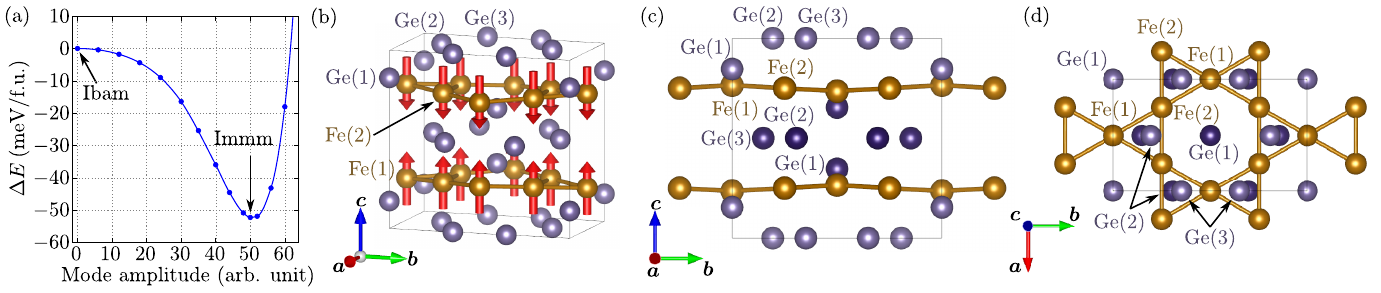}
\caption{
(a) The soft mode amplitude dependence of system's energy for the structure induced by the soft mode of the Ibam structure at the $\Gamma$ point.
Energy changes are given relative to the ground-state energy of the Ibam structure. 
Mode amplitude expressed as displacement of Ge(1) atoms, $u_{\text{Ge}} \approx 0.8$~\AA.
The system reaches the lowest energy for the Immm symmetry, with magnetic unit cell presented in (b).
The front and top view of the crystal with the Immm symmetry is presented in (c) and (d) panels, respectively.
\label{fig.71}
}
\end{figure*}

We should briefly discuss the magnitude of $U$ for Fe $d$ orbitals.
The effective Hubbard $U$ parameter of $4$~eV was used for Fe-based compounds, mainly iron oxides such as FeO~\cite{liao.carter.10}, Fe$_{2}$O$_{3}$~\cite{liao.carter.10}, Fe$_{3}$O$_{4}$~\cite{piekarz.oles.10}, or Fe$_{2}$SiO$_{4}$~\cite{piekarz.oles.10}. 
In some cases a larger effective $U$ is necessary (e.g. for iron-bearing sphalerite~\cite{feng.li.23}).
For metals, due to the screening effects the $U$ parameter is usually smaller~\cite{lazewski.piekarz.06}.
The comparison of Fe magnetic moments with the experimental value of $1.7$~$\mu_\text{B}$ also indicates a smaller value of $U$ in FeGe~\cite{zhou.yan.23}.
Here, we assumed $U = 4$~eV to demonstrate the existence of the soft mode with the frequency approaching zero and becoming imaginary. 
Such a gimmick allowed for the study of a structure with lower symmetry.
In reality, as the diffraction studies on FeGe showed, the phase transition to the CDW phase has the 1st-order character~\cite{teng.chen.22,miao.zhang.23}. 
In such a case, the soft mode does not need to go to zero frequency to trigger the structural transformation and therefore, a real value of $U$ is smaller than $4$~eV.
Nevertheless, as we mentioned earlier, for $U > U_{c}$ the phonon dispersion curves show imaginary soft modes. 
Since structural changes are defined by the polarization vector of the soft mode, the resulting space group symmetry is independent of $U$.

Let us now analyze displacements induced by the lowest energy soft mode occurring at L=(0,1/2,1/2).
Condensation of such a mode enforces doubling of the P6/mmm primitive unit cell along the $b$ and $c$ directions.
Additionally, freezing of displacements induced by its polarization vector lowers system energy because of the imaginary value of the soft mode frequency.
In fact, the system's energy as a function of the displacement amplitude directly shows the existence of a structure with the lower energy [Fig.~\ref{fig.72}(a)].
This displaced and more stable structure will be the ``base'' of our further analysis.

The structure with the P6/mmm symmetry (space group No.~191) possesses the lattice parameters $a = b = 5.163$~\AA\ and $c = 4.251$~\AA. 
The atoms are located at three nonequivalent Wyckoff positions: ({\it 3f}) Fe (1/2,0,0), ({\it 1a}) Ge(1) (0,0,0), and ({\it 2d}) Ge(2) (1/3,2/3,1/2).
The Fe atoms form an ideal kagome net, decorated by Ge(1) in the same plane. 
The Ge(2) atoms form the honeycomb lattice, located between the kagome layers.
The AFM magnetic order leads to doubling of the unit cell along the $c$ direction [see Fig.~\ref{fig.191}(d)], i.e., the magnetic unit cell contains 6 formula units.
The L-point imaginary mode leads mainly to the displacement of Ge(2) atoms (along the $a+b$ direction of the P6/mmm structure) by about $\pm 0.06$~\AA~[see Fig.~\ref{fig.disp}(a) in the SM~\cite{Note1}].
We should also mention that the Fe atoms still form a kagome-like net with two different distances between atoms: Fe(1)--Fe(2) and Fe(2)--Fe(2) equal to $2.5860$~\AA\ and $2.5809$~\AA, respectively.
The optimized structure has the Ibam symmetry (space group No.~72), with lattice constants $a = 5.158$~\AA, $b = 8.956$~\AA, and $c = 8.501$~\AA, and four nonequivalent Wyckoff positions: ({\it 4a}) Fe(1) (0,0,1/4), ({\it 8e}) Fe(2) (1/4,1/4,1/4), 
({\it 4b}) Ge(1) (1/2,0,1/4), and ({\it 8j}) Ge(2) (0.5142,0.6654,0).
Just like before, Ge(1) atoms decorate the kagome net, while Ge(2) atoms form the deformed honeycomb-like lattice.
The magnetic unit cell corresponds to the conventional cell and is presented in Fig.~\ref{fig.72}(b).
The ground state energy of the system with the magnetic moment on Fe atoms along $c$ and tilted from $c$ axis are close to each other.

The phonon dispersion curves for the optimized Ibam structure are presented in Fig.~\ref{fig.72}(c).
As we can see, the phonon spectrum still contains imaginary soft modes (marked with blue lines) and the most predominant soft mode occurs at the $\Gamma$ point. 
In that case, the structure can be stabilized by shifts of atoms, mostly Ge(1), that do not modify the size of the unit cell.
As a result, we can apply the same strategy as before to find the final crystal structure.

\begin{figure}[!b]
\centering
\includegraphics[width=\columnwidth]{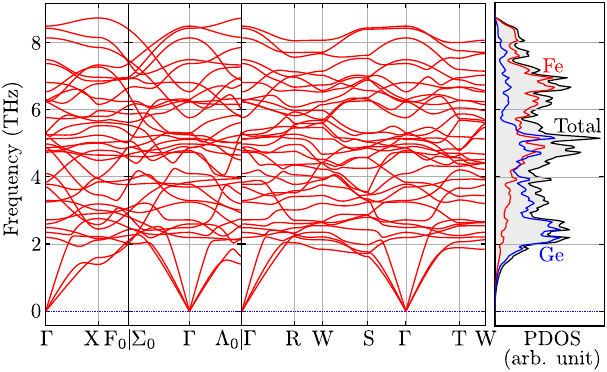}
\caption{
The phonon dispersion curves along high symmetry directions and phonon density of states for FeGe with the Immm symmetry.
The symbols of the high symmetry points are the same as in Fig.~\ref{fig.72}(d).
\label{fig.71ph}
}
\end{figure}

The dependence of the system's energy on the displacement amplitude of the polarization vector of the soft mode at the $\Gamma$ point for the Ibam structure is presented in Fig.~\ref{fig.71}(a).
The most important displacement is associated with the shift of the Ge(1) atoms along the $c$ direction [see Fig.~\ref{fig.disp}(b) in the SM~\cite{Note1}], which is in agreement with the previous theoretical studies~\cite{wang.23}, 
where the large dimerization of Ge(1) sites was studied without breaking the P6/mmm symmetry.
The soft mode of Ibam structure leads to the Immm symmetry (space group No.~71). 
After optimization, the total energy decreases and lattice parameters read $a = 5.158$~\AA, $b = 8.956$~\AA, and $c = 8.501$~\AA.
The atoms are located in the Wyckoff positions: ({\it 4j}) 
Fe(1) (1/2,0,0.2353), ({\it 8k}) Fe(2) (1/4,1/4,1/4), ({\it 4i}) Ge(1) (0,0,0.8462), ({\it 4h}) Ge(2) (0,0.3063,1/2), and({\it 4g}) Ge(3) (0,0.3515,0).
Similarly, the energies of the system with Fe magnetic moments along the $c$ axis and tilted from the $c$ direction are comparable.
In this structure, the Ge(1) atoms are shifted from their initial positions by $\pm 0.82$~\AA\ along the $c$ direction [in practice, the distance between Ge(1) atoms decreases from $4.06$~\AA~(P6/mmm) to $2.61$~\AA~(Immm), what corresponds to effective shifting of $0.725$~\AA].
That large modification of the atomic position leads indirectly to large deformation of the Fe kagome-like net and the Ge honeycomb sublattice [see Fig.~\ref{fig.71}(c) and~\ref{fig.71}(d)]. 
The kagome-like lattice is not flat, while the deviation from the plane is around $0.13$~\AA.
The distances between Fe atoms within the kagome-like net are $2.5791$~\AA\ and $2.5839$~\AA\ for Fe(2)--Fe(2) and Fe(2)--Fe(3), respectively.
Similarly, the new distances between atoms in the deformed honeycomb sublattice are $2.94$~\AA\ and $2.66$~\AA\ for Ge(2)--Ge(3) and Ge(3)--Ge(3), respectively.

The phonon dispersion curves and phonon density of states for the final structure are presented in Fig.~\ref{fig.71ph}.
As we can see, the structure with the Immm symmetry is dynamically stable.
The vibrations associated with Fe atoms are mostly realized by the phonon modes in the higher frequency range,
while the lowest modes correspond to the vibrations of Ge.
Acoustic branches around the $\Gamma$ point show well-visible linearity.
The first nearly flat bands within the Brillouin zone are located above $2$~THz ($8.27$~meV).
Furthermore, a dense complex band structure is visible around $5$~THz ($20.68$~meV), while phonon branches with the highest frequencies are around $9$~THz ($37.22$~meV).
These frequency ranges are in excellent agreement with the experimentally observed phonon spectrum~\cite{teng.oh.23}.

From the above analysis it is clear that the final stable structure
arises from the condensation of two soft modes, the first at the L point of P6/mmm structure and the second at the $\Gamma$ point of the intermediate Ibam symmetry.
The structural transformation involving two soft modes (two order parameters) leads to the first order phase transition in which the crystal structure changes discontinuously at the critical temperature.
This scenario agrees well with the experimental observation of the CDW phase in FeGe~\cite{teng.chen.22,miao.zhang.23}.
As we demonstrated, the soft mode frequency strongly depends on $U$, which modifies the Fe $d$ states, and thus influences the interatomic forces. 
Moreover,  analyzing the IFC changes introduced by $U$, we noticed that the strongest effect is observed for the diagonal component $zz$ of the on-site Ge force constant matrix.
With the increasing $U$, this force constant decreases significantly, leading to a softening of the phonon modes.
It shows that the electron-phonon coupling plays an important role in the mechanism of structural deformation and the CDW formation.

\subsection{Electronic properties}
\label{sec.ele}

The electronic band structures for the P6/mmm and Immm magnetic unit cells (see Fig.~\ref{fig.71el} in the SM~\cite{Note1}) possess several similarities.
This suggests that the electronic spectra observed within angle-resolved photoemission spectroscopy (ARPES) measurements should be similar for the system before and after the transition to the CDW phase.
The same behavior is also observed in vanadium-based kagome systems $A$V$_{3}$Sb$_{5}$~\cite{luo.gao.22,kang.fang.22,kato.li.22,jiang.ma.23}, where modification of the electronic ARPES spectra is mostly connected with transition to the CDW phase.

In the case of P6/mmm structure, all Fe atoms carry a magnetic moment of $2.96$~$\mu_{B}$.
An additional magnetic moment is also induced on the Ge(1) atoms ($0.165$~$\mu_{B}$).
As we mentioned earlier, the transition from the P6/mmm to Imma phase leads to modification of the magnetic moments. 
The deformation of the kagome net causes a small increase of the Fe(1) magnetic moment ($3.00$~$\mu_{B}$), while the Fe(2) magnetic moment remains unchanged ($2.96$~$\mu_{B}$).
Similarly, the shift of Ge(1) in the Immm structure, with respect to the position in the initial P6/mmm symmetry, leads to a small decrease of the magnetic moment on Ge(1) ($0.125$~$\mu_{B}$).
Additionally, the Ge(1) magnetic moment has an opposite direction to the nearest Fe kagome-like net.

\begin{figure}[!t]
\centering
\includegraphics[width=\columnwidth]{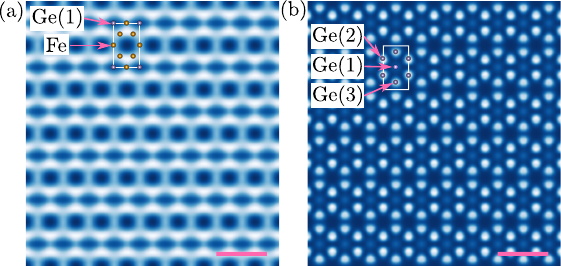}
\caption{
Theoretically obtained STM topography of FeGe with the Immm symmetry, calculated $\sim 1$~\AA\ above the surface.
Results for termination on Fe kagome-like net (a), and deformed Ge honeycomb lattice (b).
Inset presents the relative position and unit cell for the (001) surface.
Scale bar, $\sim 1$~nm.
\label{fig.71stm}
}
\end{figure}

For the Immm structure obtained here, we can simulate the STM topography for 
two different (001) surface terminations (see Fig.~\ref{fig.71stm}).
The results presented are based on DFT slab calculations, while the STM topography comes from $E_{F} \pm 0.1$~eV energy window.
In Figs~\ref{fig.71stm}(a) and~\ref{fig.71stm}(b) we present theoretically obtained STM topography for the FeGe terminated on the Fe kagome-like net and the deformed Ge honeycomb lattice, respectively.
Comparison to the experimentally reported STM pattern~\cite{teng.chen.22} suggests that FeGe has termination on the deformed Ge honeycomb lattice.
In the case of theoretically obtained results, both topographies are similar to the CDW stripe order.
However, the Fourier transform of the experimental STM topography shows that the CDW peaks have a non-equal intensity~\cite{teng.chen.22}. 
In fact, one of the CDW peaks is much stronger than the other, which can suggest a stripe type of the CDW order.
Moreover, from the theoretical point of view, the $2 \times 2$ CDW order in the P6/mmm structure due to the shift of Ge(1) out of the Fe kagome layer corresponds to the stripe order in the Immm structure, as presented in Fig.~\ref{fig.71stm}(b).

\section{Summary}
\label{sec.sum}

Starting from the initial P6/mmm structure, which contains an ideal Fe kagome net, we found that there exists a critical value of the Hubbard $U$ parameter ($\sim 3.5$~eV), for which the softened phonon mode becomes imaginary.
From analysis of the most predominant imaginary soft branch at the L point, we found a new structure with the Ibam symmetry.
This structure comes from P6/mmm due to the shift of Ge atoms within the Fe kagome net layer.
However, the lattice dynamics of the Ibam structure reveals that it is also unstable with soft mode at the $\Gamma$ point.
In this case, the displacement introduced by this soft mode associated with the shift of Ge atoms in a direction perpendicular to the Fe kagome net leads to a stable Immm structure.
Condensation of both modes leads to the first-order structural phase transition, which  is, in turn, responsible for the formation of the CDW.
It is also worth noting that the occurrence of the first-order phase transition in FeGe is confirmed by experiments.

For completeness, we performed also calculations for the slab structure with the Immm symmetry to generate STM images for the Fe-kagome-like and Ge-deformed honeycomb surface terminations. 
The calculated STM topographies indicate the existence of charge ordering. 
The observed CDW phase is associated with a deformed Ge honeycomb-like lattice.

\begin{acknowledgments}
Some figures in this work were rendered using {\sc Vesta}~\cite{momma.izumi.11} and {\sc XCrySDen}~\cite{kokalj.99} software.
We kindly acknowledge support frrom the National Science Centre (NCN, Poland) 
under Project No.~2021/43/B/ST3/02166.
\end{acknowledgments}

%\nocite{*}
\bibliography{biblio.bib}

\begin{thebibliography}{106}%
\makeatletter
\providecommand \@ifxundefined [1]{%
 \@ifx{#1\undefined}
}%
\providecommand \@ifnum [1]{%
 \ifnum #1\expandafter \@firstoftwo
 \else \expandafter \@secondoftwo
 \fi
}%
\providecommand \@ifx [1]{%
 \ifx #1\expandafter \@firstoftwo
 \else \expandafter \@secondoftwo
 \fi
}%
\providecommand \natexlab [1]{#1}%
\providecommand \enquote  [1]{``#1''}%
\providecommand \bibnamefont  [1]{#1}%
\providecommand \bibfnamefont [1]{#1}%
\providecommand \citenamefont [1]{#1}%
\providecommand \href@noop [0]{\@secondoftwo}%
\providecommand \href [0]{\begingroup \@sanitize@url \@href}%
\providecommand \@href[1]{\@@startlink{#1}\@@href}%
\providecommand \@@href[1]{\endgroup#1\@@endlink}%
\providecommand \@sanitize@url [0]{\catcode `\\12\catcode `\$12\catcode `\&12\catcode `\#12\catcode `\^12\catcode `\_12\catcode `\%12\relax}%
\providecommand \@@startlink[1]{}%
\providecommand \@@endlink[0]{}%
\providecommand \url  [0]{\begingroup\@sanitize@url \@url }%
\providecommand \@url [1]{\endgroup\@href {#1}{\urlprefix }}%
\providecommand \urlprefix  [0]{URL }%
\providecommand \Eprint [0]{\href }%
\providecommand \doibase [0]{https://doi.org/}%
\providecommand \selectlanguage [0]{\@gobble}%
\providecommand \bibinfo  [0]{\@secondoftwo}%
\providecommand \bibfield  [0]{\@secondoftwo}%
\providecommand \translation [1]{[#1]}%
\providecommand \BibitemOpen [0]{}%
\providecommand \bibitemStop [0]{}%
\providecommand \bibitemNoStop [0]{.\EOS\space}%
\providecommand \EOS [0]{\spacefactor3000\relax}%
\providecommand \BibitemShut  [1]{\csname bibitem#1\endcsname}%
\let\auto@bib@innerbib\@empty
%</preamble>
\bibitem [{\citenamefont {Yin}\ \emph {et~al.}(2022{\natexlab{a}})\citenamefont {Yin}, \citenamefont {Lian},\ and\ \citenamefont {Hasan}}]{yin.lian.22}%
  \BibitemOpen
  \bibfield  {author} {\bibinfo {author} {\bibfnamefont {J.-X.}\ \bibnamefont {Yin}}, \bibinfo {author} {\bibfnamefont {B.}~\bibnamefont {Lian}},\ and\ \bibinfo {author} {\bibfnamefont {M.~Z.}\ \bibnamefont {Hasan}},\ }\bibfield  {title} {\bibinfo {title} {Topological kagome magnets and superconductors},\ }\href {https://doi.org/10.1038/s41586-022-05516-0} {\bibfield  {journal} {\bibinfo  {journal} {Nature}\ }\textbf {\bibinfo {volume} {612}},\ \bibinfo {pages} {647} (\bibinfo {year} {2022}{\natexlab{a}})}\BibitemShut {NoStop}%
\bibitem [{\citenamefont {Guterding}\ \emph {et~al.}(2016)\citenamefont {Guterding}, \citenamefont {Jeschke},\ and\ \citenamefont {Valent{\'i}}}]{guterding.jeschke.16}%
  \BibitemOpen
  \bibfield  {author} {\bibinfo {author} {\bibfnamefont {D.}~\bibnamefont {Guterding}}, \bibinfo {author} {\bibfnamefont {H.~O.}\ \bibnamefont {Jeschke}},\ and\ \bibinfo {author} {\bibfnamefont {R.}~\bibnamefont {Valent{\'i}}},\ }\bibfield  {title} {\bibinfo {title} {Prospect of quantum anomalous {Hall} and quantum spin {Hall} effect in doped kagome lattice {Mott} insulators},\ }\href {https://doi.org/10.1038/srep25988} {\bibfield  {journal} {\bibinfo  {journal} {Sci. Rep.}\ }\textbf {\bibinfo {volume} {6}},\ \bibinfo {pages} {25988} (\bibinfo {year} {2016})}\BibitemShut {NoStop}%
\bibitem [{\citenamefont {Basak}\ and\ \citenamefont {Ptok}(2022)}]{basak.ptok.22}%
  \BibitemOpen
  \bibfield  {author} {\bibinfo {author} {\bibfnamefont {S.}~\bibnamefont {Basak}}\ and\ \bibinfo {author} {\bibfnamefont {A.}~\bibnamefont {Ptok}},\ }\bibfield  {title} {\bibinfo {title} {Shiba states in systems with density of states singularities},\ }\href {https://doi.org/10.1103/PhysRevB.105.094204} {\bibfield  {journal} {\bibinfo  {journal} {Phys. Rev. B}\ }\textbf {\bibinfo {volume} {105}},\ \bibinfo {pages} {094204} (\bibinfo {year} {2022})}\BibitemShut {NoStop}%
\bibitem [{\citenamefont {Wang}\ \emph {et~al.}(2023)\citenamefont {Wang}, \citenamefont {Wu}, \citenamefont {McCandless}, \citenamefont {Chan},\ and\ \citenamefont {Ali}}]{wang.wu.23}%
  \BibitemOpen
  \bibfield  {author} {\bibinfo {author} {\bibfnamefont {Y.}~\bibnamefont {Wang}}, \bibinfo {author} {\bibfnamefont {H.}~\bibnamefont {Wu}}, \bibinfo {author} {\bibfnamefont {G.~T.}\ \bibnamefont {McCandless}}, \bibinfo {author} {\bibfnamefont {J.~Y.}\ \bibnamefont {Chan}},\ and\ \bibinfo {author} {\bibfnamefont {M.~N.}\ \bibnamefont {Ali}},\ }\bibfield  {title} {\bibinfo {title} {Quantum states and intertwining phases in kagome materials},\ }\href {https://doi.org/10.1038/s42254-023-00635-7} {\bibfield  {journal} {\bibinfo  {journal} {Nat. Rev. Phys.}\ }\textbf {\bibinfo {volume} {5}},\ \bibinfo {pages} {635} (\bibinfo {year} {2023})}\BibitemShut {NoStop}%
\bibitem [{\citenamefont {Tanaka}\ and\ \citenamefont {Ueda}(2003)}]{tanaka.ueda.03}%
  \BibitemOpen
  \bibfield  {author} {\bibinfo {author} {\bibfnamefont {A.}~\bibnamefont {Tanaka}}\ and\ \bibinfo {author} {\bibfnamefont {H.}~\bibnamefont {Ueda}},\ }\bibfield  {title} {\bibinfo {title} {Stability of ferromagnetism in the hubbard model on the kagome lattice},\ }\href {https://doi.org/10.1103/PhysRevLett.90.067204} {\bibfield  {journal} {\bibinfo  {journal} {Phys. Rev. Lett.}\ }\textbf {\bibinfo {volume} {90}},\ \bibinfo {pages} {067204} (\bibinfo {year} {2003})}\BibitemShut {NoStop}%
\bibitem [{\citenamefont {Pollmann}\ \emph {et~al.}(2008)\citenamefont {Pollmann}, \citenamefont {Fulde},\ and\ \citenamefont {Shtengel}}]{pollmann.fulde.08}%
  \BibitemOpen
  \bibfield  {author} {\bibinfo {author} {\bibfnamefont {F.}~\bibnamefont {Pollmann}}, \bibinfo {author} {\bibfnamefont {P.}~\bibnamefont {Fulde}},\ and\ \bibinfo {author} {\bibfnamefont {K.}~\bibnamefont {Shtengel}},\ }\bibfield  {title} {\bibinfo {title} {Kinetic ferromagnetism on a kagome lattice},\ }\href {https://doi.org/10.1103/PhysRevLett.100.136404} {\bibfield  {journal} {\bibinfo  {journal} {Phys. Rev. Lett.}\ }\textbf {\bibinfo {volume} {100}},\ \bibinfo {pages} {136404} (\bibinfo {year} {2008})}\BibitemShut {NoStop}%
\bibitem [{\citenamefont {Kiesel}\ \emph {et~al.}(2013)\citenamefont {Kiesel}, \citenamefont {Platt},\ and\ \citenamefont {Thomale}}]{kiesel.platt.13}%
  \BibitemOpen
  \bibfield  {author} {\bibinfo {author} {\bibfnamefont {M.~L.}\ \bibnamefont {Kiesel}}, \bibinfo {author} {\bibfnamefont {C.}~\bibnamefont {Platt}},\ and\ \bibinfo {author} {\bibfnamefont {R.}~\bibnamefont {Thomale}},\ }\bibfield  {title} {\bibinfo {title} {Unconventional {Fermi} surface instabilities in the kagome {Hubbard} model},\ }\href {https://doi.org/10.1103/PhysRevLett.110.126405} {\bibfield  {journal} {\bibinfo  {journal} {Phys. Rev. Lett.}\ }\textbf {\bibinfo {volume} {110}},\ \bibinfo {pages} {126405} (\bibinfo {year} {2013})}\BibitemShut {NoStop}%
\bibitem [{\citenamefont {Wang}\ \emph {et~al.}(2013)\citenamefont {Wang}, \citenamefont {Li}, \citenamefont {Xiang},\ and\ \citenamefont {Wang}}]{wang.li.13}%
  \BibitemOpen
  \bibfield  {author} {\bibinfo {author} {\bibfnamefont {W.-S.}\ \bibnamefont {Wang}}, \bibinfo {author} {\bibfnamefont {Z.-Z.}\ \bibnamefont {Li}}, \bibinfo {author} {\bibfnamefont {Y.-Y.}\ \bibnamefont {Xiang}},\ and\ \bibinfo {author} {\bibfnamefont {Q.-H.}\ \bibnamefont {Wang}},\ }\bibfield  {title} {\bibinfo {title} {Competing electronic orders on kagome lattices at van {Hove} filling},\ }\href {https://doi.org/10.1103/PhysRevB.87.115135} {\bibfield  {journal} {\bibinfo  {journal} {Phys. Rev. B}\ }\textbf {\bibinfo {volume} {87}},\ \bibinfo {pages} {115135} (\bibinfo {year} {2013})}\BibitemShut {NoStop}%
\bibitem [{\citenamefont {Guo}\ and\ \citenamefont {Franz}(2009)}]{guo.franz.09}%
  \BibitemOpen
  \bibfield  {author} {\bibinfo {author} {\bibfnamefont {H.-M.}\ \bibnamefont {Guo}}\ and\ \bibinfo {author} {\bibfnamefont {M.}~\bibnamefont {Franz}},\ }\bibfield  {title} {\bibinfo {title} {Topological insulator on the kagome lattice},\ }\href {https://doi.org/10.1103/PhysRevB.80.113102} {\bibfield  {journal} {\bibinfo  {journal} {Phys. Rev. B}\ }\textbf {\bibinfo {volume} {80}},\ \bibinfo {pages} {113102} (\bibinfo {year} {2009})}\BibitemShut {NoStop}%
\bibitem [{\citenamefont {Ortiz}\ \emph {et~al.}(2020)\citenamefont {Ortiz}, \citenamefont {Teicher}, \citenamefont {Hu}, \citenamefont {Zuo}, \citenamefont {Sarte}, \citenamefont {Schueller}, \citenamefont {Abeykoon}, \citenamefont {Krogstad}, \citenamefont {Rosenkranz}, \citenamefont {Osborn}, \citenamefont {Seshadri}, \citenamefont {Balents}, \citenamefont {He},\ and\ \citenamefont {Wilson}}]{ortiz.teicher.20}%
  \BibitemOpen
  \bibfield  {author} {\bibinfo {author} {\bibfnamefont {B.~R.}\ \bibnamefont {Ortiz}}, \bibinfo {author} {\bibfnamefont {S.~M.~L.}\ \bibnamefont {Teicher}}, \bibinfo {author} {\bibfnamefont {Y.}~\bibnamefont {Hu}}, \bibinfo {author} {\bibfnamefont {J.~L.}\ \bibnamefont {Zuo}}, \bibinfo {author} {\bibfnamefont {P.~M.}\ \bibnamefont {Sarte}}, \bibinfo {author} {\bibfnamefont {E.~C.}\ \bibnamefont {Schueller}}, \bibinfo {author} {\bibfnamefont {A.~M.~M.}\ \bibnamefont {Abeykoon}}, \bibinfo {author} {\bibfnamefont {M.~J.}\ \bibnamefont {Krogstad}}, \bibinfo {author} {\bibfnamefont {S.}~\bibnamefont {Rosenkranz}}, \bibinfo {author} {\bibfnamefont {R.}~\bibnamefont {Osborn}}, \bibinfo {author} {\bibfnamefont {R.}~\bibnamefont {Seshadri}}, \bibinfo {author} {\bibfnamefont {L.}~\bibnamefont {Balents}}, \bibinfo {author} {\bibfnamefont {J.}~\bibnamefont {He}},\ and\ \bibinfo {author} {\bibfnamefont {S.~D.}\ \bibnamefont {Wilson}},\ }\bibfield  {title} {\bibinfo {title} {{CsV$_{3}$Sb$_{5}$}: A {$\mathbb{Z}_{2}$}
  topological kagome metal with a superconducting ground state},\ }\href {https://doi.org/10.1103/PhysRevLett.125.247002} {\bibfield  {journal} {\bibinfo  {journal} {Phys. Rev. Lett.}\ }\textbf {\bibinfo {volume} {125}},\ \bibinfo {pages} {247002} (\bibinfo {year} {2020})}\BibitemShut {NoStop}%
\bibitem [{\citenamefont {Hu}\ \emph {et~al.}(2022)\citenamefont {Hu}, \citenamefont {Wu}, \citenamefont {Yang}, \citenamefont {Gao}, \citenamefont {Plumb}, \citenamefont {Schnyder}, \citenamefont {Xie}, \citenamefont {Ma},\ and\ \citenamefont {Shi}}]{hu.wu.22}%
  \BibitemOpen
  \bibfield  {author} {\bibinfo {author} {\bibfnamefont {Y.}~\bibnamefont {Hu}}, \bibinfo {author} {\bibfnamefont {X.}~\bibnamefont {Wu}}, \bibinfo {author} {\bibfnamefont {Y.}~\bibnamefont {Yang}}, \bibinfo {author} {\bibfnamefont {S.}~\bibnamefont {Gao}}, \bibinfo {author} {\bibfnamefont {N.~C.}\ \bibnamefont {Plumb}}, \bibinfo {author} {\bibfnamefont {A.~P.}\ \bibnamefont {Schnyder}}, \bibinfo {author} {\bibfnamefont {W.}~\bibnamefont {Xie}}, \bibinfo {author} {\bibfnamefont {J.}~\bibnamefont {Ma}},\ and\ \bibinfo {author} {\bibfnamefont {M.}~\bibnamefont {Shi}},\ }\bibfield  {title} {\bibinfo {title} {Tunable topological {Dirac} surface states and van {Hove} singularities in kagome metal {GdV$_{6}$Sn$_{6}$}},\ }\href {https://doi.org/10.1126/sciadv.add2024} {\bibfield  {journal} {\bibinfo  {journal} {Sci. Adv.}\ }\textbf {\bibinfo {volume} {8}},\ \bibinfo {pages} {eadd2024} (\bibinfo {year} {2022})}\BibitemShut {NoStop}%
\bibitem [{\citenamefont {Chisnell}\ \emph {et~al.}(2015)\citenamefont {Chisnell}, \citenamefont {Helton}, \citenamefont {Freedman}, \citenamefont {Singh}, \citenamefont {Bewley}, \citenamefont {Nocera},\ and\ \citenamefont {Lee}}]{chisnell.helton.15}%
  \BibitemOpen
  \bibfield  {author} {\bibinfo {author} {\bibfnamefont {R.}~\bibnamefont {Chisnell}}, \bibinfo {author} {\bibfnamefont {J.~S.}\ \bibnamefont {Helton}}, \bibinfo {author} {\bibfnamefont {D.~E.}\ \bibnamefont {Freedman}}, \bibinfo {author} {\bibfnamefont {D.~K.}\ \bibnamefont {Singh}}, \bibinfo {author} {\bibfnamefont {R.~I.}\ \bibnamefont {Bewley}}, \bibinfo {author} {\bibfnamefont {D.~G.}\ \bibnamefont {Nocera}},\ and\ \bibinfo {author} {\bibfnamefont {Y.~S.}\ \bibnamefont {Lee}},\ }\bibfield  {title} {\bibinfo {title} {Topological magnon bands in a kagome lattice ferromagnet},\ }\href {https://doi.org/10.1103/PhysRevLett.115.147201} {\bibfield  {journal} {\bibinfo  {journal} {Phys. Rev. Lett.}\ }\textbf {\bibinfo {volume} {115}},\ \bibinfo {pages} {147201} (\bibinfo {year} {2015})}\BibitemShut {NoStop}%
\bibitem [{\citenamefont {Ni}\ \emph {et~al.}(2017)\citenamefont {Ni}, \citenamefont {Gorlach}, \citenamefont {Alu},\ and\ \citenamefont {Khanikaev}}]{ni.gorlach.17}%
  \BibitemOpen
  \bibfield  {author} {\bibinfo {author} {\bibfnamefont {X.}~\bibnamefont {Ni}}, \bibinfo {author} {\bibfnamefont {M.~A.}\ \bibnamefont {Gorlach}}, \bibinfo {author} {\bibfnamefont {A.}~\bibnamefont {Alu}},\ and\ \bibinfo {author} {\bibfnamefont {A.~B.}\ \bibnamefont {Khanikaev}},\ }\bibfield  {title} {\bibinfo {title} {Topological edge states in acoustic kagome lattices},\ }\href {https://doi.org/10.1088/1367-2630/aa6996} {\bibfield  {journal} {\bibinfo  {journal} {New J. Phys.}\ }\textbf {\bibinfo {volume} {19}},\ \bibinfo {pages} {055002} (\bibinfo {year} {2017})}\BibitemShut {NoStop}%
\bibitem [{\citenamefont {Chen}\ \emph {et~al.}(2018)\citenamefont {Chen}, \citenamefont {Nassar}, \citenamefont {Norris}, \citenamefont {Hu},\ and\ \citenamefont {Huang}}]{chen.nassar.18}%
  \BibitemOpen
  \bibfield  {author} {\bibinfo {author} {\bibfnamefont {H.}~\bibnamefont {Chen}}, \bibinfo {author} {\bibfnamefont {H.}~\bibnamefont {Nassar}}, \bibinfo {author} {\bibfnamefont {A.~N.}\ \bibnamefont {Norris}}, \bibinfo {author} {\bibfnamefont {G.~K.}\ \bibnamefont {Hu}},\ and\ \bibinfo {author} {\bibfnamefont {G.~L.}\ \bibnamefont {Huang}},\ }\bibfield  {title} {\bibinfo {title} {Elastic quantum spin {Hall} effect in kagome lattices},\ }\href {https://doi.org/10.1103/PhysRevB.98.094302} {\bibfield  {journal} {\bibinfo  {journal} {Phys. Rev. B}\ }\textbf {\bibinfo {volume} {98}},\ \bibinfo {pages} {094302} (\bibinfo {year} {2018})}\BibitemShut {NoStop}%
\bibitem [{\citenamefont {Xue}\ \emph {et~al.}(2019)\citenamefont {Xue}, \citenamefont {Yang}, \citenamefont {Gao}, \citenamefont {Chong},\ and\ \citenamefont {Zhang}}]{xue.yang.19}%
  \BibitemOpen
  \bibfield  {author} {\bibinfo {author} {\bibfnamefont {H.}~\bibnamefont {Xue}}, \bibinfo {author} {\bibfnamefont {Y.}~\bibnamefont {Yang}}, \bibinfo {author} {\bibfnamefont {F.}~\bibnamefont {Gao}}, \bibinfo {author} {\bibfnamefont {Y.}~\bibnamefont {Chong}},\ and\ \bibinfo {author} {\bibfnamefont {B.}~\bibnamefont {Zhang}},\ }\bibfield  {title} {\bibinfo {title} {Acoustic higher-order topological insulator on a kagome lattice},\ }\href {https://doi.org/10.1038/s41563-018-0251-x} {\bibfield  {journal} {\bibinfo  {journal} {Nature Mater.}\ }\textbf {\bibinfo {volume} {18}},\ \bibinfo {pages} {108} (\bibinfo {year} {2019})}\BibitemShut {NoStop}%
\bibitem [{\citenamefont {Xing}\ \emph {et~al.}(2022)\citenamefont {Xing}, \citenamefont {Chen}, \citenamefont {Xu}, \citenamefont {Li},\ and\ \citenamefont {Zhang}}]{xing.chen.22}%
  \BibitemOpen
  \bibfield  {author} {\bibinfo {author} {\bibfnamefont {Y.}~\bibnamefont {Xing}}, \bibinfo {author} {\bibfnamefont {H.}~\bibnamefont {Chen}}, \bibinfo {author} {\bibfnamefont {N.}~\bibnamefont {Xu}}, \bibinfo {author} {\bibfnamefont {X.}~\bibnamefont {Li}},\ and\ \bibinfo {author} {\bibfnamefont {L.}~\bibnamefont {Zhang}},\ }\bibfield  {title} {\bibinfo {title} {Valley modulation and single-edge transport of magnons in breathing kagome ferromagnets},\ }\href {https://doi.org/10.1103/PhysRevB.105.104409} {\bibfield  {journal} {\bibinfo  {journal} {Phys. Rev. B}\ }\textbf {\bibinfo {volume} {105}},\ \bibinfo {pages} {104409} (\bibinfo {year} {2022})}\BibitemShut {NoStop}%
\bibitem [{\citenamefont {He}\ \emph {et~al.}(2023)\citenamefont {He}, \citenamefont {Gao}, \citenamefont {He}, \citenamefont {Qi}, \citenamefont {Si}, \citenamefont {Yang},\ and\ \citenamefont {Zhou}}]{he.gao.23}%
  \BibitemOpen
  \bibfield  {author} {\bibinfo {author} {\bibfnamefont {Y.-H.}\ \bibnamefont {He}}, \bibinfo {author} {\bibfnamefont {Y.-F.}\ \bibnamefont {Gao}}, \bibinfo {author} {\bibfnamefont {Y.}~\bibnamefont {He}}, \bibinfo {author} {\bibfnamefont {X.-F.}\ \bibnamefont {Qi}}, \bibinfo {author} {\bibfnamefont {J.-Q.}\ \bibnamefont {Si}}, \bibinfo {author} {\bibfnamefont {M.}~\bibnamefont {Yang}},\ and\ \bibinfo {author} {\bibfnamefont {S.-Y.}\ \bibnamefont {Zhou}},\ }\bibfield  {title} {\bibinfo {title} {Realization of edge and corner states in photonic crystals with kagome lattices through topological insulator generators},\ }\href {https://doi.org/10.1016/j.optlastec.2023.109196} {\bibfield  {journal} {\bibinfo  {journal} {Optics \& Laser Technology}\ }\textbf {\bibinfo {volume} {161}},\ \bibinfo {pages} {109196} (\bibinfo {year} {2023})}\BibitemShut {NoStop}%
\bibitem [{\citenamefont {Liu}\ \emph {et~al.}(2018)\citenamefont {Liu}, \citenamefont {Sun}, \citenamefont {Kumar}, \citenamefont {Muechler}, \citenamefont {Sun}, \citenamefont {Jiao}, \citenamefont {Yang}, \citenamefont {Liu}, \citenamefont {Liang}, \citenamefont {Xu}, \citenamefont {Kroder}, \citenamefont {S{\"u}{\ss}}, \citenamefont {Borrmann}, \citenamefont {Shekhar}, \citenamefont {Wang}, \citenamefont {Xi}, \citenamefont {Wang}, \citenamefont {Schnelle}, \citenamefont {Wirth}, \citenamefont {Chen}, \citenamefont {Goennenwein},\ and\ \citenamefont {Felser}}]{liu.sun.18}%
  \BibitemOpen
  \bibfield  {author} {\bibinfo {author} {\bibfnamefont {E.}~\bibnamefont {Liu}}, \bibinfo {author} {\bibfnamefont {Y.}~\bibnamefont {Sun}}, \bibinfo {author} {\bibfnamefont {N.}~\bibnamefont {Kumar}}, \bibinfo {author} {\bibfnamefont {L.}~\bibnamefont {Muechler}}, \bibinfo {author} {\bibfnamefont {A.}~\bibnamefont {Sun}}, \bibinfo {author} {\bibfnamefont {L.}~\bibnamefont {Jiao}}, \bibinfo {author} {\bibfnamefont {S.-Y.}\ \bibnamefont {Yang}}, \bibinfo {author} {\bibfnamefont {D.}~\bibnamefont {Liu}}, \bibinfo {author} {\bibfnamefont {A.}~\bibnamefont {Liang}}, \bibinfo {author} {\bibfnamefont {Q.}~\bibnamefont {Xu}}, \bibinfo {author} {\bibfnamefont {J.}~\bibnamefont {Kroder}}, \bibinfo {author} {\bibfnamefont {V.}~\bibnamefont {S{\"u}{\ss}}}, \bibinfo {author} {\bibfnamefont {H.}~\bibnamefont {Borrmann}}, \bibinfo {author} {\bibfnamefont {C.}~\bibnamefont {Shekhar}}, \bibinfo {author} {\bibfnamefont {Z.}~\bibnamefont {Wang}}, \bibinfo {author} {\bibfnamefont {C.}~\bibnamefont {Xi}}, \bibinfo {author}
  {\bibfnamefont {W.}~\bibnamefont {Wang}}, \bibinfo {author} {\bibfnamefont {W.}~\bibnamefont {Schnelle}}, \bibinfo {author} {\bibfnamefont {S.}~\bibnamefont {Wirth}}, \bibinfo {author} {\bibfnamefont {Y.}~\bibnamefont {Chen}}, \bibinfo {author} {\bibfnamefont {S.~T.~B.}\ \bibnamefont {Goennenwein}},\ and\ \bibinfo {author} {\bibfnamefont {C.}~\bibnamefont {Felser}},\ }\bibfield  {title} {\bibinfo {title} {Giant anomalous hall effect in a ferromagnetic kagome-lattice semimetal},\ }\href {https://doi.org/10.1038/s41567-018-0234-5} {\bibfield  {journal} {\bibinfo  {journal} {Nat. Phys.}\ }\textbf {\bibinfo {volume} {14}},\ \bibinfo {pages} {1125} (\bibinfo {year} {2018})}\BibitemShut {NoStop}%
\bibitem [{\citenamefont {Liu}\ \emph {et~al.}(2019)\citenamefont {Liu}, \citenamefont {Liang}, \citenamefont {Liu}, \citenamefont {Xu}, \citenamefont {Li}, \citenamefont {Chen}, \citenamefont {Pei}, \citenamefont {Shi}, \citenamefont {Mo}, \citenamefont {Dudin}, \citenamefont {Kim}, \citenamefont {Cacho}, \citenamefont {Li}, \citenamefont {Sun}, \citenamefont {Yang}, \citenamefont {Liu}, \citenamefont {Parkin}, \citenamefont {Felser},\ and\ \citenamefont {Chen}}]{liu.liang.19}%
  \BibitemOpen
  \bibfield  {author} {\bibinfo {author} {\bibfnamefont {D.~F.}\ \bibnamefont {Liu}}, \bibinfo {author} {\bibfnamefont {A.~J.}\ \bibnamefont {Liang}}, \bibinfo {author} {\bibfnamefont {E.~K.}\ \bibnamefont {Liu}}, \bibinfo {author} {\bibfnamefont {Q.~N.}\ \bibnamefont {Xu}}, \bibinfo {author} {\bibfnamefont {Y.~W.}\ \bibnamefont {Li}}, \bibinfo {author} {\bibfnamefont {C.}~\bibnamefont {Chen}}, \bibinfo {author} {\bibfnamefont {D.}~\bibnamefont {Pei}}, \bibinfo {author} {\bibfnamefont {W.~J.}\ \bibnamefont {Shi}}, \bibinfo {author} {\bibfnamefont {S.~K.}\ \bibnamefont {Mo}}, \bibinfo {author} {\bibfnamefont {P.}~\bibnamefont {Dudin}}, \bibinfo {author} {\bibfnamefont {T.}~\bibnamefont {Kim}}, \bibinfo {author} {\bibfnamefont {C.}~\bibnamefont {Cacho}}, \bibinfo {author} {\bibfnamefont {G.}~\bibnamefont {Li}}, \bibinfo {author} {\bibfnamefont {Y.}~\bibnamefont {Sun}}, \bibinfo {author} {\bibfnamefont {L.~X.}\ \bibnamefont {Yang}}, \bibinfo {author} {\bibfnamefont {Z.~K.}\ \bibnamefont {Liu}}, \bibinfo {author}
  {\bibfnamefont {S.~S.~P.}\ \bibnamefont {Parkin}}, \bibinfo {author} {\bibfnamefont {C.}~\bibnamefont {Felser}},\ and\ \bibinfo {author} {\bibfnamefont {Y.~L.}\ \bibnamefont {Chen}},\ }\bibfield  {title} {\bibinfo {title} {Magnetic {Weyl} semimetal phase in a kagom\'{e} crystal},\ }\href {https://doi.org/10.1126/science.aav2873} {\bibfield  {journal} {\bibinfo  {journal} {Science}\ }\textbf {\bibinfo {volume} {365}},\ \bibinfo {pages} {1282} (\bibinfo {year} {2019})}\BibitemShut {NoStop}%
\bibitem [{\citenamefont {Morali}\ \emph {et~al.}(2019)\citenamefont {Morali}, \citenamefont {Batabyal}, \citenamefont {Nag}, \citenamefont {Liu}, \citenamefont {Xu}, \citenamefont {Sun}, \citenamefont {Yan}, \citenamefont {Felser}, \citenamefont {Avraham},\ and\ \citenamefont {Beidenkopf}}]{morali.batabyal.19}%
  \BibitemOpen
  \bibfield  {author} {\bibinfo {author} {\bibfnamefont {N.}~\bibnamefont {Morali}}, \bibinfo {author} {\bibfnamefont {R.}~\bibnamefont {Batabyal}}, \bibinfo {author} {\bibfnamefont {P.~K.}\ \bibnamefont {Nag}}, \bibinfo {author} {\bibfnamefont {E.}~\bibnamefont {Liu}}, \bibinfo {author} {\bibfnamefont {Q.}~\bibnamefont {Xu}}, \bibinfo {author} {\bibfnamefont {Y.}~\bibnamefont {Sun}}, \bibinfo {author} {\bibfnamefont {B.}~\bibnamefont {Yan}}, \bibinfo {author} {\bibfnamefont {C.}~\bibnamefont {Felser}}, \bibinfo {author} {\bibfnamefont {N.}~\bibnamefont {Avraham}},\ and\ \bibinfo {author} {\bibfnamefont {H.}~\bibnamefont {Beidenkopf}},\ }\bibfield  {title} {\bibinfo {title} {Fermi-arc diversity on surface terminations of the magnetic {Weyl} semimetal {Co$_{3}$Sn$_{2}$S$_{2}$}},\ }\href {https://doi.org/10.1126/science.aav2334} {\bibfield  {journal} {\bibinfo  {journal} {Science}\ }\textbf {\bibinfo {volume} {365}},\ \bibinfo {pages} {1286} (\bibinfo {year} {2019})}\BibitemShut {NoStop}%
\bibitem [{\citenamefont {Yin}\ \emph {et~al.}(2019)\citenamefont {Yin}, \citenamefont {Zhang}, \citenamefont {Chang}, \citenamefont {Wang}, \citenamefont {Tsirkin}, \citenamefont {Guguchia}, \citenamefont {Lian}, \citenamefont {Zhou}, \citenamefont {Jiang}, \citenamefont {Belopolski}, \citenamefont {Shumiya}, \citenamefont {Multer}, \citenamefont {Litskevich}, \citenamefont {Cochran}, \citenamefont {Lin}, \citenamefont {Wang}, \citenamefont {Neupert}, \citenamefont {Jia}, \citenamefont {Lei},\ and\ \citenamefont {Hasan}}]{yin.zhang.19}%
  \BibitemOpen
  \bibfield  {author} {\bibinfo {author} {\bibfnamefont {J.-X.}\ \bibnamefont {Yin}}, \bibinfo {author} {\bibfnamefont {S.~S.}\ \bibnamefont {Zhang}}, \bibinfo {author} {\bibfnamefont {G.}~\bibnamefont {Chang}}, \bibinfo {author} {\bibfnamefont {Q.}~\bibnamefont {Wang}}, \bibinfo {author} {\bibfnamefont {S.~S.}\ \bibnamefont {Tsirkin}}, \bibinfo {author} {\bibfnamefont {Z.}~\bibnamefont {Guguchia}}, \bibinfo {author} {\bibfnamefont {B.}~\bibnamefont {Lian}}, \bibinfo {author} {\bibfnamefont {H.}~\bibnamefont {Zhou}}, \bibinfo {author} {\bibfnamefont {K.}~\bibnamefont {Jiang}}, \bibinfo {author} {\bibfnamefont {I.}~\bibnamefont {Belopolski}}, \bibinfo {author} {\bibfnamefont {N.}~\bibnamefont {Shumiya}}, \bibinfo {author} {\bibfnamefont {D.}~\bibnamefont {Multer}}, \bibinfo {author} {\bibfnamefont {M.}~\bibnamefont {Litskevich}}, \bibinfo {author} {\bibfnamefont {T.~A.}\ \bibnamefont {Cochran}}, \bibinfo {author} {\bibfnamefont {H.}~\bibnamefont {Lin}}, \bibinfo {author} {\bibfnamefont {Z.}~\bibnamefont
  {Wang}}, \bibinfo {author} {\bibfnamefont {T.}~\bibnamefont {Neupert}}, \bibinfo {author} {\bibfnamefont {S.}~\bibnamefont {Jia}}, \bibinfo {author} {\bibfnamefont {H.}~\bibnamefont {Lei}},\ and\ \bibinfo {author} {\bibfnamefont {M.~Z.}\ \bibnamefont {Hasan}},\ }\bibfield  {title} {\bibinfo {title} {Negative flat band magnetism in a spin--orbit-coupled correlated kagome magnet},\ }\href {https://doi.org/10.1038/s41567-019-0426-7} {\bibfield  {journal} {\bibinfo  {journal} {Nat. Phys.}\ }\textbf {\bibinfo {volume} {15}},\ \bibinfo {pages} {443} (\bibinfo {year} {2019})}\BibitemShut {NoStop}%
\bibitem [{\citenamefont {Xu}\ \emph {et~al.}(2020)\citenamefont {Xu}, \citenamefont {Zhao}, \citenamefont {Yi}, \citenamefont {Wang}, \citenamefont {Yin}, \citenamefont {Wang}, \citenamefont {Hu}, \citenamefont {Wang}, \citenamefont {Liu}, \citenamefont {Xu}, \citenamefont {Lu}, \citenamefont {Soluyanov}, \citenamefont {Lei}, \citenamefont {Shi}, \citenamefont {Luo},\ and\ \citenamefont {Chen}}]{xu.zhao.20}%
  \BibitemOpen
  \bibfield  {author} {\bibinfo {author} {\bibfnamefont {Y.}~\bibnamefont {Xu}}, \bibinfo {author} {\bibfnamefont {J.}~\bibnamefont {Zhao}}, \bibinfo {author} {\bibfnamefont {C.}~\bibnamefont {Yi}}, \bibinfo {author} {\bibfnamefont {Q.}~\bibnamefont {Wang}}, \bibinfo {author} {\bibfnamefont {Q.}~\bibnamefont {Yin}}, \bibinfo {author} {\bibfnamefont {Y.}~\bibnamefont {Wang}}, \bibinfo {author} {\bibfnamefont {X.}~\bibnamefont {Hu}}, \bibinfo {author} {\bibfnamefont {L.}~\bibnamefont {Wang}}, \bibinfo {author} {\bibfnamefont {E.}~\bibnamefont {Liu}}, \bibinfo {author} {\bibfnamefont {G.}~\bibnamefont {Xu}}, \bibinfo {author} {\bibfnamefont {L.}~\bibnamefont {Lu}}, \bibinfo {author} {\bibfnamefont {A.~A.}\ \bibnamefont {Soluyanov}}, \bibinfo {author} {\bibfnamefont {H.}~\bibnamefont {Lei}}, \bibinfo {author} {\bibfnamefont {Y.}~\bibnamefont {Shi}}, \bibinfo {author} {\bibfnamefont {J.}~\bibnamefont {Luo}},\ and\ \bibinfo {author} {\bibfnamefont {Z.-G.}\ \bibnamefont {Chen}},\ }\bibfield  {title} {\bibinfo
  {title} {Electronic correlations and flattened band in magnetic {Weyl} semimetal candidate {Co$_{3}$Sn$_{2}$S$_{2}$}},\ }\href {https://doi.org/10.1038/s41467-020-17234-0} {\bibfield  {journal} {\bibinfo  {journal} {Nat. Commun.}\ }\textbf {\bibinfo {volume} {11}},\ \bibinfo {pages} {3985} (\bibinfo {year} {2020})}\BibitemShut {NoStop}%
\bibitem [{\citenamefont {Kanagaraj}\ \emph {et~al.}(2022)\citenamefont {Kanagaraj}, \citenamefont {Ning},\ and\ \citenamefont {He}}]{kanagaraj.ning.22}%
  \BibitemOpen
  \bibfield  {author} {\bibinfo {author} {\bibfnamefont {M.}~\bibnamefont {Kanagaraj}}, \bibinfo {author} {\bibfnamefont {J.}~\bibnamefont {Ning}},\ and\ \bibinfo {author} {\bibfnamefont {L.}~\bibnamefont {He}},\ }\bibfield  {title} {\bibinfo {title} {Topological {Co$_{3}$Sn$_{2}$S$_{2}$} magnetic {Weyl} semimetal: From fundamental understanding to diverse fields of study},\ }\href {https://doi.org/10.1016/j.revip.2022.100072} {\bibfield  {journal} {\bibinfo  {journal} {Reviews in Physics}\ }\textbf {\bibinfo {volume} {8}},\ \bibinfo {pages} {100072} (\bibinfo {year} {2022})}\BibitemShut {NoStop}%
\bibitem [{\citenamefont {Wang}\ \emph {et~al.}(2018)\citenamefont {Wang}, \citenamefont {Xu}, \citenamefont {Lou}, \citenamefont {Liu}, \citenamefont {Li}, \citenamefont {Huang}, \citenamefont {Shen}, \citenamefont {Weng}, \citenamefont {Wang},\ and\ \citenamefont {Lei}}]{wang.xu.18}%
  \BibitemOpen
  \bibfield  {author} {\bibinfo {author} {\bibfnamefont {Q.}~\bibnamefont {Wang}}, \bibinfo {author} {\bibfnamefont {Y.}~\bibnamefont {Xu}}, \bibinfo {author} {\bibfnamefont {R.}~\bibnamefont {Lou}}, \bibinfo {author} {\bibfnamefont {Z.}~\bibnamefont {Liu}}, \bibinfo {author} {\bibfnamefont {M.}~\bibnamefont {Li}}, \bibinfo {author} {\bibfnamefont {Y.}~\bibnamefont {Huang}}, \bibinfo {author} {\bibfnamefont {D.}~\bibnamefont {Shen}}, \bibinfo {author} {\bibfnamefont {H.}~\bibnamefont {Weng}}, \bibinfo {author} {\bibfnamefont {S.}~\bibnamefont {Wang}},\ and\ \bibinfo {author} {\bibfnamefont {H.}~\bibnamefont {Lei}},\ }\bibfield  {title} {\bibinfo {title} {Large intrinsic anomalous {Hall} effect in half-metallic ferromagnet {Co$_{3}$Sn$_{2}$S$_{2}$} with magnetic {Weyl} fermions},\ }\href {https://doi.org/10.1038/s41467-018-06088-2} {\bibfield  {journal} {\bibinfo  {journal} {Nat. Commun.}\ }\textbf {\bibinfo {volume} {9}},\ \bibinfo {pages} {3681} (\bibinfo {year} {2018})}\BibitemShut {NoStop}%
\bibitem [{\citenamefont {Yang}\ \emph {et~al.}(2020{\natexlab{a}})\citenamefont {Yang}, \citenamefont {Noky}, \citenamefont {Gayles}, \citenamefont {Dejene}, \citenamefont {Sun}, \citenamefont {D{\"o}rr}, \citenamefont {Skourski}, \citenamefont {Felser}, \citenamefont {Ali}, \citenamefont {Liu},\ and\ \citenamefont {Parkin}}]{yang.noky.20}%
  \BibitemOpen
  \bibfield  {author} {\bibinfo {author} {\bibfnamefont {S.-Y.}\ \bibnamefont {Yang}}, \bibinfo {author} {\bibfnamefont {J.}~\bibnamefont {Noky}}, \bibinfo {author} {\bibfnamefont {J.}~\bibnamefont {Gayles}}, \bibinfo {author} {\bibfnamefont {F.~K.}\ \bibnamefont {Dejene}}, \bibinfo {author} {\bibfnamefont {Y.}~\bibnamefont {Sun}}, \bibinfo {author} {\bibfnamefont {M.}~\bibnamefont {D{\"o}rr}}, \bibinfo {author} {\bibfnamefont {Y.}~\bibnamefont {Skourski}}, \bibinfo {author} {\bibfnamefont {C.}~\bibnamefont {Felser}}, \bibinfo {author} {\bibfnamefont {M.~N.}\ \bibnamefont {Ali}}, \bibinfo {author} {\bibfnamefont {E.}~\bibnamefont {Liu}},\ and\ \bibinfo {author} {\bibfnamefont {S.~S.~P.}\ \bibnamefont {Parkin}},\ }\bibfield  {title} {\bibinfo {title} {Field-modulated anomalous {Hall} conductivity and planar hall effect in {Co$_{3}$Sn$_{2}$S$_{2}$} nanoflakes},\ }\href {https://doi.org/10.1021/acs.nanolett.0c02219} {\bibfield  {journal} {\bibinfo  {journal} {Nano Letters}\ }\textbf {\bibinfo {volume} {20}},\
  \bibinfo {pages} {7860} (\bibinfo {year} {2020}{\natexlab{a}})}\BibitemShut {NoStop}%
\bibitem [{\citenamefont {Ortiz}\ \emph {et~al.}(2019)\citenamefont {Ortiz}, \citenamefont {Gomes}, \citenamefont {Morey}, \citenamefont {Winiarski}, \citenamefont {Bordelon}, \citenamefont {Mangum}, \citenamefont {Oswald}, \citenamefont {Rodriguez-Rivera}, \citenamefont {Neilson}, \citenamefont {Wilson}, \citenamefont {Ertekin}, \citenamefont {McQueen},\ and\ \citenamefont {Toberer}}]{ortiz.gomes.19}%
  \BibitemOpen
  \bibfield  {author} {\bibinfo {author} {\bibfnamefont {B.~R.}\ \bibnamefont {Ortiz}}, \bibinfo {author} {\bibfnamefont {L.~C.}\ \bibnamefont {Gomes}}, \bibinfo {author} {\bibfnamefont {J.~R.}\ \bibnamefont {Morey}}, \bibinfo {author} {\bibfnamefont {M.}~\bibnamefont {Winiarski}}, \bibinfo {author} {\bibfnamefont {M.}~\bibnamefont {Bordelon}}, \bibinfo {author} {\bibfnamefont {J.~S.}\ \bibnamefont {Mangum}}, \bibinfo {author} {\bibfnamefont {I.~W.~H.}\ \bibnamefont {Oswald}}, \bibinfo {author} {\bibfnamefont {J.~A.}\ \bibnamefont {Rodriguez-Rivera}}, \bibinfo {author} {\bibfnamefont {J.~R.}\ \bibnamefont {Neilson}}, \bibinfo {author} {\bibfnamefont {S.~D.}\ \bibnamefont {Wilson}}, \bibinfo {author} {\bibfnamefont {E.}~\bibnamefont {Ertekin}}, \bibinfo {author} {\bibfnamefont {T.~M.}\ \bibnamefont {McQueen}},\ and\ \bibinfo {author} {\bibfnamefont {E.~S.}\ \bibnamefont {Toberer}},\ }\bibfield  {title} {\bibinfo {title} {New kagome prototype materials: discovery of {KV$_{3}$Sb$_{5}$}, {RbV$_{3}$Sb$_{5}$}, and
  {CsV$_{3}$Sb$_{5}$}},\ }\href {https://doi.org/10.1103/PhysRevMaterials.3.094407} {\bibfield  {journal} {\bibinfo  {journal} {Phys. Rev. Mater.}\ }\textbf {\bibinfo {volume} {3}},\ \bibinfo {pages} {094407} (\bibinfo {year} {2019})}\BibitemShut {NoStop}%
\bibitem [{\citenamefont {Li}\ \emph {et~al.}(2021{\natexlab{a}})\citenamefont {Li}, \citenamefont {Zhang}, \citenamefont {Yilmaz}, \citenamefont {Pai}, \citenamefont {Marvinney}, \citenamefont {Said}, \citenamefont {Yin}, \citenamefont {Gong}, \citenamefont {Tu}, \citenamefont {Vescovo}, \citenamefont {Nelson}, \citenamefont {Moore}, \citenamefont {Murakami}, \citenamefont {Lei}, \citenamefont {Lee}, \citenamefont {Lawrie},\ and\ \citenamefont {Miao}}]{li.zhabg.21}%
  \BibitemOpen
  \bibfield  {author} {\bibinfo {author} {\bibfnamefont {H.}~\bibnamefont {Li}}, \bibinfo {author} {\bibfnamefont {T.~T.}\ \bibnamefont {Zhang}}, \bibinfo {author} {\bibfnamefont {T.}~\bibnamefont {Yilmaz}}, \bibinfo {author} {\bibfnamefont {Y.~Y.}\ \bibnamefont {Pai}}, \bibinfo {author} {\bibfnamefont {C.~E.}\ \bibnamefont {Marvinney}}, \bibinfo {author} {\bibfnamefont {A.}~\bibnamefont {Said}}, \bibinfo {author} {\bibfnamefont {Q.~W.}\ \bibnamefont {Yin}}, \bibinfo {author} {\bibfnamefont {C.~S.}\ \bibnamefont {Gong}}, \bibinfo {author} {\bibfnamefont {Z.~J.}\ \bibnamefont {Tu}}, \bibinfo {author} {\bibfnamefont {E.}~\bibnamefont {Vescovo}}, \bibinfo {author} {\bibfnamefont {C.~S.}\ \bibnamefont {Nelson}}, \bibinfo {author} {\bibfnamefont {R.~G.}\ \bibnamefont {Moore}}, \bibinfo {author} {\bibfnamefont {S.}~\bibnamefont {Murakami}}, \bibinfo {author} {\bibfnamefont {H.~C.}\ \bibnamefont {Lei}}, \bibinfo {author} {\bibfnamefont {H.~N.}\ \bibnamefont {Lee}}, \bibinfo {author} {\bibfnamefont {B.~J.}\
  \bibnamefont {Lawrie}},\ and\ \bibinfo {author} {\bibfnamefont {H.}~\bibnamefont {Miao}},\ }\bibfield  {title} {\bibinfo {title} {Observation of unconventional charge density wave without acoustic phonon anomaly in kagome superconductors {$A$V$_{3}$Sb$_{5}$} ({$A=$Rb, Cs})},\ }\href {https://doi.org/10.1103/PhysRevX.11.031050} {\bibfield  {journal} {\bibinfo  {journal} {Phys. Rev. X}\ }\textbf {\bibinfo {volume} {11}},\ \bibinfo {pages} {031050} (\bibinfo {year} {2021}{\natexlab{a}})}\BibitemShut {NoStop}%
\bibitem [{\citenamefont {Liang}\ \emph {et~al.}(2021)\citenamefont {Liang}, \citenamefont {Hou}, \citenamefont {Zhang}, \citenamefont {Ma}, \citenamefont {Wu}, \citenamefont {Zhang}, \citenamefont {Yu}, \citenamefont {Ying}, \citenamefont {Jiang}, \citenamefont {Shan}, \citenamefont {Wang},\ and\ \citenamefont {Chen}}]{liang.hou.21}%
  \BibitemOpen
  \bibfield  {author} {\bibinfo {author} {\bibfnamefont {Z.}~\bibnamefont {Liang}}, \bibinfo {author} {\bibfnamefont {X.}~\bibnamefont {Hou}}, \bibinfo {author} {\bibfnamefont {F.}~\bibnamefont {Zhang}}, \bibinfo {author} {\bibfnamefont {W.}~\bibnamefont {Ma}}, \bibinfo {author} {\bibfnamefont {P.}~\bibnamefont {Wu}}, \bibinfo {author} {\bibfnamefont {Z.}~\bibnamefont {Zhang}}, \bibinfo {author} {\bibfnamefont {F.}~\bibnamefont {Yu}}, \bibinfo {author} {\bibfnamefont {J.-J.}\ \bibnamefont {Ying}}, \bibinfo {author} {\bibfnamefont {K.}~\bibnamefont {Jiang}}, \bibinfo {author} {\bibfnamefont {L.}~\bibnamefont {Shan}}, \bibinfo {author} {\bibfnamefont {Z.}~\bibnamefont {Wang}},\ and\ \bibinfo {author} {\bibfnamefont {X.-H.}\ \bibnamefont {Chen}},\ }\bibfield  {title} {\bibinfo {title} {Three-dimensional charge density wave and surface-dependent vortex-core states in a kagome superconductor {CsV$_{3}$Sb$_{5}$}},\ }\href {https://doi.org/10.1103/PhysRevX.11.031026} {\bibfield  {journal} {\bibinfo  {journal} {Phys.
  Rev. X}\ }\textbf {\bibinfo {volume} {11}},\ \bibinfo {pages} {031026} (\bibinfo {year} {2021})}\BibitemShut {NoStop}%
\bibitem [{\citenamefont {Ortiz}\ \emph {et~al.}(2021)\citenamefont {Ortiz}, \citenamefont {Sarte}, \citenamefont {Kenney}, \citenamefont {Graf}, \citenamefont {Teicher}, \citenamefont {Seshadri},\ and\ \citenamefont {Wilson}}]{ortiz.sarite.21}%
  \BibitemOpen
  \bibfield  {author} {\bibinfo {author} {\bibfnamefont {B.~R.}\ \bibnamefont {Ortiz}}, \bibinfo {author} {\bibfnamefont {P.~M.}\ \bibnamefont {Sarte}}, \bibinfo {author} {\bibfnamefont {E.~M.}\ \bibnamefont {Kenney}}, \bibinfo {author} {\bibfnamefont {M.~J.}\ \bibnamefont {Graf}}, \bibinfo {author} {\bibfnamefont {S.~M.~L.}\ \bibnamefont {Teicher}}, \bibinfo {author} {\bibfnamefont {R.}~\bibnamefont {Seshadri}},\ and\ \bibinfo {author} {\bibfnamefont {S.~D.}\ \bibnamefont {Wilson}},\ }\bibfield  {title} {\bibinfo {title} {Superconductivity in the {$\mathbb{Z}_{2}$} kagome metal {KV$_{3}$Sb$_{5}$}},\ }\href {https://doi.org/10.1103/PhysRevMaterials.5.034801} {\bibfield  {journal} {\bibinfo  {journal} {Phys. Rev. Mater.}\ }\textbf {\bibinfo {volume} {5}},\ \bibinfo {pages} {034801} (\bibinfo {year} {2021})}\BibitemShut {NoStop}%
\bibitem [{\citenamefont {Gupta}\ \emph {et~al.}(2022)\citenamefont {Gupta}, \citenamefont {Das}, \citenamefont {Mielke~III}, \citenamefont {Guguchia}, \citenamefont {Shiroka}, \citenamefont {Baines}, \citenamefont {Bartkowiak}, \citenamefont {Luetkens}, \citenamefont {Khasanov}, \citenamefont {Yin}, \citenamefont {Tu}, \citenamefont {Gong},\ and\ \citenamefont {Lei}}]{gupta.das.22}%
  \BibitemOpen
  \bibfield  {author} {\bibinfo {author} {\bibfnamefont {R.}~\bibnamefont {Gupta}}, \bibinfo {author} {\bibfnamefont {D.}~\bibnamefont {Das}}, \bibinfo {author} {\bibfnamefont {C.~H.}\ \bibnamefont {Mielke~III}}, \bibinfo {author} {\bibfnamefont {Z.}~\bibnamefont {Guguchia}}, \bibinfo {author} {\bibfnamefont {T.}~\bibnamefont {Shiroka}}, \bibinfo {author} {\bibfnamefont {C.}~\bibnamefont {Baines}}, \bibinfo {author} {\bibfnamefont {M.}~\bibnamefont {Bartkowiak}}, \bibinfo {author} {\bibfnamefont {H.}~\bibnamefont {Luetkens}}, \bibinfo {author} {\bibfnamefont {R.}~\bibnamefont {Khasanov}}, \bibinfo {author} {\bibfnamefont {Q.}~\bibnamefont {Yin}}, \bibinfo {author} {\bibfnamefont {Z.}~\bibnamefont {Tu}}, \bibinfo {author} {\bibfnamefont {C.}~\bibnamefont {Gong}},\ and\ \bibinfo {author} {\bibfnamefont {H.}~\bibnamefont {Lei}},\ }\bibfield  {title} {\bibinfo {title} {Microscopic evidence for anisotropic multigap superconductivity in the {CsV$_{3}$Sb$_{5}$} kagome superconductor},\ }\href
  {https://doi.org/10.1038/s41535-022-00453-7} {\bibfield  {journal} {\bibinfo  {journal} {npj Quantum Materials}\ }\textbf {\bibinfo {volume} {7}},\ \bibinfo {pages} {49} (\bibinfo {year} {2022})}\BibitemShut {NoStop}%
\bibitem [{\citenamefont {Yang}\ \emph {et~al.}(2020{\natexlab{b}})\citenamefont {Yang}, \citenamefont {Wang}, \citenamefont {Ortiz}, \citenamefont {Liu}, \citenamefont {Gayles}, \citenamefont {Derunova}, \citenamefont {Gonzalez-Hernandez}, \citenamefont {Šmejkal}, \citenamefont {Chen}, \citenamefont {Parkin}, \citenamefont {Wilson}, \citenamefont {Toberer}, \citenamefont {McQueen},\ and\ \citenamefont {Ali}}]{yang.wang.20}%
  \BibitemOpen
  \bibfield  {author} {\bibinfo {author} {\bibfnamefont {S.-Y.}\ \bibnamefont {Yang}}, \bibinfo {author} {\bibfnamefont {Y.}~\bibnamefont {Wang}}, \bibinfo {author} {\bibfnamefont {B.~R.}\ \bibnamefont {Ortiz}}, \bibinfo {author} {\bibfnamefont {D.}~\bibnamefont {Liu}}, \bibinfo {author} {\bibfnamefont {J.}~\bibnamefont {Gayles}}, \bibinfo {author} {\bibfnamefont {E.}~\bibnamefont {Derunova}}, \bibinfo {author} {\bibfnamefont {R.}~\bibnamefont {Gonzalez-Hernandez}}, \bibinfo {author} {\bibfnamefont {L.}~\bibnamefont {Šmejkal}}, \bibinfo {author} {\bibfnamefont {Y.}~\bibnamefont {Chen}}, \bibinfo {author} {\bibfnamefont {S.~S.~P.}\ \bibnamefont {Parkin}}, \bibinfo {author} {\bibfnamefont {S.~D.}\ \bibnamefont {Wilson}}, \bibinfo {author} {\bibfnamefont {E.~S.}\ \bibnamefont {Toberer}}, \bibinfo {author} {\bibfnamefont {T.}~\bibnamefont {McQueen}},\ and\ \bibinfo {author} {\bibfnamefont {M.~N.}\ \bibnamefont {Ali}},\ }\bibfield  {title} {\bibinfo {title} {Giant, unconventional anomalous hall effect in the
  metallic frustrated magnet candidate, {KV$_{3}$Sb$_{5}$}},\ }\href {https://doi.org/10.1126/sciadv.abb6003} {\bibfield  {journal} {\bibinfo  {journal} {Sci. Adv.}\ }\textbf {\bibinfo {volume} {6}},\ \bibinfo {pages} {eabb6003} (\bibinfo {year} {2020}{\natexlab{b}})}\BibitemShut {NoStop}%
\bibitem [{\citenamefont {Yu}\ \emph {et~al.}(2021)\citenamefont {Yu}, \citenamefont {Wu}, \citenamefont {Wang}, \citenamefont {Lei}, \citenamefont {Zhuo}, \citenamefont {Ying},\ and\ \citenamefont {Chen}}]{yu.wu.21}%
  \BibitemOpen
  \bibfield  {author} {\bibinfo {author} {\bibfnamefont {F.~H.}\ \bibnamefont {Yu}}, \bibinfo {author} {\bibfnamefont {T.}~\bibnamefont {Wu}}, \bibinfo {author} {\bibfnamefont {Z.~Y.}\ \bibnamefont {Wang}}, \bibinfo {author} {\bibfnamefont {B.}~\bibnamefont {Lei}}, \bibinfo {author} {\bibfnamefont {W.~Z.}\ \bibnamefont {Zhuo}}, \bibinfo {author} {\bibfnamefont {J.~J.}\ \bibnamefont {Ying}},\ and\ \bibinfo {author} {\bibfnamefont {X.~H.}\ \bibnamefont {Chen}},\ }\bibfield  {title} {\bibinfo {title} {Concurrence of anomalous {Hall} effect and charge density wave in a superconducting topological kagome metal},\ }\href {https://doi.org/10.1103/PhysRevB.104.L041103} {\bibfield  {journal} {\bibinfo  {journal} {Phys. Rev. B}\ }\textbf {\bibinfo {volume} {104}},\ \bibinfo {pages} {L041103} (\bibinfo {year} {2021})}\BibitemShut {NoStop}%
\bibitem [{\citenamefont {Kang}\ \emph {et~al.}(2020{\natexlab{a}})\citenamefont {Kang}, \citenamefont {Ye}, \citenamefont {Fang}, \citenamefont {You}, \citenamefont {Levitan}, \citenamefont {Han}, \citenamefont {Facio}, \citenamefont {Jozwiak}, \citenamefont {Bostwick}, \citenamefont {Rotenberg}, \citenamefont {Chan}, \citenamefont {McDonald}, \citenamefont {Graf}, \citenamefont {Kaznatcheev}, \citenamefont {Vescovo}, \citenamefont {Bell}, \citenamefont {Kaxiras}, \citenamefont {van~den Brink}, \citenamefont {Richter}, \citenamefont {Prasad~Ghimire}, \citenamefont {Checkelsky},\ and\ \citenamefont {Comin}}]{kang.ye.20}%
  \BibitemOpen
  \bibfield  {author} {\bibinfo {author} {\bibfnamefont {M.}~\bibnamefont {Kang}}, \bibinfo {author} {\bibfnamefont {L.}~\bibnamefont {Ye}}, \bibinfo {author} {\bibfnamefont {S.}~\bibnamefont {Fang}}, \bibinfo {author} {\bibfnamefont {J.-S.}\ \bibnamefont {You}}, \bibinfo {author} {\bibfnamefont {A.}~\bibnamefont {Levitan}}, \bibinfo {author} {\bibfnamefont {M.}~\bibnamefont {Han}}, \bibinfo {author} {\bibfnamefont {J.~I.}\ \bibnamefont {Facio}}, \bibinfo {author} {\bibfnamefont {C.}~\bibnamefont {Jozwiak}}, \bibinfo {author} {\bibfnamefont {A.}~\bibnamefont {Bostwick}}, \bibinfo {author} {\bibfnamefont {E.}~\bibnamefont {Rotenberg}}, \bibinfo {author} {\bibfnamefont {M.~K.}\ \bibnamefont {Chan}}, \bibinfo {author} {\bibfnamefont {R.~D.}\ \bibnamefont {McDonald}}, \bibinfo {author} {\bibfnamefont {D.}~\bibnamefont {Graf}}, \bibinfo {author} {\bibfnamefont {K.}~\bibnamefont {Kaznatcheev}}, \bibinfo {author} {\bibfnamefont {E.}~\bibnamefont {Vescovo}}, \bibinfo {author} {\bibfnamefont {D.~C.}\ \bibnamefont
  {Bell}}, \bibinfo {author} {\bibfnamefont {E.}~\bibnamefont {Kaxiras}}, \bibinfo {author} {\bibfnamefont {J.}~\bibnamefont {van~den Brink}}, \bibinfo {author} {\bibfnamefont {M.}~\bibnamefont {Richter}}, \bibinfo {author} {\bibfnamefont {M.}~\bibnamefont {Prasad~Ghimire}}, \bibinfo {author} {\bibfnamefont {J.~G.}\ \bibnamefont {Checkelsky}},\ and\ \bibinfo {author} {\bibfnamefont {R.}~\bibnamefont {Comin}},\ }\bibfield  {title} {\bibinfo {title} {Dirac fermions and flat bands in the ideal kagome metal {FeSn}},\ }\href {https://doi.org/10.1038/s41563-019-0531-0} {\bibfield  {journal} {\bibinfo  {journal} {Nat. Mater.}\ }\textbf {\bibinfo {volume} {19}},\ \bibinfo {pages} {163} (\bibinfo {year} {2020}{\natexlab{a}})}\BibitemShut {NoStop}%
\bibitem [{\citenamefont {Han}\ \emph {et~al.}(2021)\citenamefont {Han}, \citenamefont {Inoue}, \citenamefont {Fang}, \citenamefont {John}, \citenamefont {Ye}, \citenamefont {Chan}, \citenamefont {Graf}, \citenamefont {Suzuki}, \citenamefont {Ghimire}, \citenamefont {Cho}, \citenamefont {Kaxiras},\ and\ \citenamefont {Checkelsky}}]{han.inoue.21}%
  \BibitemOpen
  \bibfield  {author} {\bibinfo {author} {\bibfnamefont {M.}~\bibnamefont {Han}}, \bibinfo {author} {\bibfnamefont {H.}~\bibnamefont {Inoue}}, \bibinfo {author} {\bibfnamefont {S.}~\bibnamefont {Fang}}, \bibinfo {author} {\bibfnamefont {C.}~\bibnamefont {John}}, \bibinfo {author} {\bibfnamefont {L.}~\bibnamefont {Ye}}, \bibinfo {author} {\bibfnamefont {M.~K.}\ \bibnamefont {Chan}}, \bibinfo {author} {\bibfnamefont {D.}~\bibnamefont {Graf}}, \bibinfo {author} {\bibfnamefont {T.}~\bibnamefont {Suzuki}}, \bibinfo {author} {\bibfnamefont {M.~P.}\ \bibnamefont {Ghimire}}, \bibinfo {author} {\bibfnamefont {W.~J.}\ \bibnamefont {Cho}}, \bibinfo {author} {\bibfnamefont {E.}~\bibnamefont {Kaxiras}},\ and\ \bibinfo {author} {\bibfnamefont {J.~G.}\ \bibnamefont {Checkelsky}},\ }\bibfield  {title} {\bibinfo {title} {Evidence of two-dimensional flat band at the surface of antiferromagnetic kagome metal {FeSn}},\ }\href {https://doi.org/10.1038/s41467-021-25705-1} {\bibfield  {journal} {\bibinfo  {journal} {Nat. Commun.}\
  }\textbf {\bibinfo {volume} {12}},\ \bibinfo {pages} {5345} (\bibinfo {year} {2021})}\BibitemShut {NoStop}%
\bibitem [{\citenamefont {Zhang}\ \emph {et~al.}(2023)\citenamefont {Zhang}, \citenamefont {Oli}, \citenamefont {Zou}, \citenamefont {Guo}, \citenamefont {Wang},\ and\ \citenamefont {Li}}]{zhang.oli.23}%
  \BibitemOpen
  \bibfield  {author} {\bibinfo {author} {\bibfnamefont {H.}~\bibnamefont {Zhang}}, \bibinfo {author} {\bibfnamefont {B.~D.}\ \bibnamefont {Oli}}, \bibinfo {author} {\bibfnamefont {Q.}~\bibnamefont {Zou}}, \bibinfo {author} {\bibfnamefont {X.}~\bibnamefont {Guo}}, \bibinfo {author} {\bibfnamefont {Z.}~\bibnamefont {Wang}},\ and\ \bibinfo {author} {\bibfnamefont {L.}~\bibnamefont {Li}},\ }\bibfield  {title} {\bibinfo {title} {Visualizing symmetry-breaking electronic orders in epitaxial kagome magnet {FeSn} films},\ }\href {https://doi.org/10.1038/s41467-023-41831-4} {\bibfield  {journal} {\bibinfo  {journal} {Nat. Commun.}\ }\textbf {\bibinfo {volume} {14}},\ \bibinfo {pages} {6167} (\bibinfo {year} {2023})}\BibitemShut {NoStop}%
\bibitem [{\citenamefont {Meier}\ \emph {et~al.}(2020)\citenamefont {Meier}, \citenamefont {Du}, \citenamefont {Okamoto}, \citenamefont {Mohanta}, \citenamefont {May}, \citenamefont {McGuire}, \citenamefont {Bridges}, \citenamefont {Samolyuk},\ and\ \citenamefont {Sales}}]{meier.du.20}%
  \BibitemOpen
  \bibfield  {author} {\bibinfo {author} {\bibfnamefont {W.~R.}\ \bibnamefont {Meier}}, \bibinfo {author} {\bibfnamefont {M.-H.}\ \bibnamefont {Du}}, \bibinfo {author} {\bibfnamefont {S.}~\bibnamefont {Okamoto}}, \bibinfo {author} {\bibfnamefont {N.}~\bibnamefont {Mohanta}}, \bibinfo {author} {\bibfnamefont {A.~F.}\ \bibnamefont {May}}, \bibinfo {author} {\bibfnamefont {M.~A.}\ \bibnamefont {McGuire}}, \bibinfo {author} {\bibfnamefont {C.~A.}\ \bibnamefont {Bridges}}, \bibinfo {author} {\bibfnamefont {G.~D.}\ \bibnamefont {Samolyuk}},\ and\ \bibinfo {author} {\bibfnamefont {B.~C.}\ \bibnamefont {Sales}},\ }\bibfield  {title} {\bibinfo {title} {Flat bands in the {CoSn}-type compounds},\ }\href {https://doi.org/10.1103/PhysRevB.102.075148} {\bibfield  {journal} {\bibinfo  {journal} {Phys. Rev. B}\ }\textbf {\bibinfo {volume} {102}},\ \bibinfo {pages} {075148} (\bibinfo {year} {2020})}\BibitemShut {NoStop}%
\bibitem [{\citenamefont {Kang}\ \emph {et~al.}(2020{\natexlab{b}})\citenamefont {Kang}, \citenamefont {Fang}, \citenamefont {Ye}, \citenamefont {Po}, \citenamefont {Denlinger}, \citenamefont {Jozwiak}, \citenamefont {Bostwick}, \citenamefont {Rotenberg}, \citenamefont {Kaxiras}, \citenamefont {Checkelsky},\ and\ \citenamefont {Comin}}]{kang.fang.20}%
  \BibitemOpen
  \bibfield  {author} {\bibinfo {author} {\bibfnamefont {M.}~\bibnamefont {Kang}}, \bibinfo {author} {\bibfnamefont {S.}~\bibnamefont {Fang}}, \bibinfo {author} {\bibfnamefont {L.}~\bibnamefont {Ye}}, \bibinfo {author} {\bibfnamefont {H.~C.}\ \bibnamefont {Po}}, \bibinfo {author} {\bibfnamefont {J.}~\bibnamefont {Denlinger}}, \bibinfo {author} {\bibfnamefont {C.}~\bibnamefont {Jozwiak}}, \bibinfo {author} {\bibfnamefont {A.}~\bibnamefont {Bostwick}}, \bibinfo {author} {\bibfnamefont {E.}~\bibnamefont {Rotenberg}}, \bibinfo {author} {\bibfnamefont {E.}~\bibnamefont {Kaxiras}}, \bibinfo {author} {\bibfnamefont {J.~G.}\ \bibnamefont {Checkelsky}},\ and\ \bibinfo {author} {\bibfnamefont {R.}~\bibnamefont {Comin}},\ }\bibfield  {title} {\bibinfo {title} {Topological flat bands in frustrated kagome lattice {CoSn}},\ }\href {https://doi.org/10.1038/s41467-020-17465-1} {\bibfield  {journal} {\bibinfo  {journal} {Nat. Commun.}\ }\textbf {\bibinfo {volume} {11}},\ \bibinfo {pages} {4004} (\bibinfo {year}
  {2020}{\natexlab{b}})}\BibitemShut {NoStop}%
\bibitem [{\citenamefont {Liu}\ \emph {et~al.}(2020)\citenamefont {Liu}, \citenamefont {Li}, \citenamefont {Wang}, \citenamefont {Wang}, \citenamefont {Wen}, \citenamefont {Jiang}, \citenamefont {Lu}, \citenamefont {Yan}, \citenamefont {Huang}, \citenamefont {Shen}, \citenamefont {Yin}, \citenamefont {Wang}, \citenamefont {Yin}, \citenamefont {Lei},\ and\ \citenamefont {Wang}}]{liu.li.20}%
  \BibitemOpen
  \bibfield  {author} {\bibinfo {author} {\bibfnamefont {Z.}~\bibnamefont {Liu}}, \bibinfo {author} {\bibfnamefont {M.}~\bibnamefont {Li}}, \bibinfo {author} {\bibfnamefont {Q.}~\bibnamefont {Wang}}, \bibinfo {author} {\bibfnamefont {G.}~\bibnamefont {Wang}}, \bibinfo {author} {\bibfnamefont {C.}~\bibnamefont {Wen}}, \bibinfo {author} {\bibfnamefont {K.}~\bibnamefont {Jiang}}, \bibinfo {author} {\bibfnamefont {X.}~\bibnamefont {Lu}}, \bibinfo {author} {\bibfnamefont {S.}~\bibnamefont {Yan}}, \bibinfo {author} {\bibfnamefont {Y.}~\bibnamefont {Huang}}, \bibinfo {author} {\bibfnamefont {D.}~\bibnamefont {Shen}}, \bibinfo {author} {\bibfnamefont {J.-X.}\ \bibnamefont {Yin}}, \bibinfo {author} {\bibfnamefont {Z.}~\bibnamefont {Wang}}, \bibinfo {author} {\bibfnamefont {Z.}~\bibnamefont {Yin}}, \bibinfo {author} {\bibfnamefont {H.}~\bibnamefont {Lei}},\ and\ \bibinfo {author} {\bibfnamefont {S.}~\bibnamefont {Wang}},\ }\bibfield  {title} {\bibinfo {title} {Orbital-selective {Dirac} fermions and extremely flat bands
  in frustrated kagome-lattice metal {CoSn}},\ }\href {https://doi.org/10.1038/s41467-020-17462-4} {\bibfield  {journal} {\bibinfo  {journal} {Nat. Commun.}\ }\textbf {\bibinfo {volume} {11}},\ \bibinfo {pages} {4002} (\bibinfo {year} {2020})}\BibitemShut {NoStop}%
\bibitem [{\citenamefont {Ye}\ \emph {et~al.}(2018)\citenamefont {Ye}, \citenamefont {Kang}, \citenamefont {Liu}, \citenamefont {von Cube}, \citenamefont {Wicker}, \citenamefont {Suzuki}, \citenamefont {Jozwiak}, \citenamefont {Bostwick}, \citenamefont {Rotenberg}, \citenamefont {Bell}, \citenamefont {Fu}, \citenamefont {Comin},\ and\ \citenamefont {Checkelsky}}]{ye.kang.18}%
  \BibitemOpen
  \bibfield  {author} {\bibinfo {author} {\bibfnamefont {L.}~\bibnamefont {Ye}}, \bibinfo {author} {\bibfnamefont {M.}~\bibnamefont {Kang}}, \bibinfo {author} {\bibfnamefont {J.}~\bibnamefont {Liu}}, \bibinfo {author} {\bibfnamefont {F.}~\bibnamefont {von Cube}}, \bibinfo {author} {\bibfnamefont {C.~R.}\ \bibnamefont {Wicker}}, \bibinfo {author} {\bibfnamefont {T.}~\bibnamefont {Suzuki}}, \bibinfo {author} {\bibfnamefont {C.}~\bibnamefont {Jozwiak}}, \bibinfo {author} {\bibfnamefont {A.}~\bibnamefont {Bostwick}}, \bibinfo {author} {\bibfnamefont {E.}~\bibnamefont {Rotenberg}}, \bibinfo {author} {\bibfnamefont {D.~C.}\ \bibnamefont {Bell}}, \bibinfo {author} {\bibfnamefont {L.}~\bibnamefont {Fu}}, \bibinfo {author} {\bibfnamefont {R.}~\bibnamefont {Comin}},\ and\ \bibinfo {author} {\bibfnamefont {J.~G.}\ \bibnamefont {Checkelsky}},\ }\bibfield  {title} {\bibinfo {title} {Massive {Dirac} fermions in a ferromagnetic kagome metal},\ }\href {https://doi.org/10.1038/nature25987} {\bibfield  {journal} {\bibinfo
  {journal} {Nature}\ }\textbf {\bibinfo {volume} {555}},\ \bibinfo {pages} {638} (\bibinfo {year} {2018})}\BibitemShut {NoStop}%
\bibitem [{\citenamefont {Lin}\ \emph {et~al.}(2018)\citenamefont {Lin}, \citenamefont {Choi}, \citenamefont {Zhang}, \citenamefont {Qin}, \citenamefont {Yi}, \citenamefont {Wang}, \citenamefont {Li}, \citenamefont {Wang}, \citenamefont {Zhang}, \citenamefont {Sun}, \citenamefont {Wei}, \citenamefont {Zhang}, \citenamefont {Guo}, \citenamefont {Lu}, \citenamefont {Cho}, \citenamefont {Zeng},\ and\ \citenamefont {Zhang}}]{lin.choi.18}%
  \BibitemOpen
  \bibfield  {author} {\bibinfo {author} {\bibfnamefont {Z.}~\bibnamefont {Lin}}, \bibinfo {author} {\bibfnamefont {J.-H.}\ \bibnamefont {Choi}}, \bibinfo {author} {\bibfnamefont {Q.}~\bibnamefont {Zhang}}, \bibinfo {author} {\bibfnamefont {W.}~\bibnamefont {Qin}}, \bibinfo {author} {\bibfnamefont {S.}~\bibnamefont {Yi}}, \bibinfo {author} {\bibfnamefont {P.}~\bibnamefont {Wang}}, \bibinfo {author} {\bibfnamefont {L.}~\bibnamefont {Li}}, \bibinfo {author} {\bibfnamefont {Y.}~\bibnamefont {Wang}}, \bibinfo {author} {\bibfnamefont {H.}~\bibnamefont {Zhang}}, \bibinfo {author} {\bibfnamefont {Z.}~\bibnamefont {Sun}}, \bibinfo {author} {\bibfnamefont {L.}~\bibnamefont {Wei}}, \bibinfo {author} {\bibfnamefont {S.}~\bibnamefont {Zhang}}, \bibinfo {author} {\bibfnamefont {T.}~\bibnamefont {Guo}}, \bibinfo {author} {\bibfnamefont {Q.}~\bibnamefont {Lu}}, \bibinfo {author} {\bibfnamefont {J.-H.}\ \bibnamefont {Cho}}, \bibinfo {author} {\bibfnamefont {C.}~\bibnamefont {Zeng}},\ and\ \bibinfo {author} {\bibfnamefont
  {Z.}~\bibnamefont {Zhang}},\ }\bibfield  {title} {\bibinfo {title} {Flatbands and emergent ferromagnetic ordering in {Fe$_{3}$Sn$_{2}$} kagome lattices},\ }\href {https://doi.org/10.1103/PhysRevLett.121.096401} {\bibfield  {journal} {\bibinfo  {journal} {Phys. Rev. Lett.}\ }\textbf {\bibinfo {volume} {121}},\ \bibinfo {pages} {096401} (\bibinfo {year} {2018})}\BibitemShut {NoStop}%
\bibitem [{\citenamefont {Yin}\ \emph {et~al.}(2018)\citenamefont {Yin}, \citenamefont {Zhang}, \citenamefont {Li}, \citenamefont {Jiang}, \citenamefont {Chang}, \citenamefont {Zhang}, \citenamefont {Lian}, \citenamefont {Xiang}, \citenamefont {Belopolski}, \citenamefont {Zheng}, \citenamefont {Cochran}, \citenamefont {Xu}, \citenamefont {Bian}, \citenamefont {Liu}, \citenamefont {Chang}, \citenamefont {Lin}, \citenamefont {Lu}, \citenamefont {Wang}, \citenamefont {Jia}, \citenamefont {Wang},\ and\ \citenamefont {Hasan}}]{yin.zhang.18}%
  \BibitemOpen
  \bibfield  {author} {\bibinfo {author} {\bibfnamefont {J.-X.}\ \bibnamefont {Yin}}, \bibinfo {author} {\bibfnamefont {S.~S.}\ \bibnamefont {Zhang}}, \bibinfo {author} {\bibfnamefont {H.}~\bibnamefont {Li}}, \bibinfo {author} {\bibfnamefont {K.}~\bibnamefont {Jiang}}, \bibinfo {author} {\bibfnamefont {G.}~\bibnamefont {Chang}}, \bibinfo {author} {\bibfnamefont {B.}~\bibnamefont {Zhang}}, \bibinfo {author} {\bibfnamefont {B.}~\bibnamefont {Lian}}, \bibinfo {author} {\bibfnamefont {C.}~\bibnamefont {Xiang}}, \bibinfo {author} {\bibfnamefont {I.}~\bibnamefont {Belopolski}}, \bibinfo {author} {\bibfnamefont {H.}~\bibnamefont {Zheng}}, \bibinfo {author} {\bibfnamefont {T.~A.}\ \bibnamefont {Cochran}}, \bibinfo {author} {\bibfnamefont {S.-Y.}\ \bibnamefont {Xu}}, \bibinfo {author} {\bibfnamefont {G.}~\bibnamefont {Bian}}, \bibinfo {author} {\bibfnamefont {K.}~\bibnamefont {Liu}}, \bibinfo {author} {\bibfnamefont {T.-R.}\ \bibnamefont {Chang}}, \bibinfo {author} {\bibfnamefont {H.}~\bibnamefont {Lin}}, \bibinfo
  {author} {\bibfnamefont {Z.-Y.}\ \bibnamefont {Lu}}, \bibinfo {author} {\bibfnamefont {Z.}~\bibnamefont {Wang}}, \bibinfo {author} {\bibfnamefont {S.}~\bibnamefont {Jia}}, \bibinfo {author} {\bibfnamefont {W.}~\bibnamefont {Wang}},\ and\ \bibinfo {author} {\bibfnamefont {M.~Z.}\ \bibnamefont {Hasan}},\ }\bibfield  {title} {\bibinfo {title} {Giant and anisotropic many-body spin--orbit tunability in a strongly correlated kagome magnet},\ }\href {https://doi.org/10.1038/s41586-018-0502-7} {\bibfield  {journal} {\bibinfo  {journal} {Nature}\ }\textbf {\bibinfo {volume} {562}},\ \bibinfo {pages} {91} (\bibinfo {year} {2018})}\BibitemShut {NoStop}%
\bibitem [{\citenamefont {Chen}\ \emph {et~al.}(2023{\natexlab{a}})\citenamefont {Chen}, \citenamefont {Zhou}, \citenamefont {Zhang}, \citenamefont {Ji}, \citenamefont {Liao}, \citenamefont {Ji}, \citenamefont {Li}, \citenamefont {Guo}, \citenamefont {Shen}, \citenamefont {Yu}, \citenamefont {Yu}, \citenamefont {Weng},\ and\ \citenamefont {Wang}}]{chen.zhou.23}%
  \BibitemOpen
  \bibfield  {author} {\bibinfo {author} {\bibfnamefont {L.}~\bibnamefont {Chen}}, \bibinfo {author} {\bibfnamefont {Y.}~\bibnamefont {Zhou}}, \bibinfo {author} {\bibfnamefont {H.}~\bibnamefont {Zhang}}, \bibinfo {author} {\bibfnamefont {X.}~\bibnamefont {Ji}}, \bibinfo {author} {\bibfnamefont {K.}~\bibnamefont {Liao}}, \bibinfo {author} {\bibfnamefont {Y.}~\bibnamefont {Ji}}, \bibinfo {author} {\bibfnamefont {Y.}~\bibnamefont {Li}}, \bibinfo {author} {\bibfnamefont {Z.}~\bibnamefont {Guo}}, \bibinfo {author} {\bibfnamefont {X.}~\bibnamefont {Shen}}, \bibinfo {author} {\bibfnamefont {R.}~\bibnamefont {Yu}}, \bibinfo {author} {\bibfnamefont {X.}~\bibnamefont {Yu}}, \bibinfo {author} {\bibfnamefont {H.}~\bibnamefont {Weng}},\ and\ \bibinfo {author} {\bibfnamefont {G.}~\bibnamefont {Wang}},\ }\href@noop {} {\bibinfo {title} {Tunable magnetism and electron correlation in titanium-based kagome metals {$RE$Ti$_{3}$Bi$_{4}$} ({$RE =$ Yb, Pr, and Nd}) by rare-earth engineering}} (\bibinfo {year}
  {2023}{\natexlab{a}}),\ \Eprint {https://arxiv.org/abs/arXiv:2307.02942} {arXiv:2307.02942} \BibitemShut {NoStop}%
\bibitem [{\citenamefont {Sakhya}\ \emph {et~al.}(2023)\citenamefont {Sakhya}, \citenamefont {Ortiz}, \citenamefont {Ghosh}, \citenamefont {Sprague}, \citenamefont {Mondal}, \citenamefont {Matzelle}, \citenamefont {Elius}, \citenamefont {Valadez}, \citenamefont {Mandrus}, \citenamefont {Bansil},\ and\ \citenamefont {Neupane}}]{sakhya.ortiz.23}%
  \BibitemOpen
  \bibfield  {author} {\bibinfo {author} {\bibfnamefont {A.~P.}\ \bibnamefont {Sakhya}}, \bibinfo {author} {\bibfnamefont {B.~R.}\ \bibnamefont {Ortiz}}, \bibinfo {author} {\bibfnamefont {B.}~\bibnamefont {Ghosh}}, \bibinfo {author} {\bibfnamefont {M.}~\bibnamefont {Sprague}}, \bibinfo {author} {\bibfnamefont {M.~I.}\ \bibnamefont {Mondal}}, \bibinfo {author} {\bibfnamefont {M.}~\bibnamefont {Matzelle}}, \bibinfo {author} {\bibfnamefont {I.~B.}\ \bibnamefont {Elius}}, \bibinfo {author} {\bibfnamefont {N.}~\bibnamefont {Valadez}}, \bibinfo {author} {\bibfnamefont {D.~G.}\ \bibnamefont {Mandrus}}, \bibinfo {author} {\bibfnamefont {A.}~\bibnamefont {Bansil}},\ and\ \bibinfo {author} {\bibfnamefont {M.}~\bibnamefont {Neupane}},\ }\href@noop {} {\bibinfo {title} {Observation of multiple flat bands and topological dirac states in a new titanium based slightly distorted kagome metal {YbTi$_{3}$Bi$_{4}$}}} (\bibinfo {year} {2023}),\ \Eprint {https://arxiv.org/abs/arXiv:2309.01176} {arXiv:2309.01176} \BibitemShut
  {NoStop}%
\bibitem [{\citenamefont {Mondal}\ \emph {et~al.}(2023)\citenamefont {Mondal}, \citenamefont {Sakhya}, \citenamefont {Sprague}, \citenamefont {Ortiz}, \citenamefont {Matzelle}, \citenamefont {Ghosh}, \citenamefont {Valadez}, \citenamefont {Elius}, \citenamefont {Bansil},\ and\ \citenamefont {Neupane}}]{mondal.sakhya.23}%
  \BibitemOpen
  \bibfield  {author} {\bibinfo {author} {\bibfnamefont {M.~I.}\ \bibnamefont {Mondal}}, \bibinfo {author} {\bibfnamefont {A.~P.}\ \bibnamefont {Sakhya}}, \bibinfo {author} {\bibfnamefont {M.}~\bibnamefont {Sprague}}, \bibinfo {author} {\bibfnamefont {B.~R.}\ \bibnamefont {Ortiz}}, \bibinfo {author} {\bibfnamefont {M.}~\bibnamefont {Matzelle}}, \bibinfo {author} {\bibfnamefont {B.}~\bibnamefont {Ghosh}}, \bibinfo {author} {\bibfnamefont {N.}~\bibnamefont {Valadez}}, \bibinfo {author} {\bibfnamefont {I.~B.}\ \bibnamefont {Elius}}, \bibinfo {author} {\bibfnamefont {A.}~\bibnamefont {Bansil}},\ and\ \bibinfo {author} {\bibfnamefont {M.}~\bibnamefont {Neupane}},\ }\href@noop {} {\bibinfo {title} {Observation of multiple van hove singularities and correlated electronic states in a new topological ferromagnetic kagome metal {NdTi$_{3}$Bi$_{4}$}}} (\bibinfo {year} {2023}),\ \Eprint {https://arxiv.org/abs/arXiv:2311.11488} {arXiv:2311.11488} \BibitemShut {NoStop}%
\bibitem [{\citenamefont {Yin}\ \emph {et~al.}(2020)\citenamefont {Yin}, \citenamefont {Ma}, \citenamefont {Cochran}, \citenamefont {Xu}, \citenamefont {Zhang}, \citenamefont {Tien}, \citenamefont {Shumiya}, \citenamefont {Cheng}, \citenamefont {Jiang}, \citenamefont {Lian}, \citenamefont {Song}, \citenamefont {Chang}, \citenamefont {Belopolski}, \citenamefont {Multer}, \citenamefont {Litskevich}, \citenamefont {Cheng}, \citenamefont {Yang}, \citenamefont {Swidler}, \citenamefont {Zhou}, \citenamefont {Lin}, \citenamefont {Neupert}, \citenamefont {Wang}, \citenamefont {Yao}, \citenamefont {Chang}, \citenamefont {Jia},\ and\ \citenamefont {Zahid~Hasan}}]{yin.ma.20}%
  \BibitemOpen
  \bibfield  {author} {\bibinfo {author} {\bibfnamefont {J.-X.}\ \bibnamefont {Yin}}, \bibinfo {author} {\bibfnamefont {W.}~\bibnamefont {Ma}}, \bibinfo {author} {\bibfnamefont {T.~A.}\ \bibnamefont {Cochran}}, \bibinfo {author} {\bibfnamefont {X.}~\bibnamefont {Xu}}, \bibinfo {author} {\bibfnamefont {S.~S.}\ \bibnamefont {Zhang}}, \bibinfo {author} {\bibfnamefont {H.-J.}\ \bibnamefont {Tien}}, \bibinfo {author} {\bibfnamefont {N.}~\bibnamefont {Shumiya}}, \bibinfo {author} {\bibfnamefont {G.}~\bibnamefont {Cheng}}, \bibinfo {author} {\bibfnamefont {K.}~\bibnamefont {Jiang}}, \bibinfo {author} {\bibfnamefont {B.}~\bibnamefont {Lian}}, \bibinfo {author} {\bibfnamefont {Z.}~\bibnamefont {Song}}, \bibinfo {author} {\bibfnamefont {G.}~\bibnamefont {Chang}}, \bibinfo {author} {\bibfnamefont {I.}~\bibnamefont {Belopolski}}, \bibinfo {author} {\bibfnamefont {D.}~\bibnamefont {Multer}}, \bibinfo {author} {\bibfnamefont {M.}~\bibnamefont {Litskevich}}, \bibinfo {author} {\bibfnamefont {Z.-J.}\ \bibnamefont {Cheng}},
  \bibinfo {author} {\bibfnamefont {X.~P.}\ \bibnamefont {Yang}}, \bibinfo {author} {\bibfnamefont {B.}~\bibnamefont {Swidler}}, \bibinfo {author} {\bibfnamefont {H.}~\bibnamefont {Zhou}}, \bibinfo {author} {\bibfnamefont {H.}~\bibnamefont {Lin}}, \bibinfo {author} {\bibfnamefont {T.}~\bibnamefont {Neupert}}, \bibinfo {author} {\bibfnamefont {Z.}~\bibnamefont {Wang}}, \bibinfo {author} {\bibfnamefont {N.}~\bibnamefont {Yao}}, \bibinfo {author} {\bibfnamefont {T.-R.}\ \bibnamefont {Chang}}, \bibinfo {author} {\bibfnamefont {S.}~\bibnamefont {Jia}},\ and\ \bibinfo {author} {\bibfnamefont {M.}~\bibnamefont {Zahid~Hasan}},\ }\bibfield  {title} {\bibinfo {title} {Quantum-limit {Chern} topological magnetism in {TbMn$_{6}$Sn$_{6}$}},\ }\href {https://doi.org/10.1038/s41586-020-2482-7} {\bibfield  {journal} {\bibinfo  {journal} {Nature}\ }\textbf {\bibinfo {volume} {583}},\ \bibinfo {pages} {533} (\bibinfo {year} {2020})}\BibitemShut {NoStop}%
\bibitem [{\citenamefont {Ghimire}\ \emph {et~al.}(2020)\citenamefont {Ghimire}, \citenamefont {Dally}, \citenamefont {Poudel}, \citenamefont {Jones}, \citenamefont {Michel}, \citenamefont {Magar}, \citenamefont {Bleuel}, \citenamefont {McGuire}, \citenamefont {Jiang}, \citenamefont {Mitchell}, \citenamefont {Lynn},\ and\ \citenamefont {Mazin}}]{ghimire.dally.20}%
  \BibitemOpen
  \bibfield  {author} {\bibinfo {author} {\bibfnamefont {N.~J.}\ \bibnamefont {Ghimire}}, \bibinfo {author} {\bibfnamefont {R.~L.}\ \bibnamefont {Dally}}, \bibinfo {author} {\bibfnamefont {L.}~\bibnamefont {Poudel}}, \bibinfo {author} {\bibfnamefont {D.~C.}\ \bibnamefont {Jones}}, \bibinfo {author} {\bibfnamefont {D.}~\bibnamefont {Michel}}, \bibinfo {author} {\bibfnamefont {N.~T.}\ \bibnamefont {Magar}}, \bibinfo {author} {\bibfnamefont {M.}~\bibnamefont {Bleuel}}, \bibinfo {author} {\bibfnamefont {M.~A.}\ \bibnamefont {McGuire}}, \bibinfo {author} {\bibfnamefont {J.~S.}\ \bibnamefont {Jiang}}, \bibinfo {author} {\bibfnamefont {J.~F.}\ \bibnamefont {Mitchell}}, \bibinfo {author} {\bibfnamefont {J.~W.}\ \bibnamefont {Lynn}},\ and\ \bibinfo {author} {\bibfnamefont {I.~I.}\ \bibnamefont {Mazin}},\ }\bibfield  {title} {\bibinfo {title} {Competing magnetic phases and fluctuation-driven scalar spin chirality in the kagome metal {YMn$_{6}$Sn$_{6}$}},\ }\href {https://doi.org/10.1126/sciadv.abe2680} {\bibfield
  {journal} {\bibinfo  {journal} {Sci. Adv.}\ }\textbf {\bibinfo {volume} {6}},\ \bibinfo {pages} {eabe2680} (\bibinfo {year} {2020})}\BibitemShut {NoStop}%
\bibitem [{\citenamefont {Li}\ \emph {et~al.}(2021{\natexlab{b}})\citenamefont {Li}, \citenamefont {Wang}, \citenamefont {Wang}, \citenamefont {Yuan}, \citenamefont {Song}, \citenamefont {Lou}, \citenamefont {Liu}, \citenamefont {Huang}, \citenamefont {Liu}, \citenamefont {Lei}, \citenamefont {Yin},\ and\ \citenamefont {Wang}}]{li.wang.21}%
  \BibitemOpen
  \bibfield  {author} {\bibinfo {author} {\bibfnamefont {M.}~\bibnamefont {Li}}, \bibinfo {author} {\bibfnamefont {Q.}~\bibnamefont {Wang}}, \bibinfo {author} {\bibfnamefont {G.}~\bibnamefont {Wang}}, \bibinfo {author} {\bibfnamefont {Z.}~\bibnamefont {Yuan}}, \bibinfo {author} {\bibfnamefont {W.}~\bibnamefont {Song}}, \bibinfo {author} {\bibfnamefont {R.}~\bibnamefont {Lou}}, \bibinfo {author} {\bibfnamefont {Z.}~\bibnamefont {Liu}}, \bibinfo {author} {\bibfnamefont {Y.}~\bibnamefont {Huang}}, \bibinfo {author} {\bibfnamefont {Z.}~\bibnamefont {Liu}}, \bibinfo {author} {\bibfnamefont {H.}~\bibnamefont {Lei}}, \bibinfo {author} {\bibfnamefont {Z.}~\bibnamefont {Yin}},\ and\ \bibinfo {author} {\bibfnamefont {S.}~\bibnamefont {Wang}},\ }\bibfield  {title} {\bibinfo {title} {Dirac cone, flat band and saddle point in kagome magnet {YMn$_{6}$Sn$_{6}$}},\ }\href {https://doi.org/10.1038/s41467-021-23536-8} {\bibfield  {journal} {\bibinfo  {journal} {Nat. Commun.}\ }\textbf {\bibinfo {volume} {12}},\ \bibinfo
  {pages} {3129} (\bibinfo {year} {2021}{\natexlab{b}})}\BibitemShut {NoStop}%
\bibitem [{\citenamefont {Fruhling}\ \emph {et~al.}(2024)\citenamefont {Fruhling}, \citenamefont {Streeter}, \citenamefont {Mardanya}, \citenamefont {Wang}, \citenamefont {Baral}, \citenamefont {Zaharko}, \citenamefont {Mazin}, \citenamefont {Chowdhury}, \citenamefont {Ratcliff},\ and\ \citenamefont {Tafti}}]{fruhling.streeter.21}%
  \BibitemOpen
  \bibfield  {author} {\bibinfo {author} {\bibfnamefont {K.}~\bibnamefont {Fruhling}}, \bibinfo {author} {\bibfnamefont {A.}~\bibnamefont {Streeter}}, \bibinfo {author} {\bibfnamefont {S.}~\bibnamefont {Mardanya}}, \bibinfo {author} {\bibfnamefont {X.}~\bibnamefont {Wang}}, \bibinfo {author} {\bibfnamefont {P.}~\bibnamefont {Baral}}, \bibinfo {author} {\bibfnamefont {O.}~\bibnamefont {Zaharko}}, \bibinfo {author} {\bibfnamefont {I.~I.}\ \bibnamefont {Mazin}}, \bibinfo {author} {\bibfnamefont {S.}~\bibnamefont {Chowdhury}}, \bibinfo {author} {\bibfnamefont {W.~D.}\ \bibnamefont {Ratcliff}},\ and\ \bibinfo {author} {\bibfnamefont {F.}~\bibnamefont {Tafti}},\ }\href@noop {} {\bibinfo {title} {Topological {Hall} effect induced by chiral fluctuations in a kagome lattice}} (\bibinfo {year} {2024}),\ \Eprint {https://arxiv.org/abs/arXiv:2401.17449} {arXiv:2401.17449} \BibitemShut {NoStop}%
\bibitem [{\citenamefont {Ma}\ \emph {et~al.}(2021)\citenamefont {Ma}, \citenamefont {Xu}, \citenamefont {Yin}, \citenamefont {Yang}, \citenamefont {Zhou}, \citenamefont {Cheng}, \citenamefont {Huang}, \citenamefont {Qu}, \citenamefont {Wang}, \citenamefont {Hasan},\ and\ \citenamefont {Jia}}]{ma.xu.21}%
  \BibitemOpen
  \bibfield  {author} {\bibinfo {author} {\bibfnamefont {W.}~\bibnamefont {Ma}}, \bibinfo {author} {\bibfnamefont {X.}~\bibnamefont {Xu}}, \bibinfo {author} {\bibfnamefont {J.-X.}\ \bibnamefont {Yin}}, \bibinfo {author} {\bibfnamefont {H.}~\bibnamefont {Yang}}, \bibinfo {author} {\bibfnamefont {H.}~\bibnamefont {Zhou}}, \bibinfo {author} {\bibfnamefont {Z.-J.}\ \bibnamefont {Cheng}}, \bibinfo {author} {\bibfnamefont {Y.}~\bibnamefont {Huang}}, \bibinfo {author} {\bibfnamefont {Z.}~\bibnamefont {Qu}}, \bibinfo {author} {\bibfnamefont {F.}~\bibnamefont {Wang}}, \bibinfo {author} {\bibfnamefont {M.~Z.}\ \bibnamefont {Hasan}},\ and\ \bibinfo {author} {\bibfnamefont {S.}~\bibnamefont {Jia}},\ }\bibfield  {title} {\bibinfo {title} {Rare earth engineering in {$R$Mn$_{6}$Sn$_{6}$} ({$R=$Gd-Tm, Lu}) topological kagome magnets},\ }\href {https://doi.org/10.1103/PhysRevLett.126.246602} {\bibfield  {journal} {\bibinfo  {journal} {Phys. Rev. Lett.}\ }\textbf {\bibinfo {volume} {126}},\ \bibinfo {pages} {246602} (\bibinfo
  {year} {2021})}\BibitemShut {NoStop}%
\bibitem [{\citenamefont {Zhang}\ \emph {et~al.}(2022)\citenamefont {Zhang}, \citenamefont {Koo}, \citenamefont {Xu}, \citenamefont {Sretenovic}, \citenamefont {Yan},\ and\ \citenamefont {Ke}}]{zhang.koo.22}%
  \BibitemOpen
  \bibfield  {author} {\bibinfo {author} {\bibfnamefont {H.}~\bibnamefont {Zhang}}, \bibinfo {author} {\bibfnamefont {J.}~\bibnamefont {Koo}}, \bibinfo {author} {\bibfnamefont {C.}~\bibnamefont {Xu}}, \bibinfo {author} {\bibfnamefont {M.}~\bibnamefont {Sretenovic}}, \bibinfo {author} {\bibfnamefont {B.}~\bibnamefont {Yan}},\ and\ \bibinfo {author} {\bibfnamefont {X.}~\bibnamefont {Ke}},\ }\bibfield  {title} {\bibinfo {title} {Exchange-biased topological transverse thermoelectric effects in a kagome ferrimagnet},\ }\href {https://doi.org/10.1038/s41467-022-28733-7} {\bibfield  {journal} {\bibinfo  {journal} {Nat. Commun.}\ }\textbf {\bibinfo {volume} {13}},\ \bibinfo {pages} {1091} (\bibinfo {year} {2022})}\BibitemShut {NoStop}%
\bibitem [{\citenamefont {Mielke~III}\ \emph {et~al.}(2022)\citenamefont {Mielke~III}, \citenamefont {Ma}, \citenamefont {Pomjakushin}, \citenamefont {Zaharko}, \citenamefont {Sturniolo}, \citenamefont {Liu}, \citenamefont {Ukleev}, \citenamefont {White}, \citenamefont {Yin}, \citenamefont {Tsirkin}, \citenamefont {Larsen}, \citenamefont {Cochran}, \citenamefont {Medarde}, \citenamefont {Por{\'e}e}, \citenamefont {Das}, \citenamefont {Gupta}, \citenamefont {Wang}, \citenamefont {Chang}, \citenamefont {Wang}, \citenamefont {Khasanov}, \citenamefont {Neupert}, \citenamefont {Amato}, \citenamefont {Liborio}, \citenamefont {Jia}, \citenamefont {Hasan}, \citenamefont {Luetkens},\ and\ \citenamefont {Guguchia}}]{mielke.ma.22}%
  \BibitemOpen
  \bibfield  {author} {\bibinfo {author} {\bibfnamefont {C.}~\bibnamefont {Mielke~III}}, \bibinfo {author} {\bibfnamefont {W.~L.}\ \bibnamefont {Ma}}, \bibinfo {author} {\bibfnamefont {V.}~\bibnamefont {Pomjakushin}}, \bibinfo {author} {\bibfnamefont {O.}~\bibnamefont {Zaharko}}, \bibinfo {author} {\bibfnamefont {S.}~\bibnamefont {Sturniolo}}, \bibinfo {author} {\bibfnamefont {X.}~\bibnamefont {Liu}}, \bibinfo {author} {\bibfnamefont {V.}~\bibnamefont {Ukleev}}, \bibinfo {author} {\bibfnamefont {J.~S.}\ \bibnamefont {White}}, \bibinfo {author} {\bibfnamefont {J.-X.}\ \bibnamefont {Yin}}, \bibinfo {author} {\bibfnamefont {S.~S.}\ \bibnamefont {Tsirkin}}, \bibinfo {author} {\bibfnamefont {C.~B.}\ \bibnamefont {Larsen}}, \bibinfo {author} {\bibfnamefont {T.~A.}\ \bibnamefont {Cochran}}, \bibinfo {author} {\bibfnamefont {M.}~\bibnamefont {Medarde}}, \bibinfo {author} {\bibfnamefont {V.}~\bibnamefont {Por{\'e}e}}, \bibinfo {author} {\bibfnamefont {D.}~\bibnamefont {Das}}, \bibinfo {author} {\bibfnamefont
  {R.}~\bibnamefont {Gupta}}, \bibinfo {author} {\bibfnamefont {C.~N.}\ \bibnamefont {Wang}}, \bibinfo {author} {\bibfnamefont {J.}~\bibnamefont {Chang}}, \bibinfo {author} {\bibfnamefont {Z.~Q.}\ \bibnamefont {Wang}}, \bibinfo {author} {\bibfnamefont {R.}~\bibnamefont {Khasanov}}, \bibinfo {author} {\bibfnamefont {T.}~\bibnamefont {Neupert}}, \bibinfo {author} {\bibfnamefont {A.}~\bibnamefont {Amato}}, \bibinfo {author} {\bibfnamefont {L.}~\bibnamefont {Liborio}}, \bibinfo {author} {\bibfnamefont {S.}~\bibnamefont {Jia}}, \bibinfo {author} {\bibfnamefont {M.~Z.}\ \bibnamefont {Hasan}}, \bibinfo {author} {\bibfnamefont {H.}~\bibnamefont {Luetkens}},\ and\ \bibinfo {author} {\bibfnamefont {Z.}~\bibnamefont {Guguchia}},\ }\bibfield  {title} {\bibinfo {title} {Low-temperature magnetic crossover in the topological kagome magnet {TbMn$_{6}$Sn$_{6}$}},\ }\href {https://doi.org/10.1038/s42005-022-00885-4} {\bibfield  {journal} {\bibinfo  {journal} {Commun. Phys.}\ }\textbf {\bibinfo {volume} {5}},\ \bibinfo {pages}
  {107} (\bibinfo {year} {2022})}\BibitemShut {NoStop}%
\bibitem [{\citenamefont {Riberolles}\ \emph {et~al.}(2022)\citenamefont {Riberolles}, \citenamefont {Slade}, \citenamefont {Abernathy}, \citenamefont {Granroth}, \citenamefont {Li}, \citenamefont {Lee}, \citenamefont {Canfield}, \citenamefont {Ueland}, \citenamefont {Ke},\ and\ \citenamefont {McQueeney}}]{riberolles.slade.22}%
  \BibitemOpen
  \bibfield  {author} {\bibinfo {author} {\bibfnamefont {S.~X.~M.}\ \bibnamefont {Riberolles}}, \bibinfo {author} {\bibfnamefont {T.~J.}\ \bibnamefont {Slade}}, \bibinfo {author} {\bibfnamefont {D.~L.}\ \bibnamefont {Abernathy}}, \bibinfo {author} {\bibfnamefont {G.~E.}\ \bibnamefont {Granroth}}, \bibinfo {author} {\bibfnamefont {B.}~\bibnamefont {Li}}, \bibinfo {author} {\bibfnamefont {Y.}~\bibnamefont {Lee}}, \bibinfo {author} {\bibfnamefont {P.~C.}\ \bibnamefont {Canfield}}, \bibinfo {author} {\bibfnamefont {B.~G.}\ \bibnamefont {Ueland}}, \bibinfo {author} {\bibfnamefont {L.}~\bibnamefont {Ke}},\ and\ \bibinfo {author} {\bibfnamefont {R.~J.}\ \bibnamefont {McQueeney}},\ }\bibfield  {title} {\bibinfo {title} {Low-temperature competing magnetic energy scales in the topological ferrimagnet {TbMn$_{6}$Sn$_{6}$}},\ }\href {https://doi.org/10.1103/PhysRevX.12.021043} {\bibfield  {journal} {\bibinfo  {journal} {Phys. Rev. X}\ }\textbf {\bibinfo {volume} {12}},\ \bibinfo {pages} {021043} (\bibinfo {year}
  {2022})}\BibitemShut {NoStop}%
\bibitem [{\citenamefont {Arachchige}\ \emph {et~al.}(2022)\citenamefont {Arachchige}, \citenamefont {Meier}, \citenamefont {Marshall}, \citenamefont {Matsuoka}, \citenamefont {Xue}, \citenamefont {McGuire}, \citenamefont {Hermann}, \citenamefont {Cao},\ and\ \citenamefont {Mandrus}}]{suriya.meier.22}%
  \BibitemOpen
  \bibfield  {author} {\bibinfo {author} {\bibfnamefont {H.~W.~S.}\ \bibnamefont {Arachchige}}, \bibinfo {author} {\bibfnamefont {W.~R.}\ \bibnamefont {Meier}}, \bibinfo {author} {\bibfnamefont {M.}~\bibnamefont {Marshall}}, \bibinfo {author} {\bibfnamefont {T.}~\bibnamefont {Matsuoka}}, \bibinfo {author} {\bibfnamefont {R.}~\bibnamefont {Xue}}, \bibinfo {author} {\bibfnamefont {M.~A.}\ \bibnamefont {McGuire}}, \bibinfo {author} {\bibfnamefont {R.~P.}\ \bibnamefont {Hermann}}, \bibinfo {author} {\bibfnamefont {H.}~\bibnamefont {Cao}},\ and\ \bibinfo {author} {\bibfnamefont {D.}~\bibnamefont {Mandrus}},\ }\bibfield  {title} {\bibinfo {title} {Charge density wave in kagome lattice intermetallic {ScV$_{6}$Sn$_{6}$}},\ }\href {https://doi.org/10.1103/PhysRevLett.129.216402} {\bibfield  {journal} {\bibinfo  {journal} {Phys. Rev. Lett.}\ }\textbf {\bibinfo {volume} {129}},\ \bibinfo {pages} {216402} (\bibinfo {year} {2022})}\BibitemShut {NoStop}%
\bibitem [{\citenamefont {Cao}\ \emph {et~al.}(2023)\citenamefont {Cao}, \citenamefont {Xu}, \citenamefont {Fukui}, \citenamefont {Manjo}, \citenamefont {Dong}, \citenamefont {Shi}, \citenamefont {Liu}, \citenamefont {Cao},\ and\ \citenamefont {Song}}]{cao.xu.23}%
  \BibitemOpen
  \bibfield  {author} {\bibinfo {author} {\bibfnamefont {S.}~\bibnamefont {Cao}}, \bibinfo {author} {\bibfnamefont {C.}~\bibnamefont {Xu}}, \bibinfo {author} {\bibfnamefont {H.}~\bibnamefont {Fukui}}, \bibinfo {author} {\bibfnamefont {T.}~\bibnamefont {Manjo}}, \bibinfo {author} {\bibfnamefont {Y.}~\bibnamefont {Dong}}, \bibinfo {author} {\bibfnamefont {M.}~\bibnamefont {Shi}}, \bibinfo {author} {\bibfnamefont {Y.}~\bibnamefont {Liu}}, \bibinfo {author} {\bibfnamefont {C.}~\bibnamefont {Cao}},\ and\ \bibinfo {author} {\bibfnamefont {Y.}~\bibnamefont {Song}},\ }\bibfield  {title} {\bibinfo {title} {Competing charge-density wave instabilities in the kagome metal {ScV$_{6}$Sn$_{6}$}},\ }\href {https://doi.org/10.1038/s41467-023-43454-1} {\bibfield  {journal} {\bibinfo  {journal} {Nat. Commun.}\ }\textbf {\bibinfo {volume} {14}},\ \bibinfo {pages} {7671} (\bibinfo {year} {2023})}\BibitemShut {NoStop}%
\bibitem [{\citenamefont {Yang}\ \emph {et~al.}(2021)\citenamefont {Yang}, \citenamefont {Fan}, \citenamefont {Zhang}, \citenamefont {Chen}, \citenamefont {Chen}, \citenamefont {Ying}, \citenamefont {Wu}, \citenamefont {Yang}, \citenamefont {Meng}, \citenamefont {Li}, \citenamefont {Li}, \citenamefont {Gu}, \citenamefont {Qian}, \citenamefont {Schnyder}, \citenamefont {gang Guo},\ and\ \citenamefont {Chen}}]{yang.fan.21}%
  \BibitemOpen
  \bibfield  {author} {\bibinfo {author} {\bibfnamefont {Y.}~\bibnamefont {Yang}}, \bibinfo {author} {\bibfnamefont {W.}~\bibnamefont {Fan}}, \bibinfo {author} {\bibfnamefont {Q.}~\bibnamefont {Zhang}}, \bibinfo {author} {\bibfnamefont {Z.}~\bibnamefont {Chen}}, \bibinfo {author} {\bibfnamefont {X.}~\bibnamefont {Chen}}, \bibinfo {author} {\bibfnamefont {T.}~\bibnamefont {Ying}}, \bibinfo {author} {\bibfnamefont {X.}~\bibnamefont {Wu}}, \bibinfo {author} {\bibfnamefont {X.}~\bibnamefont {Yang}}, \bibinfo {author} {\bibfnamefont {F.}~\bibnamefont {Meng}}, \bibinfo {author} {\bibfnamefont {G.}~\bibnamefont {Li}}, \bibinfo {author} {\bibfnamefont {S.}~\bibnamefont {Li}}, \bibinfo {author} {\bibfnamefont {L.}~\bibnamefont {Gu}}, \bibinfo {author} {\bibfnamefont {T.}~\bibnamefont {Qian}}, \bibinfo {author} {\bibfnamefont {A.~P.}\ \bibnamefont {Schnyder}}, \bibinfo {author} {\bibfnamefont {J.}~\bibnamefont {gang Guo}},\ and\ \bibinfo {author} {\bibfnamefont {X.}~\bibnamefont {Chen}},\ }\bibfield  {title} {\bibinfo
  {title} {Discovery of two families of vsb-based compounds with v-kagome lattice},\ }\href {https://doi.org/10.1088/0256-307X/38/12/127102} {\bibfield  {journal} {\bibinfo  {journal} {Chinese Phys. Lett.}\ }\textbf {\bibinfo {volume} {38}},\ \bibinfo {pages} {127102} (\bibinfo {year} {2021})}\BibitemShut {NoStop}%
\bibitem [{\citenamefont {Yin}\ \emph {et~al.}(2021)\citenamefont {Yin}, \citenamefont {Tu}, \citenamefont {Gong}, \citenamefont {Tian},\ and\ \citenamefont {Lei}}]{yin.tu.21}%
  \BibitemOpen
  \bibfield  {author} {\bibinfo {author} {\bibfnamefont {Q.}~\bibnamefont {Yin}}, \bibinfo {author} {\bibfnamefont {Z.}~\bibnamefont {Tu}}, \bibinfo {author} {\bibfnamefont {C.}~\bibnamefont {Gong}}, \bibinfo {author} {\bibfnamefont {S.}~\bibnamefont {Tian}},\ and\ \bibinfo {author} {\bibfnamefont {H.}~\bibnamefont {Lei}},\ }\bibfield  {title} {\bibinfo {title} {Structures and physical properties of {V}-based kagome metals {CsV$_{6}$Sb$_{6}$} and {CsV$_{8}$Sb$_{12}$}},\ }\href {https://doi.org/10.1088/0256-307X/38/12/127401} {\bibfield  {journal} {\bibinfo  {journal} {Chinese Phys. Lett.}\ }\textbf {\bibinfo {volume} {38}},\ \bibinfo {pages} {127401} (\bibinfo {year} {2021})}\BibitemShut {NoStop}%
\bibitem [{\citenamefont {Shi}\ \emph {et~al.}(2022)\citenamefont {Shi}, \citenamefont {Yu}, \citenamefont {Yang}, \citenamefont {Meng}, \citenamefont {Lei}, \citenamefont {Luo}, \citenamefont {Sun}, \citenamefont {He}, \citenamefont {Wang}, \citenamefont {Jiang}, \citenamefont {Liu}, \citenamefont {Shen}, \citenamefont {Wu}, \citenamefont {Wang}, \citenamefont {Xiang}, \citenamefont {Ying},\ and\ \citenamefont {Chen}}]{shi.yu.22}%
  \BibitemOpen
  \bibfield  {author} {\bibinfo {author} {\bibfnamefont {M.}~\bibnamefont {Shi}}, \bibinfo {author} {\bibfnamefont {F.}~\bibnamefont {Yu}}, \bibinfo {author} {\bibfnamefont {Y.}~\bibnamefont {Yang}}, \bibinfo {author} {\bibfnamefont {F.}~\bibnamefont {Meng}}, \bibinfo {author} {\bibfnamefont {B.}~\bibnamefont {Lei}}, \bibinfo {author} {\bibfnamefont {Y.}~\bibnamefont {Luo}}, \bibinfo {author} {\bibfnamefont {Z.}~\bibnamefont {Sun}}, \bibinfo {author} {\bibfnamefont {J.}~\bibnamefont {He}}, \bibinfo {author} {\bibfnamefont {R.}~\bibnamefont {Wang}}, \bibinfo {author} {\bibfnamefont {Z.}~\bibnamefont {Jiang}}, \bibinfo {author} {\bibfnamefont {Z.}~\bibnamefont {Liu}}, \bibinfo {author} {\bibfnamefont {D.}~\bibnamefont {Shen}}, \bibinfo {author} {\bibfnamefont {T.}~\bibnamefont {Wu}}, \bibinfo {author} {\bibfnamefont {Z.}~\bibnamefont {Wang}}, \bibinfo {author} {\bibfnamefont {Z.}~\bibnamefont {Xiang}}, \bibinfo {author} {\bibfnamefont {J.}~\bibnamefont {Ying}},\ and\ \bibinfo {author} {\bibfnamefont
  {X.}~\bibnamefont {Chen}},\ }\bibfield  {title} {\bibinfo {title} {A new class of bilayer kagome lattice compounds with dirac nodal lines and pressure-induced superconductivity},\ }\href {https://doi.org/10.1038/s41467-022-30442-0} {\bibfield  {journal} {\bibinfo  {journal} {Nat. Commun.}\ }\textbf {\bibinfo {volume} {13}},\ \bibinfo {pages} {2773} (\bibinfo {year} {2022})}\BibitemShut {NoStop}%
\bibitem [{\citenamefont {Mantravadi}\ \emph {et~al.}(2023)\citenamefont {Mantravadi}, \citenamefont {Gvozdetskyi}, \citenamefont {Sarkar}, \citenamefont {Mudryk},\ and\ \citenamefont {Zaikina}}]{mantravadi.gvozdetskyi.23}%
  \BibitemOpen
  \bibfield  {author} {\bibinfo {author} {\bibfnamefont {A.}~\bibnamefont {Mantravadi}}, \bibinfo {author} {\bibfnamefont {V.}~\bibnamefont {Gvozdetskyi}}, \bibinfo {author} {\bibfnamefont {A.}~\bibnamefont {Sarkar}}, \bibinfo {author} {\bibfnamefont {Y.}~\bibnamefont {Mudryk}},\ and\ \bibinfo {author} {\bibfnamefont {J.~V.}\ \bibnamefont {Zaikina}},\ }\bibfield  {title} {\bibinfo {title} {Exploring the {$A$-V-Sb} landscape beyond {$A$V$_{3}$Sb$_{5}$}: A case study on the {KV$_{6}$Sb$_{6}$} kagome compound},\ }\href {https://doi.org/10.1103/PhysRevMaterials.7.115002} {\bibfield  {journal} {\bibinfo  {journal} {Phys. Rev. Mater.}\ }\textbf {\bibinfo {volume} {7}},\ \bibinfo {pages} {115002} (\bibinfo {year} {2023})}\BibitemShut {NoStop}%
\bibitem [{\citenamefont {Teng}\ \emph {et~al.}(2022)\citenamefont {Teng}, \citenamefont {Chen}, \citenamefont {Ye}, \citenamefont {Rosenberg}, \citenamefont {Liu}, \citenamefont {Yin}, \citenamefont {Jiang}, \citenamefont {Oh}, \citenamefont {Hasan}, \citenamefont {Neubauer}, \citenamefont {Gao}, \citenamefont {Xie}, \citenamefont {Hashimoto}, \citenamefont {Lu}, \citenamefont {Jozwiak}, \citenamefont {Bostwick}, \citenamefont {Rotenberg}, \citenamefont {Birgeneau}, \citenamefont {Chu}, \citenamefont {Yi},\ and\ \citenamefont {Dai}}]{teng.chen.22}%
  \BibitemOpen
  \bibfield  {author} {\bibinfo {author} {\bibfnamefont {X.}~\bibnamefont {Teng}}, \bibinfo {author} {\bibfnamefont {L.}~\bibnamefont {Chen}}, \bibinfo {author} {\bibfnamefont {F.}~\bibnamefont {Ye}}, \bibinfo {author} {\bibfnamefont {E.}~\bibnamefont {Rosenberg}}, \bibinfo {author} {\bibfnamefont {Z.}~\bibnamefont {Liu}}, \bibinfo {author} {\bibfnamefont {J.-X.}\ \bibnamefont {Yin}}, \bibinfo {author} {\bibfnamefont {Y.-X.}\ \bibnamefont {Jiang}}, \bibinfo {author} {\bibfnamefont {J.~S.}\ \bibnamefont {Oh}}, \bibinfo {author} {\bibfnamefont {M.~Z.}\ \bibnamefont {Hasan}}, \bibinfo {author} {\bibfnamefont {K.~J.}\ \bibnamefont {Neubauer}}, \bibinfo {author} {\bibfnamefont {B.}~\bibnamefont {Gao}}, \bibinfo {author} {\bibfnamefont {Y.}~\bibnamefont {Xie}}, \bibinfo {author} {\bibfnamefont {M.}~\bibnamefont {Hashimoto}}, \bibinfo {author} {\bibfnamefont {D.}~\bibnamefont {Lu}}, \bibinfo {author} {\bibfnamefont {C.}~\bibnamefont {Jozwiak}}, \bibinfo {author} {\bibfnamefont {A.}~\bibnamefont {Bostwick}}, \bibinfo
  {author} {\bibfnamefont {E.}~\bibnamefont {Rotenberg}}, \bibinfo {author} {\bibfnamefont {R.~J.}\ \bibnamefont {Birgeneau}}, \bibinfo {author} {\bibfnamefont {J.-H.}\ \bibnamefont {Chu}}, \bibinfo {author} {\bibfnamefont {M.}~\bibnamefont {Yi}},\ and\ \bibinfo {author} {\bibfnamefont {P.}~\bibnamefont {Dai}},\ }\bibfield  {title} {\bibinfo {title} {Discovery of charge density wave in a kagome lattice antiferromagnet},\ }\href {https://doi.org/10.1038/s41586-022-05034-z} {\bibfield  {journal} {\bibinfo  {journal} {Nature}\ }\textbf {\bibinfo {volume} {609}},\ \bibinfo {pages} {490} (\bibinfo {year} {2022})}\BibitemShut {NoStop}%
\bibitem [{\citenamefont {Yin}\ \emph {et~al.}(2022{\natexlab{b}})\citenamefont {Yin}, \citenamefont {Jiang}, \citenamefont {Teng}, \citenamefont {Hossain}, \citenamefont {Mardanya}, \citenamefont {Chang}, \citenamefont {Ye}, \citenamefont {Xu}, \citenamefont {Denner}, \citenamefont {Neupert}, \citenamefont {Lienhard}, \citenamefont {Deng}, \citenamefont {Setty}, \citenamefont {Si}, \citenamefont {Chang}, \citenamefont {Guguchia}, \citenamefont {Gao}, \citenamefont {Shumiya}, \citenamefont {Zhang}, \citenamefont {Cochran}, \citenamefont {Multer}, \citenamefont {Yi}, \citenamefont {Dai},\ and\ \citenamefont {Hasan}}]{yin.jiang.22}%
  \BibitemOpen
  \bibfield  {author} {\bibinfo {author} {\bibfnamefont {J.-X.}\ \bibnamefont {Yin}}, \bibinfo {author} {\bibfnamefont {Y.-X.}\ \bibnamefont {Jiang}}, \bibinfo {author} {\bibfnamefont {X.}~\bibnamefont {Teng}}, \bibinfo {author} {\bibfnamefont {M.~S.}\ \bibnamefont {Hossain}}, \bibinfo {author} {\bibfnamefont {S.}~\bibnamefont {Mardanya}}, \bibinfo {author} {\bibfnamefont {T.-R.}\ \bibnamefont {Chang}}, \bibinfo {author} {\bibfnamefont {Z.}~\bibnamefont {Ye}}, \bibinfo {author} {\bibfnamefont {G.}~\bibnamefont {Xu}}, \bibinfo {author} {\bibfnamefont {M.~M.}\ \bibnamefont {Denner}}, \bibinfo {author} {\bibfnamefont {T.}~\bibnamefont {Neupert}}, \bibinfo {author} {\bibfnamefont {B.}~\bibnamefont {Lienhard}}, \bibinfo {author} {\bibfnamefont {H.-B.}\ \bibnamefont {Deng}}, \bibinfo {author} {\bibfnamefont {C.}~\bibnamefont {Setty}}, \bibinfo {author} {\bibfnamefont {Q.}~\bibnamefont {Si}}, \bibinfo {author} {\bibfnamefont {G.}~\bibnamefont {Chang}}, \bibinfo {author} {\bibfnamefont {Z.}~\bibnamefont {Guguchia}},
  \bibinfo {author} {\bibfnamefont {B.}~\bibnamefont {Gao}}, \bibinfo {author} {\bibfnamefont {N.}~\bibnamefont {Shumiya}}, \bibinfo {author} {\bibfnamefont {Q.}~\bibnamefont {Zhang}}, \bibinfo {author} {\bibfnamefont {T.~A.}\ \bibnamefont {Cochran}}, \bibinfo {author} {\bibfnamefont {D.}~\bibnamefont {Multer}}, \bibinfo {author} {\bibfnamefont {M.}~\bibnamefont {Yi}}, \bibinfo {author} {\bibfnamefont {P.}~\bibnamefont {Dai}},\ and\ \bibinfo {author} {\bibfnamefont {M.~Z.}\ \bibnamefont {Hasan}},\ }\bibfield  {title} {\bibinfo {title} {Discovery of charge order and corresponding edge state in kagome magnet {FeGe}},\ }\href {https://doi.org/10.1103/PhysRevLett.129.166401} {\bibfield  {journal} {\bibinfo  {journal} {Phys. Rev. Lett.}\ }\textbf {\bibinfo {volume} {129}},\ \bibinfo {pages} {166401} (\bibinfo {year} {2022}{\natexlab{b}})}\BibitemShut {NoStop}%
\bibitem [{\citenamefont {Watanabe}\ and\ \citenamefont {Kunitomi}(1966)}]{watanabe.kunitomi.66}%
  \BibitemOpen
  \bibfield  {author} {\bibinfo {author} {\bibfnamefont {H.}~\bibnamefont {Watanabe}}\ and\ \bibinfo {author} {\bibfnamefont {N.}~\bibnamefont {Kunitomi}},\ }\bibfield  {title} {\bibinfo {title} {On the neutron diffraction study of {FeGe}},\ }\href {https://doi.org/10.1143/JPSJ.21.1932} {\bibfield  {journal} {\bibinfo  {journal} {J. Phys. Soc. Jpn.}\ }\textbf {\bibinfo {volume} {21}},\ \bibinfo {pages} {1932} (\bibinfo {year} {1966})}\BibitemShut {NoStop}%
\bibitem [{\citenamefont {Tomiyoshi}\ \emph {et~al.}(1966)\citenamefont {Tomiyoshi}, \citenamefont {Yamamoto},\ and\ \citenamefont {Watanabe}}]{tomiyoshi.yamamoto.66}%
  \BibitemOpen
  \bibfield  {author} {\bibinfo {author} {\bibfnamefont {S.}~\bibnamefont {Tomiyoshi}}, \bibinfo {author} {\bibfnamefont {H.}~\bibnamefont {Yamamoto}},\ and\ \bibinfo {author} {\bibfnamefont {H.}~\bibnamefont {Watanabe}},\ }\bibfield  {title} {\bibinfo {title} {The m\"{o}ssbauer study of {FeGe}},\ }\href {https://doi.org/10.1143/JPSJ.21.709} {\bibfield  {journal} {\bibinfo  {journal} {J. Phys. Soc. Jpn.}\ }\textbf {\bibinfo {volume} {21}},\ \bibinfo {pages} {709} (\bibinfo {year} {1966})}\BibitemShut {NoStop}%
\bibitem [{\citenamefont {Beckman}\ \emph {et~al.}(1972)\citenamefont {Beckman}, \citenamefont {Carrander}, \citenamefont {Lundgren},\ and\ \citenamefont {Richardson}}]{beckma.carrander.72}%
  \BibitemOpen
  \bibfield  {author} {\bibinfo {author} {\bibfnamefont {O.}~\bibnamefont {Beckman}}, \bibinfo {author} {\bibfnamefont {K.}~\bibnamefont {Carrander}}, \bibinfo {author} {\bibfnamefont {L.}~\bibnamefont {Lundgren}},\ and\ \bibinfo {author} {\bibfnamefont {M.}~\bibnamefont {Richardson}},\ }\bibfield  {title} {\bibinfo {title} {Susceptibility measurements and magnetic ordering of hexagonal {FeGe}},\ }\href {https://doi.org/10.1088/0031-8949/6/2-3/009} {\bibfield  {journal} {\bibinfo  {journal} {Phys. Scr.}\ }\textbf {\bibinfo {volume} {6}},\ \bibinfo {pages} {151} (\bibinfo {year} {1972})}\BibitemShut {NoStop}%
\bibitem [{\citenamefont {H\"{a}ggstr\"{o}m}\ \emph {et~al.}(1975)\citenamefont {H\"{a}ggstr\"{o}m}, \citenamefont {Ericsson}, \citenamefont {Wäppling},\ and\ \citenamefont {Karlsson}}]{haggstrom.ericsson.75}%
  \BibitemOpen
  \bibfield  {author} {\bibinfo {author} {\bibfnamefont {L.}~\bibnamefont {H\"{a}ggstr\"{o}m}}, \bibinfo {author} {\bibfnamefont {T.}~\bibnamefont {Ericsson}}, \bibinfo {author} {\bibfnamefont {R.}~\bibnamefont {Wäppling}},\ and\ \bibinfo {author} {\bibfnamefont {E.}~\bibnamefont {Karlsson}},\ }\bibfield  {title} {\bibinfo {title} {Mössbauer study of hexagonal {FeGe}},\ }\href {https://doi.org/10.1088/0031-8949/11/1/009} {\bibfield  {journal} {\bibinfo  {journal} {Phys. Scr.}\ }\textbf {\bibinfo {volume} {11}},\ \bibinfo {pages} {55} (\bibinfo {year} {1975})}\BibitemShut {NoStop}%
\bibitem [{\citenamefont {Forsyth}\ \emph {et~al.}(1978)\citenamefont {Forsyth}, \citenamefont {Wilkinson},\ and\ \citenamefont {Gardner}}]{forsyth.wilkinson.78}%
  \BibitemOpen
  \bibfield  {author} {\bibinfo {author} {\bibfnamefont {J.~B.}\ \bibnamefont {Forsyth}}, \bibinfo {author} {\bibfnamefont {C.}~\bibnamefont {Wilkinson}},\ and\ \bibinfo {author} {\bibfnamefont {P.}~\bibnamefont {Gardner}},\ }\bibfield  {title} {\bibinfo {title} {The low-temperature magnetic structure of hexagonal {FeGe}},\ }\href {https://doi.org/10.1088/0305-4608/8/10/019} {\bibfield  {journal} {\bibinfo  {journal} {J. Phys. F: Met. Phys.}\ }\textbf {\bibinfo {volume} {8}},\ \bibinfo {pages} {2195} (\bibinfo {year} {1978})}\BibitemShut {NoStop}%
\bibitem [{\citenamefont {Bernhard}\ \emph {et~al.}(1984)\citenamefont {Bernhard}, \citenamefont {Lebech},\ and\ \citenamefont {Beckman}}]{bernhard.lebech.84}%
  \BibitemOpen
  \bibfield  {author} {\bibinfo {author} {\bibfnamefont {J.}~\bibnamefont {Bernhard}}, \bibinfo {author} {\bibfnamefont {B.}~\bibnamefont {Lebech}},\ and\ \bibinfo {author} {\bibfnamefont {O.}~\bibnamefont {Beckman}},\ }\bibfield  {title} {\bibinfo {title} {Neutron diffraction studies of the low-temperature magnetic structure of hexagonal {FeGe}},\ }\href {https://doi.org/10.1088/0305-4608/14/10/017} {\bibfield  {journal} {\bibinfo  {journal} {J. Phys. F: Met. Phys.}\ }\textbf {\bibinfo {volume} {14}},\ \bibinfo {pages} {2379} (\bibinfo {year} {1984})}\BibitemShut {NoStop}%
\bibitem [{\citenamefont {Bernhard}\ \emph {et~al.}(1988)\citenamefont {Bernhard}, \citenamefont {Lebech},\ and\ \citenamefont {Beckman}}]{bernhard.lebech.88}%
  \BibitemOpen
  \bibfield  {author} {\bibinfo {author} {\bibfnamefont {J.}~\bibnamefont {Bernhard}}, \bibinfo {author} {\bibfnamefont {B.}~\bibnamefont {Lebech}},\ and\ \bibinfo {author} {\bibfnamefont {O.}~\bibnamefont {Beckman}},\ }\bibfield  {title} {\bibinfo {title} {Magnetic phase diagram of hexagonal {FeGe} determined by neutron diffraction},\ }\href {https://doi.org/10.1088/0305-4608/18/3/023} {\bibfield  {journal} {\bibinfo  {journal} {J. Phys. F: Met. Phys.}\ }\textbf {\bibinfo {volume} {18}},\ \bibinfo {pages} {539} (\bibinfo {year} {1988})}\BibitemShut {NoStop}%
\bibitem [{\citenamefont {Teng}\ \emph {et~al.}(2023)\citenamefont {Teng}, \citenamefont {Oh}, \citenamefont {Tan}, \citenamefont {Chen}, \citenamefont {Huang}, \citenamefont {Gao}, \citenamefont {Yin}, \citenamefont {Chu}, \citenamefont {Hashimoto}, \citenamefont {Lu}, \citenamefont {Jozwiak}, \citenamefont {Bostwick}, \citenamefont {Rotenberg}, \citenamefont {Granroth}, \citenamefont {Yan}, \citenamefont {Birgeneau}, \citenamefont {Dai},\ and\ \citenamefont {Yi}}]{teng.oh.23}%
  \BibitemOpen
  \bibfield  {author} {\bibinfo {author} {\bibfnamefont {X.}~\bibnamefont {Teng}}, \bibinfo {author} {\bibfnamefont {J.~S.}\ \bibnamefont {Oh}}, \bibinfo {author} {\bibfnamefont {H.}~\bibnamefont {Tan}}, \bibinfo {author} {\bibfnamefont {L.}~\bibnamefont {Chen}}, \bibinfo {author} {\bibfnamefont {J.}~\bibnamefont {Huang}}, \bibinfo {author} {\bibfnamefont {B.}~\bibnamefont {Gao}}, \bibinfo {author} {\bibfnamefont {J.-X.}\ \bibnamefont {Yin}}, \bibinfo {author} {\bibfnamefont {J.-H.}\ \bibnamefont {Chu}}, \bibinfo {author} {\bibfnamefont {M.}~\bibnamefont {Hashimoto}}, \bibinfo {author} {\bibfnamefont {D.}~\bibnamefont {Lu}}, \bibinfo {author} {\bibfnamefont {C.}~\bibnamefont {Jozwiak}}, \bibinfo {author} {\bibfnamefont {A.}~\bibnamefont {Bostwick}}, \bibinfo {author} {\bibfnamefont {E.}~\bibnamefont {Rotenberg}}, \bibinfo {author} {\bibfnamefont {G.~E.}\ \bibnamefont {Granroth}}, \bibinfo {author} {\bibfnamefont {B.}~\bibnamefont {Yan}}, \bibinfo {author} {\bibfnamefont {R.~J.}\ \bibnamefont {Birgeneau}},
  \bibinfo {author} {\bibfnamefont {P.}~\bibnamefont {Dai}},\ and\ \bibinfo {author} {\bibfnamefont {M.}~\bibnamefont {Yi}},\ }\bibfield  {title} {\bibinfo {title} {Magnetism and charge density wave order in kagome {FeGe}},\ }\href {https://doi.org/10.1038/s41567-023-01985-w} {\bibfield  {journal} {\bibinfo  {journal} {Nat. Phys.}\ }\textbf {\bibinfo {volume} {19}},\ \bibinfo {pages} {814} (\bibinfo {year} {2023})}\BibitemShut {NoStop}%
\bibitem [{\citenamefont {Shao}\ \emph {et~al.}(2023)\citenamefont {Shao}, \citenamefont {Yin}, \citenamefont {Belopolski}, \citenamefont {You}, \citenamefont {Hou}, \citenamefont {Chen}, \citenamefont {Jiang}, \citenamefont {Hossain}, \citenamefont {Yahyavi}, \citenamefont {Hsu}, \citenamefont {Feng}, \citenamefont {Bansil}, \citenamefont {Hasan},\ and\ \citenamefont {Chang}}]{shao.yin.23}%
  \BibitemOpen
  \bibfield  {author} {\bibinfo {author} {\bibfnamefont {S.}~\bibnamefont {Shao}}, \bibinfo {author} {\bibfnamefont {J.-X.}\ \bibnamefont {Yin}}, \bibinfo {author} {\bibfnamefont {I.}~\bibnamefont {Belopolski}}, \bibinfo {author} {\bibfnamefont {J.-Y.}\ \bibnamefont {You}}, \bibinfo {author} {\bibfnamefont {T.}~\bibnamefont {Hou}}, \bibinfo {author} {\bibfnamefont {H.}~\bibnamefont {Chen}}, \bibinfo {author} {\bibfnamefont {Y.}~\bibnamefont {Jiang}}, \bibinfo {author} {\bibfnamefont {M.~S.}\ \bibnamefont {Hossain}}, \bibinfo {author} {\bibfnamefont {M.}~\bibnamefont {Yahyavi}}, \bibinfo {author} {\bibfnamefont {C.-H.}\ \bibnamefont {Hsu}}, \bibinfo {author} {\bibfnamefont {Y.~P.}\ \bibnamefont {Feng}}, \bibinfo {author} {\bibfnamefont {A.}~\bibnamefont {Bansil}}, \bibinfo {author} {\bibfnamefont {M.~Z.}\ \bibnamefont {Hasan}},\ and\ \bibinfo {author} {\bibfnamefont {G.}~\bibnamefont {Chang}},\ }\bibfield  {title} {\bibinfo {title} {Intertwining of magnetism and charge ordering in kagome {FeGe}},\ }\href
  {https://doi.org/10.1021/acsnano.3c00229} {\bibfield  {journal} {\bibinfo  {journal} {ACS Nano}\ }\textbf {\bibinfo {volume} {17}},\ \bibinfo {pages} {10164} (\bibinfo {year} {2023})}\BibitemShut {NoStop}%
\bibitem [{\citenamefont {Chen}\ \emph {et~al.}(2023{\natexlab{b}})\citenamefont {Chen}, \citenamefont {Wu}, \citenamefont {Yin}, \citenamefont {Zhang}, \citenamefont {Wang}, \citenamefont {Li}, \citenamefont {Li}, \citenamefont {Wang}, \citenamefont {Wang}, \citenamefont {Yan},\ and\ \citenamefont {Feng}}]{chen.wu.23}%
  \BibitemOpen
  \bibfield  {author} {\bibinfo {author} {\bibfnamefont {Z.}~\bibnamefont {Chen}}, \bibinfo {author} {\bibfnamefont {X.}~\bibnamefont {Wu}}, \bibinfo {author} {\bibfnamefont {R.}~\bibnamefont {Yin}}, \bibinfo {author} {\bibfnamefont {J.}~\bibnamefont {Zhang}}, \bibinfo {author} {\bibfnamefont {S.}~\bibnamefont {Wang}}, \bibinfo {author} {\bibfnamefont {Y.}~\bibnamefont {Li}}, \bibinfo {author} {\bibfnamefont {M.}~\bibnamefont {Li}}, \bibinfo {author} {\bibfnamefont {A.}~\bibnamefont {Wang}}, \bibinfo {author} {\bibfnamefont {Y.}~\bibnamefont {Wang}}, \bibinfo {author} {\bibfnamefont {Y.-J.}\ \bibnamefont {Yan}},\ and\ \bibinfo {author} {\bibfnamefont {D.-L.}\ \bibnamefont {Feng}},\ }\href@noop {} {\bibinfo {title} {Charge density wave with strong quantum phase fluctuations in kagome magnet {FeGe}}} (\bibinfo {year} {2023}{\natexlab{b}}),\ \Eprint {https://arxiv.org/abs/arXiv:2302.04490} {arXiv:2302.04490} \BibitemShut {NoStop}%
\bibitem [{\citenamefont {Tan}\ \emph {et~al.}(2021)\citenamefont {Tan}, \citenamefont {Liu}, \citenamefont {Wang},\ and\ \citenamefont {Yan}}]{tan.liu.21}%
  \BibitemOpen
  \bibfield  {author} {\bibinfo {author} {\bibfnamefont {H.}~\bibnamefont {Tan}}, \bibinfo {author} {\bibfnamefont {Y.}~\bibnamefont {Liu}}, \bibinfo {author} {\bibfnamefont {Z.}~\bibnamefont {Wang}},\ and\ \bibinfo {author} {\bibfnamefont {B.}~\bibnamefont {Yan}},\ }\bibfield  {title} {\bibinfo {title} {Charge density waves and electronic properties of superconducting kagome metals},\ }\href {https://doi.org/10.1103/PhysRevLett.127.046401} {\bibfield  {journal} {\bibinfo  {journal} {Phys. Rev. Lett.}\ }\textbf {\bibinfo {volume} {127}},\ \bibinfo {pages} {046401} (\bibinfo {year} {2021})}\BibitemShut {NoStop}%
\bibitem [{\citenamefont {Ptok}\ \emph {et~al.}(2022)\citenamefont {Ptok}, \citenamefont {Kobia\l{}ka}, \citenamefont {Sternik}, \citenamefont {\L{}a\.{z}ewski}, \citenamefont {Jochym}, \citenamefont {Ole\'{s}},\ and\ \citenamefont {Piekarz}}]{ptok.kobialka.22}%
  \BibitemOpen
  \bibfield  {author} {\bibinfo {author} {\bibfnamefont {A.}~\bibnamefont {Ptok}}, \bibinfo {author} {\bibfnamefont {A.}~\bibnamefont {Kobia\l{}ka}}, \bibinfo {author} {\bibfnamefont {M.}~\bibnamefont {Sternik}}, \bibinfo {author} {\bibfnamefont {J.}~\bibnamefont {\L{}a\.{z}ewski}}, \bibinfo {author} {\bibfnamefont {P.~T.}\ \bibnamefont {Jochym}}, \bibinfo {author} {\bibfnamefont {A.~M.}\ \bibnamefont {Ole\'{s}}},\ and\ \bibinfo {author} {\bibfnamefont {P.}~\bibnamefont {Piekarz}},\ }\bibfield  {title} {\bibinfo {title} {Dynamical study of the origin of the charge density wave in {$A$V$_{3}$Sb$_{5}$} ({$A=$K, Rb, Cs}) compounds},\ }\href {https://doi.org/10.1103/PhysRevB.105.235134} {\bibfield  {journal} {\bibinfo  {journal} {Phys. Rev. B}\ }\textbf {\bibinfo {volume} {105}},\ \bibinfo {pages} {235134} (\bibinfo {year} {2022})}\BibitemShut {NoStop}%
\bibitem [{\citenamefont {Subedi}(2022)}]{subedi.22}%
  \BibitemOpen
  \bibfield  {author} {\bibinfo {author} {\bibfnamefont {A.}~\bibnamefont {Subedi}},\ }\bibfield  {title} {\bibinfo {title} {Hexagonal-to-base-centered-orthorhombic {$4Q$} charge density wave order in kagome metals {KV$_{3}$Sb$_{5}$}, {RbV$_{3}$Sb$_{5}$}, and {CsV$_{3}$Sb$_{5}$}},\ }\href {https://doi.org/10.1103/PhysRevMaterials.6.015001} {\bibfield  {journal} {\bibinfo  {journal} {Phys. Rev. Mater.}\ }\textbf {\bibinfo {volume} {6}},\ \bibinfo {pages} {015001} (\bibinfo {year} {2022})}\BibitemShut {NoStop}%
\bibitem [{\citenamefont {Subires}\ \emph {et~al.}(2023)\citenamefont {Subires}, \citenamefont {Korshunov}, \citenamefont {Said}, \citenamefont {S{\'a}nchez}, \citenamefont {Ortiz}, \citenamefont {Wilson}, \citenamefont {Bosak},\ and\ \citenamefont {Blanco-Canosa}}]{subires.korshunov.23}%
  \BibitemOpen
  \bibfield  {author} {\bibinfo {author} {\bibfnamefont {D.}~\bibnamefont {Subires}}, \bibinfo {author} {\bibfnamefont {A.}~\bibnamefont {Korshunov}}, \bibinfo {author} {\bibfnamefont {A.~H.}\ \bibnamefont {Said}}, \bibinfo {author} {\bibfnamefont {L.}~\bibnamefont {S{\'a}nchez}}, \bibinfo {author} {\bibfnamefont {B.~R.}\ \bibnamefont {Ortiz}}, \bibinfo {author} {\bibfnamefont {S.~D.}\ \bibnamefont {Wilson}}, \bibinfo {author} {\bibfnamefont {A.}~\bibnamefont {Bosak}},\ and\ \bibinfo {author} {\bibfnamefont {S.}~\bibnamefont {Blanco-Canosa}},\ }\bibfield  {title} {\bibinfo {title} {Order-disorder charge density wave instability in the kagome metal {(Cs,Rb)V$_{3}$Sb$_{5}$}},\ }\href {https://doi.org/10.1038/s41467-023-36668-w} {\bibfield  {journal} {\bibinfo  {journal} {Nat. Commun.}\ }\textbf {\bibinfo {volume} {14}},\ \bibinfo {pages} {1015} (\bibinfo {year} {2023})}\BibitemShut {NoStop}%
\bibitem [{\citenamefont {Gutierrez-Amigo}\ \emph {et~al.}(2023)\citenamefont {Gutierrez-Amigo}, \citenamefont {Dangi\'{c}}, \citenamefont {Guo}, \citenamefont {Felser}, \citenamefont {Moll}, \citenamefont {Vergniory},\ and\ \citenamefont {Errea}}]{gutierrez.dangic.23}%
  \BibitemOpen
  \bibfield  {author} {\bibinfo {author} {\bibfnamefont {M.}~\bibnamefont {Gutierrez-Amigo}}, \bibinfo {author} {\bibfnamefont {D.}~\bibnamefont {Dangi\'{c}}}, \bibinfo {author} {\bibfnamefont {C.}~\bibnamefont {Guo}}, \bibinfo {author} {\bibfnamefont {C.}~\bibnamefont {Felser}}, \bibinfo {author} {\bibfnamefont {P.~J.~W.}\ \bibnamefont {Moll}}, \bibinfo {author} {\bibfnamefont {M.~G.}\ \bibnamefont {Vergniory}},\ and\ \bibinfo {author} {\bibfnamefont {I.}~\bibnamefont {Errea}},\ }\href@noop {} {\bibinfo {title} {Phonon collapse and anharmonic melting of the {3D} charge-density wave in kagome metals}} (\bibinfo {year} {2023}),\ \Eprint {https://arxiv.org/abs/arXiv:2311.14112} {arXiv:2311.14112} \BibitemShut {NoStop}%
\bibitem [{\citenamefont {Ptok}\ \emph {et~al.}(2021)\citenamefont {Ptok}, \citenamefont {Kobia\l{}ka}, \citenamefont {Sternik}, \citenamefont {\L{}a\.{z}ewski}, \citenamefont {Jochym}, \citenamefont {Ole\'{s}}, \citenamefont {Stankov},\ and\ \citenamefont {Piekarz}}]{ptok.kobialka.21}%
  \BibitemOpen
  \bibfield  {author} {\bibinfo {author} {\bibfnamefont {A.}~\bibnamefont {Ptok}}, \bibinfo {author} {\bibfnamefont {A.}~\bibnamefont {Kobia\l{}ka}}, \bibinfo {author} {\bibfnamefont {M.}~\bibnamefont {Sternik}}, \bibinfo {author} {\bibfnamefont {J.}~\bibnamefont {\L{}a\.{z}ewski}}, \bibinfo {author} {\bibfnamefont {P.~T.}\ \bibnamefont {Jochym}}, \bibinfo {author} {\bibfnamefont {A.~M.}\ \bibnamefont {Ole\'{s}}}, \bibinfo {author} {\bibfnamefont {S.}~\bibnamefont {Stankov}},\ and\ \bibinfo {author} {\bibfnamefont {P.}~\bibnamefont {Piekarz}},\ }\bibfield  {title} {\bibinfo {title} {Chiral phonons in the honeycomb sublattice of layered {CoSn}-like compounds},\ }\href {https://doi.org/10.1103/PhysRevB.104.054305} {\bibfield  {journal} {\bibinfo  {journal} {Phys. Rev. B}\ }\textbf {\bibinfo {volume} {104}},\ \bibinfo {pages} {054305} (\bibinfo {year} {2021})}\BibitemShut {NoStop}%
\bibitem [{\citenamefont {Miao}\ \emph {et~al.}(2023)\citenamefont {Miao}, \citenamefont {Zhang}, \citenamefont {Li}, \citenamefont {Fabbris}, \citenamefont {Said}, \citenamefont {Tartaglia}, \citenamefont {Yilmaz}, \citenamefont {Vescovo}, \citenamefont {Yin}, \citenamefont {Murakami}, \citenamefont {Feng}, \citenamefont {Jiang}, \citenamefont {Wu}, \citenamefont {Wang}, \citenamefont {Okamoto}, \citenamefont {Wang},\ and\ \citenamefont {Lee}}]{miao.zhang.23}%
  \BibitemOpen
  \bibfield  {author} {\bibinfo {author} {\bibfnamefont {H.}~\bibnamefont {Miao}}, \bibinfo {author} {\bibfnamefont {T.~T.}\ \bibnamefont {Zhang}}, \bibinfo {author} {\bibfnamefont {H.~X.}\ \bibnamefont {Li}}, \bibinfo {author} {\bibfnamefont {G.}~\bibnamefont {Fabbris}}, \bibinfo {author} {\bibfnamefont {A.~H.}\ \bibnamefont {Said}}, \bibinfo {author} {\bibfnamefont {R.}~\bibnamefont {Tartaglia}}, \bibinfo {author} {\bibfnamefont {T.}~\bibnamefont {Yilmaz}}, \bibinfo {author} {\bibfnamefont {E.}~\bibnamefont {Vescovo}}, \bibinfo {author} {\bibfnamefont {J.-X.}\ \bibnamefont {Yin}}, \bibinfo {author} {\bibfnamefont {S.}~\bibnamefont {Murakami}}, \bibinfo {author} {\bibfnamefont {X.~L.}\ \bibnamefont {Feng}}, \bibinfo {author} {\bibfnamefont {K.}~\bibnamefont {Jiang}}, \bibinfo {author} {\bibfnamefont {X.~L.}\ \bibnamefont {Wu}}, \bibinfo {author} {\bibfnamefont {A.~F.}\ \bibnamefont {Wang}}, \bibinfo {author} {\bibfnamefont {S.}~\bibnamefont {Okamoto}}, \bibinfo {author} {\bibfnamefont {Y.~L.}\ \bibnamefont
  {Wang}},\ and\ \bibinfo {author} {\bibfnamefont {H.~N.}\ \bibnamefont {Lee}},\ }\bibfield  {title} {\bibinfo {title} {Signature of spin-phonon coupling driven charge density wave in a kagome magnet},\ }\href {https://doi.org/10.1038/s41467-023-41957-5} {\bibfield  {journal} {\bibinfo  {journal} {Nat. Commun.}\ }\textbf {\bibinfo {volume} {14}},\ \bibinfo {pages} {6183} (\bibinfo {year} {2023})}\BibitemShut {NoStop}%
\bibitem [{\citenamefont {Ma}\ \emph {et~al.}(2023)\citenamefont {Ma}, \citenamefont {Yin}, \citenamefont {Hasan},\ and\ \citenamefont {Liu}}]{ma.yin.23}%
  \BibitemOpen
  \bibfield  {author} {\bibinfo {author} {\bibfnamefont {H.-Y.}\ \bibnamefont {Ma}}, \bibinfo {author} {\bibfnamefont {J.-X.}\ \bibnamefont {Yin}}, \bibinfo {author} {\bibfnamefont {M.~Z.}\ \bibnamefont {Hasan}},\ and\ \bibinfo {author} {\bibfnamefont {J.}~\bibnamefont {Liu}},\ }\href@noop {} {\bibinfo {title} {Theory for charge density wave and orbital-flux state in antiferromagnetic kagome metal {FeGe}}} (\bibinfo {year} {2023}),\ \Eprint {https://arxiv.org/abs/arXiv:2303.02824} {arXiv:2303.02824} \BibitemShut {NoStop}%
\bibitem [{\citenamefont {Wang}(2023)}]{wang.23}%
  \BibitemOpen
  \bibfield  {author} {\bibinfo {author} {\bibfnamefont {Y.}~\bibnamefont {Wang}},\ }\bibfield  {title} {\bibinfo {title} {Enhanced spin-polarization via partial {Ge}-dimerization as the driving force of the charge density wave in {FeGe}},\ }\href {https://doi.org/10.1103/PhysRevMaterials.7.104006} {\bibfield  {journal} {\bibinfo  {journal} {Phys. Rev. Mater.}\ }\textbf {\bibinfo {volume} {7}},\ \bibinfo {pages} {104006} (\bibinfo {year} {2023})}\BibitemShut {NoStop}%
\bibitem [{\citenamefont {Bl\"ochl}(1994)}]{blochl.94}%
  \BibitemOpen
  \bibfield  {author} {\bibinfo {author} {\bibfnamefont {P.~E.}\ \bibnamefont {Bl\"ochl}},\ }\bibfield  {title} {\bibinfo {title} {Projector augmented-wave method},\ }\href {https://doi.org/10.1103/PhysRevB.50.17953} {\bibfield  {journal} {\bibinfo  {journal} {Phys. Rev. B}\ }\textbf {\bibinfo {volume} {50}},\ \bibinfo {pages} {17953} (\bibinfo {year} {1994})}\BibitemShut {NoStop}%
\bibitem [{\citenamefont {Kresse}\ and\ \citenamefont {Hafner}(1994)}]{kresse.hafner.94}%
  \BibitemOpen
  \bibfield  {author} {\bibinfo {author} {\bibfnamefont {G.}~\bibnamefont {Kresse}}\ and\ \bibinfo {author} {\bibfnamefont {J.}~\bibnamefont {Hafner}},\ }\bibfield  {title} {\bibinfo {title} {Ab initio molecular-dynamics simulation of the liquid-metal--amorphous-semiconductor transition in germanium},\ }\href {https://doi.org/10.1103/PhysRevB.49.14251} {\bibfield  {journal} {\bibinfo  {journal} {Phys. Rev. B}\ }\textbf {\bibinfo {volume} {49}},\ \bibinfo {pages} {14251} (\bibinfo {year} {1994})}\BibitemShut {NoStop}%
\bibitem [{\citenamefont {Kresse}\ and\ \citenamefont {Furthm\"uller}(1996)}]{kresse.furthmuller.96}%
  \BibitemOpen
  \bibfield  {author} {\bibinfo {author} {\bibfnamefont {G.}~\bibnamefont {Kresse}}\ and\ \bibinfo {author} {\bibfnamefont {J.}~\bibnamefont {Furthm\"uller}},\ }\bibfield  {title} {\bibinfo {title} {Efficient iterative schemes for ab initio total-energy calculations using a plane-wave basis set},\ }\href {https://doi.org/10.1103/PhysRevB.54.11169} {\bibfield  {journal} {\bibinfo  {journal} {Phys. Rev. B}\ }\textbf {\bibinfo {volume} {54}},\ \bibinfo {pages} {11169} (\bibinfo {year} {1996})}\BibitemShut {NoStop}%
\bibitem [{\citenamefont {Kresse}\ and\ \citenamefont {Joubert}(1999)}]{kresse.joubert.99}%
  \BibitemOpen
  \bibfield  {author} {\bibinfo {author} {\bibfnamefont {G.}~\bibnamefont {Kresse}}\ and\ \bibinfo {author} {\bibfnamefont {D.}~\bibnamefont {Joubert}},\ }\bibfield  {title} {\bibinfo {title} {From ultrasoft pseudopotentials to the projector augmented-wave method},\ }\href {https://doi.org/10.1103/PhysRevB.59.1758} {\bibfield  {journal} {\bibinfo  {journal} {Phys. Rev. B}\ }\textbf {\bibinfo {volume} {59}},\ \bibinfo {pages} {1758} (\bibinfo {year} {1999})}\BibitemShut {NoStop}%
\bibitem [{\citenamefont {Perdew}\ \emph {et~al.}(1996)\citenamefont {Perdew}, \citenamefont {Burke},\ and\ \citenamefont {Ernzerhof}}]{perdew.burke.96}%
  \BibitemOpen
  \bibfield  {author} {\bibinfo {author} {\bibfnamefont {J.~P.}\ \bibnamefont {Perdew}}, \bibinfo {author} {\bibfnamefont {K.}~\bibnamefont {Burke}},\ and\ \bibinfo {author} {\bibfnamefont {M.}~\bibnamefont {Ernzerhof}},\ }\bibfield  {title} {\bibinfo {title} {Generalized gradient approximation made simple},\ }\href {https://doi.org/10.1103/PhysRevLett.77.3865} {\bibfield  {journal} {\bibinfo  {journal} {Phys. Rev. Lett.}\ }\textbf {\bibinfo {volume} {77}},\ \bibinfo {pages} {3865} (\bibinfo {year} {1996})}\BibitemShut {NoStop}%
\bibitem [{\citenamefont {Dudarev}\ \emph {et~al.}(1998)\citenamefont {Dudarev}, \citenamefont {Botton}, \citenamefont {Savrasov}, \citenamefont {Humphreys},\ and\ \citenamefont {Sutton}}]{dudarev.botton.98}%
  \BibitemOpen
  \bibfield  {author} {\bibinfo {author} {\bibfnamefont {S.~L.}\ \bibnamefont {Dudarev}}, \bibinfo {author} {\bibfnamefont {G.~A.}\ \bibnamefont {Botton}}, \bibinfo {author} {\bibfnamefont {S.~Y.}\ \bibnamefont {Savrasov}}, \bibinfo {author} {\bibfnamefont {C.~J.}\ \bibnamefont {Humphreys}},\ and\ \bibinfo {author} {\bibfnamefont {A.~P.}\ \bibnamefont {Sutton}},\ }\bibfield  {title} {\bibinfo {title} {Electron-energy-loss spectra and the structural stability of nickel oxide: An {LSDA+U} study},\ }\href {https://doi.org/10.1103/PhysRevB.57.1505} {\bibfield  {journal} {\bibinfo  {journal} {Phys. Rev. B}\ }\textbf {\bibinfo {volume} {57}},\ \bibinfo {pages} {1505} (\bibinfo {year} {1998})}\BibitemShut {NoStop}%
\bibitem [{\citenamefont {Liechtenstein}\ \emph {et~al.}(1995)\citenamefont {Liechtenstein}, \citenamefont {Anisimov},\ and\ \citenamefont {Zaanen}}]{liechtenstein.anisimov.95}%
  \BibitemOpen
  \bibfield  {author} {\bibinfo {author} {\bibfnamefont {A.~I.}\ \bibnamefont {Liechtenstein}}, \bibinfo {author} {\bibfnamefont {V.~I.}\ \bibnamefont {Anisimov}},\ and\ \bibinfo {author} {\bibfnamefont {J.}~\bibnamefont {Zaanen}},\ }\bibfield  {title} {\bibinfo {title} {Density-functional theory and strong interactions: Orbital ordering in {Mott-Hubbard} insulators},\ }\href {https://doi.org/10.1103/PhysRevB.52.R5467} {\bibfield  {journal} {\bibinfo  {journal} {Phys. Rev. B}\ }\textbf {\bibinfo {volume} {52}},\ \bibinfo {pages} {R5467} (\bibinfo {year} {1995})}\BibitemShut {NoStop}%
\bibitem [{Note1()}]{Note1}%
  \BibitemOpen
  \bibinfo {note} {The Supplemental Material at [URL will be inserted by publisher] for additional theoretical results. We present the atom displacement induced by the soft modes, electronic band structure, and role of the correlation effects withing DFT+U.}\BibitemShut {Stop}%
\bibitem [{\citenamefont {Monkhorst}\ and\ \citenamefont {Pack}(1976)}]{monkhorst.pack.76}%
  \BibitemOpen
  \bibfield  {author} {\bibinfo {author} {\bibfnamefont {H.~J.}\ \bibnamefont {Monkhorst}}\ and\ \bibinfo {author} {\bibfnamefont {J.~D.}\ \bibnamefont {Pack}},\ }\bibfield  {title} {\bibinfo {title} {Special points for {Brillouin}-zone integrations},\ }\href {https://doi.org/10.1103/PhysRevB.13.5188} {\bibfield  {journal} {\bibinfo  {journal} {Phys. Rev. B}\ }\textbf {\bibinfo {volume} {13}},\ \bibinfo {pages} {5188} (\bibinfo {year} {1976})}\BibitemShut {NoStop}%
\bibitem [{\citenamefont {Stokes}\ and\ \citenamefont {Hatch}(2005)}]{stokes.hatch.05}%
  \BibitemOpen
  \bibfield  {author} {\bibinfo {author} {\bibfnamefont {H.~T.}\ \bibnamefont {Stokes}}\ and\ \bibinfo {author} {\bibfnamefont {D.~M.}\ \bibnamefont {Hatch}},\ }\bibfield  {title} {\bibinfo {title} {{{\sc FindSym}: program for identifying the space-group symmetry of a crystal}},\ }\href {https://doi.org/10.1107/S0021889804031528} {\bibfield  {journal} {\bibinfo  {journal} {J. Appl. Cryst.}\ }\textbf {\bibinfo {volume} {38}},\ \bibinfo {pages} {237} (\bibinfo {year} {2005})}\BibitemShut {NoStop}%
\bibitem [{\citenamefont {Togo}\ and\ \citenamefont {Tanaka}(2018)}]{togo.tanaka.18}%
  \BibitemOpen
  \bibfield  {author} {\bibinfo {author} {\bibfnamefont {A.}~\bibnamefont {Togo}}\ and\ \bibinfo {author} {\bibfnamefont {I.}~\bibnamefont {Tanaka}},\ }\href@noop {} {\bibinfo {title} {{\sc Spglib}: a software library for crystal symmetry search}} (\bibinfo {year} {2018}),\ \Eprint {https://arxiv.org/abs/arXiv:1808.01590} {arXiv:1808.01590} \BibitemShut {NoStop}%
\bibitem [{\citenamefont {Hinuma}\ \emph {et~al.}(2017)\citenamefont {Hinuma}, \citenamefont {Pizzi}, \citenamefont {Kumagai}, \citenamefont {Oba},\ and\ \citenamefont {Tanaka}}]{hinuma.pizzi.17}%
  \BibitemOpen
  \bibfield  {author} {\bibinfo {author} {\bibfnamefont {Y.}~\bibnamefont {Hinuma}}, \bibinfo {author} {\bibfnamefont {G.}~\bibnamefont {Pizzi}}, \bibinfo {author} {\bibfnamefont {Y.}~\bibnamefont {Kumagai}}, \bibinfo {author} {\bibfnamefont {F.}~\bibnamefont {Oba}},\ and\ \bibinfo {author} {\bibfnamefont {I.}~\bibnamefont {Tanaka}},\ }\bibfield  {title} {\bibinfo {title} {Band structure diagram paths based on crystallography},\ }\href {https://doi.org/10.1016/j.commatsci.2016.10.015} {\bibfield  {journal} {\bibinfo  {journal} {Comput. Mater. Sci.}\ }\textbf {\bibinfo {volume} {128}},\ \bibinfo {pages} {140} (\bibinfo {year} {2017})}\BibitemShut {NoStop}%
\bibitem [{\citenamefont {Parlinski}\ \emph {et~al.}(1997)\citenamefont {Parlinski}, \citenamefont {Li},\ and\ \citenamefont {Kawazoe}}]{parlinski.li.97}%
  \BibitemOpen
  \bibfield  {author} {\bibinfo {author} {\bibfnamefont {K.}~\bibnamefont {Parlinski}}, \bibinfo {author} {\bibfnamefont {Z.~Q.}\ \bibnamefont {Li}},\ and\ \bibinfo {author} {\bibfnamefont {Y.}~\bibnamefont {Kawazoe}},\ }\bibfield  {title} {\bibinfo {title} {First-principles determination of the soft mode in cubic {ZrO$_{2}$}},\ }\href {https://doi.org/10.1103/PhysRevLett.78.4063} {\bibfield  {journal} {\bibinfo  {journal} {Phys. Rev. Lett.}\ }\textbf {\bibinfo {volume} {78}},\ \bibinfo {pages} {4063} (\bibinfo {year} {1997})}\BibitemShut {NoStop}%
\bibitem [{\citenamefont {Togo}\ \emph {et~al.}(2023)\citenamefont {Togo}, \citenamefont {Chaput}, \citenamefont {Tadano},\ and\ \citenamefont {Tanaka}}]{togo.chaput.23}%
  \BibitemOpen
  \bibfield  {author} {\bibinfo {author} {\bibfnamefont {A.}~\bibnamefont {Togo}}, \bibinfo {author} {\bibfnamefont {L.}~\bibnamefont {Chaput}}, \bibinfo {author} {\bibfnamefont {T.}~\bibnamefont {Tadano}},\ and\ \bibinfo {author} {\bibfnamefont {I.}~\bibnamefont {Tanaka}},\ }\bibfield  {title} {\bibinfo {title} {Implementation strategies in phonopy and phono3py},\ }\href {https://doi.org/10.1088/1361-648X/acd831} {\bibfield  {journal} {\bibinfo  {journal} {J. Phys. Condens. Matter}\ }\textbf {\bibinfo {volume} {35}},\ \bibinfo {pages} {353001} (\bibinfo {year} {2023})}\BibitemShut {NoStop}%
\bibitem [{\citenamefont {Togo}(2023)}]{togo.23}%
  \BibitemOpen
  \bibfield  {author} {\bibinfo {author} {\bibfnamefont {A.}~\bibnamefont {Togo}},\ }\bibfield  {title} {\bibinfo {title} {First-principles phonon calculations with phonopy and phono3py},\ }\href {https://doi.org/10.7566/JPSJ.92.012001} {\bibfield  {journal} {\bibinfo  {journal} {J. Phys. Soc. Jpn.}\ }\textbf {\bibinfo {volume} {92}},\ \bibinfo {pages} {012001} (\bibinfo {year} {2023})}\BibitemShut {NoStop}%
\bibitem [{\citenamefont {Tersoff}\ and\ \citenamefont {Hamann}(1985)}]{tersoff.hamann.85}%
  \BibitemOpen
  \bibfield  {author} {\bibinfo {author} {\bibfnamefont {J.}~\bibnamefont {Tersoff}}\ and\ \bibinfo {author} {\bibfnamefont {D.~R.}\ \bibnamefont {Hamann}},\ }\bibfield  {title} {\bibinfo {title} {Theory of the scanning tunneling microscope},\ }\href {https://doi.org/10.1103/PhysRevB.31.805} {\bibfield  {journal} {\bibinfo  {journal} {Phys. Rev. B}\ }\textbf {\bibinfo {volume} {31}},\ \bibinfo {pages} {805} (\bibinfo {year} {1985})}\BibitemShut {NoStop}%
\bibitem [{\citenamefont {Liao}\ and\ \citenamefont {Carter}(2010)}]{liao.carter.10}%
  \BibitemOpen
  \bibfield  {author} {\bibinfo {author} {\bibfnamefont {P.}~\bibnamefont {Liao}}\ and\ \bibinfo {author} {\bibfnamefont {E.~A.}\ \bibnamefont {Carter}},\ }\bibfield  {title} {\bibinfo {title} {Ab initio {DFT+U} predictions of tensile properties of iron oxides},\ }\href {https://doi.org/10.1039/C0JM01199A} {\bibfield  {journal} {\bibinfo  {journal} {J. Mater. Chem.}\ }\textbf {\bibinfo {volume} {20}},\ \bibinfo {pages} {6703} (\bibinfo {year} {2010})}\BibitemShut {NoStop}%
\bibitem [{\citenamefont {Piekarz}\ \emph {et~al.}(2010)\citenamefont {Piekarz}, \citenamefont {Ole\'{s}},\ and\ \citenamefont {Parlinski}}]{piekarz.oles.10}%
  \BibitemOpen
  \bibfield  {author} {\bibinfo {author} {\bibfnamefont {P.}~\bibnamefont {Piekarz}}, \bibinfo {author} {\bibfnamefont {A.~M.}\ \bibnamefont {Ole\'{s}}},\ and\ \bibinfo {author} {\bibfnamefont {K.}~\bibnamefont {Parlinski}},\ }\bibfield  {title} {\bibinfo {title} {Comparative study of the electronic structures of {Fe$_{3}$O$_{4}$} and {Fe$_{2}$SiO$_{4}$}},\ }\href {https://doi.org/10.12693/APhysPolA.118.307} {\bibfield  {journal} {\bibinfo  {journal} {Acta Phys. Pol. A}\ }\textbf {\bibinfo {volume} {118}},\ \bibinfo {pages} {307} (\bibinfo {year} {2010})}\BibitemShut {NoStop}%
\bibitem [{\citenamefont {Feng}\ \emph {et~al.}(2023)\citenamefont {Feng}, \citenamefont {Li}, \citenamefont {Chen},\ and\ \citenamefont {Chen}}]{feng.li.23}%
  \BibitemOpen
  \bibfield  {author} {\bibinfo {author} {\bibfnamefont {Y.}~\bibnamefont {Feng}}, \bibinfo {author} {\bibfnamefont {Z.}~\bibnamefont {Li}}, \bibinfo {author} {\bibfnamefont {J.}~\bibnamefont {Chen}},\ and\ \bibinfo {author} {\bibfnamefont {Y.}~\bibnamefont {Chen}},\ }\bibfield  {title} {\bibinfo {title} {Effect of content and spin state of iron on electronic properties and floatability of iron-bearing sphalerite: A {DFT+U} study},\ }\href {https://doi.org/10.1016/j.ijmst.2023.09.005} {\bibfield  {journal} {\bibinfo  {journal} {J. Min. Sci. Tech.}\ }\textbf {\bibinfo {volume} {33}},\ \bibinfo {pages} {1563} (\bibinfo {year} {2023})}\BibitemShut {NoStop}%
\bibitem [{\citenamefont {\L{}a\ifmmode~\dot{z}\else \.{z}\fi{}ewski}\ \emph {et~al.}(2006)\citenamefont {\L{}a\ifmmode~\dot{z}\else \.{z}\fi{}ewski}, \citenamefont {Piekarz}, \citenamefont {Ole\ifmmode~\acute{s}\else \'{s}\fi{}},\ and\ \citenamefont {Parlinski}}]{lazewski.piekarz.06}%
  \BibitemOpen
  \bibfield  {author} {\bibinfo {author} {\bibfnamefont {J.}~\bibnamefont {\L{}a\ifmmode~\dot{z}\else \.{z}\fi{}ewski}}, \bibinfo {author} {\bibfnamefont {P.}~\bibnamefont {Piekarz}}, \bibinfo {author} {\bibfnamefont {A.~M.}\ \bibnamefont {Ole\ifmmode~\acute{s}\else \'{s}\fi{}}},\ and\ \bibinfo {author} {\bibfnamefont {K.}~\bibnamefont {Parlinski}},\ }\bibfield  {title} {\bibinfo {title} {Influence of local electron interactions on phonon spectrum in iron},\ }\href {https://doi.org/10.1103/PhysRevB.74.174304} {\bibfield  {journal} {\bibinfo  {journal} {Phys. Rev. B}\ }\textbf {\bibinfo {volume} {74}},\ \bibinfo {pages} {174304} (\bibinfo {year} {2006})}\BibitemShut {NoStop}%
\bibitem [{\citenamefont {Zhou}\ \emph {et~al.}(2023)\citenamefont {Zhou}, \citenamefont {Yan}, \citenamefont {Fan}, \citenamefont {Wang},\ and\ \citenamefont {Wan}}]{zhou.yan.23}%
  \BibitemOpen
  \bibfield  {author} {\bibinfo {author} {\bibfnamefont {H.}~\bibnamefont {Zhou}}, \bibinfo {author} {\bibfnamefont {S.}~\bibnamefont {Yan}}, \bibinfo {author} {\bibfnamefont {D.}~\bibnamefont {Fan}}, \bibinfo {author} {\bibfnamefont {D.}~\bibnamefont {Wang}},\ and\ \bibinfo {author} {\bibfnamefont {X.}~\bibnamefont {Wan}},\ }\bibfield  {title} {\bibinfo {title} {Magnetic interactions and possible structural distortion in kagome {FeGe} from first-principles calculations and symmetry analysis},\ }\href {https://doi.org/10.1103/PhysRevB.108.035138} {\bibfield  {journal} {\bibinfo  {journal} {Phys. Rev. B}\ }\textbf {\bibinfo {volume} {108}},\ \bibinfo {pages} {035138} (\bibinfo {year} {2023})}\BibitemShut {NoStop}%
\bibitem [{\citenamefont {Luo}\ \emph {et~al.}(2022)\citenamefont {Luo}, \citenamefont {Gao}, \citenamefont {Liu}, \citenamefont {Gu}, \citenamefont {Wu}, \citenamefont {Yi}, \citenamefont {Jia}, \citenamefont {Wu}, \citenamefont {Luo}, \citenamefont {Xu}, \citenamefont {Zhao}, \citenamefont {Wang}, \citenamefont {Mao}, \citenamefont {Liu}, \citenamefont {Zhu}, \citenamefont {Shi}, \citenamefont {Jiang}, \citenamefont {Hu}, \citenamefont {Xu},\ and\ \citenamefont {Zhou}}]{luo.gao.22}%
  \BibitemOpen
  \bibfield  {author} {\bibinfo {author} {\bibfnamefont {H.}~\bibnamefont {Luo}}, \bibinfo {author} {\bibfnamefont {Q.}~\bibnamefont {Gao}}, \bibinfo {author} {\bibfnamefont {H.}~\bibnamefont {Liu}}, \bibinfo {author} {\bibfnamefont {Y.}~\bibnamefont {Gu}}, \bibinfo {author} {\bibfnamefont {D.}~\bibnamefont {Wu}}, \bibinfo {author} {\bibfnamefont {C.}~\bibnamefont {Yi}}, \bibinfo {author} {\bibfnamefont {J.}~\bibnamefont {Jia}}, \bibinfo {author} {\bibfnamefont {S.}~\bibnamefont {Wu}}, \bibinfo {author} {\bibfnamefont {X.}~\bibnamefont {Luo}}, \bibinfo {author} {\bibfnamefont {Y.}~\bibnamefont {Xu}}, \bibinfo {author} {\bibfnamefont {L.}~\bibnamefont {Zhao}}, \bibinfo {author} {\bibfnamefont {Q.}~\bibnamefont {Wang}}, \bibinfo {author} {\bibfnamefont {H.}~\bibnamefont {Mao}}, \bibinfo {author} {\bibfnamefont {G.}~\bibnamefont {Liu}}, \bibinfo {author} {\bibfnamefont {Z.}~\bibnamefont {Zhu}}, \bibinfo {author} {\bibfnamefont {Y.}~\bibnamefont {Shi}}, \bibinfo {author} {\bibfnamefont {K.}~\bibnamefont {Jiang}},
  \bibinfo {author} {\bibfnamefont {J.}~\bibnamefont {Hu}}, \bibinfo {author} {\bibfnamefont {Z.}~\bibnamefont {Xu}},\ and\ \bibinfo {author} {\bibfnamefont {X.~J.}\ \bibnamefont {Zhou}},\ }\bibfield  {title} {\bibinfo {title} {Electronic nature of charge density wave and electron-phonon coupling in kagome superconductor {KV$_{3}$Sb$_{5}$}},\ }\href {https://doi.org/10.1038/s41467-021-27946-6} {\bibfield  {journal} {\bibinfo  {journal} {Nat. Commun.}\ }\textbf {\bibinfo {volume} {13}},\ \bibinfo {pages} {273} (\bibinfo {year} {2022})}\BibitemShut {NoStop}%
\bibitem [{\citenamefont {Kang}\ \emph {et~al.}(2022)\citenamefont {Kang}, \citenamefont {Fang}, \citenamefont {Kim}, \citenamefont {Ortiz}, \citenamefont {Ryu}, \citenamefont {Kim}, \citenamefont {Yoo}, \citenamefont {Sangiovanni}, \citenamefont {Di~Sante}, \citenamefont {Park}, \citenamefont {Jozwiak}, \citenamefont {Bostwick}, \citenamefont {Rotenberg}, \citenamefont {Kaxiras}, \citenamefont {Wilson}, \citenamefont {Park},\ and\ \citenamefont {Comin}}]{kang.fang.22}%
  \BibitemOpen
  \bibfield  {author} {\bibinfo {author} {\bibfnamefont {M.}~\bibnamefont {Kang}}, \bibinfo {author} {\bibfnamefont {S.}~\bibnamefont {Fang}}, \bibinfo {author} {\bibfnamefont {J.-K.}\ \bibnamefont {Kim}}, \bibinfo {author} {\bibfnamefont {B.~R.}\ \bibnamefont {Ortiz}}, \bibinfo {author} {\bibfnamefont {S.~H.}\ \bibnamefont {Ryu}}, \bibinfo {author} {\bibfnamefont {J.}~\bibnamefont {Kim}}, \bibinfo {author} {\bibfnamefont {J.}~\bibnamefont {Yoo}}, \bibinfo {author} {\bibfnamefont {G.}~\bibnamefont {Sangiovanni}}, \bibinfo {author} {\bibfnamefont {D.}~\bibnamefont {Di~Sante}}, \bibinfo {author} {\bibfnamefont {B.-G.}\ \bibnamefont {Park}}, \bibinfo {author} {\bibfnamefont {C.}~\bibnamefont {Jozwiak}}, \bibinfo {author} {\bibfnamefont {A.}~\bibnamefont {Bostwick}}, \bibinfo {author} {\bibfnamefont {E.}~\bibnamefont {Rotenberg}}, \bibinfo {author} {\bibfnamefont {E.}~\bibnamefont {Kaxiras}}, \bibinfo {author} {\bibfnamefont {S.~D.}\ \bibnamefont {Wilson}}, \bibinfo {author} {\bibfnamefont {J.-H.}\ \bibnamefont
  {Park}},\ and\ \bibinfo {author} {\bibfnamefont {R.}~\bibnamefont {Comin}},\ }\bibfield  {title} {\bibinfo {title} {Twofold van hove singularity and origin of charge order in topological kagome superconductor {CsV$_{3}$Sb$_{5}$}},\ }\href {https://doi.org/10.1038/s41567-021-01451-5} {\bibfield  {journal} {\bibinfo  {journal} {Nat. Phys.}\ }\textbf {\bibinfo {volume} {18}},\ \bibinfo {pages} {301} (\bibinfo {year} {2022})}\BibitemShut {NoStop}%
\bibitem [{\citenamefont {Kato}\ \emph {et~al.}(2022)\citenamefont {Kato}, \citenamefont {Li}, \citenamefont {Kawakami}, \citenamefont {Liu}, \citenamefont {Nakayama}, \citenamefont {Wang}, \citenamefont {Moriya}, \citenamefont {Tanaka}, \citenamefont {Takahashi}, \citenamefont {Yao},\ and\ \citenamefont {Sato}}]{kato.li.22}%
  \BibitemOpen
  \bibfield  {author} {\bibinfo {author} {\bibfnamefont {T.}~\bibnamefont {Kato}}, \bibinfo {author} {\bibfnamefont {Y.}~\bibnamefont {Li}}, \bibinfo {author} {\bibfnamefont {T.}~\bibnamefont {Kawakami}}, \bibinfo {author} {\bibfnamefont {M.}~\bibnamefont {Liu}}, \bibinfo {author} {\bibfnamefont {K.}~\bibnamefont {Nakayama}}, \bibinfo {author} {\bibfnamefont {Z.}~\bibnamefont {Wang}}, \bibinfo {author} {\bibfnamefont {A.}~\bibnamefont {Moriya}}, \bibinfo {author} {\bibfnamefont {K.}~\bibnamefont {Tanaka}}, \bibinfo {author} {\bibfnamefont {T.}~\bibnamefont {Takahashi}}, \bibinfo {author} {\bibfnamefont {Y.}~\bibnamefont {Yao}},\ and\ \bibinfo {author} {\bibfnamefont {T.}~\bibnamefont {Sato}},\ }\bibfield  {title} {\bibinfo {title} {Three-dimensional energy gap and origin of charge-density wave in kagome superconductor {KV$_{3}$Sb$_{5}$}},\ }\href {https://doi.org/10.1038/s43246-022-00255-1} {\bibfield  {journal} {\bibinfo  {journal} {Commun. Mater.}\ }\textbf {\bibinfo {volume} {3}},\ \bibinfo {pages} {30}
  (\bibinfo {year} {2022})}\BibitemShut {NoStop}%
\bibitem [{\citenamefont {Jiang}\ \emph {et~al.}(2023)\citenamefont {Jiang}, \citenamefont {Ma}, \citenamefont {Xia}, \citenamefont {Liu}, \citenamefont {Xiao}, \citenamefont {Liu}, \citenamefont {Yang}, \citenamefont {Ding}, \citenamefont {Huang}, \citenamefont {Liu}, \citenamefont {Qiao}, \citenamefont {Liu}, \citenamefont {Peng}, \citenamefont {Cho}, \citenamefont {Guo}, \citenamefont {Liu},\ and\ \citenamefont {Shen}}]{jiang.ma.23}%
  \BibitemOpen
  \bibfield  {author} {\bibinfo {author} {\bibfnamefont {Z.}~\bibnamefont {Jiang}}, \bibinfo {author} {\bibfnamefont {H.}~\bibnamefont {Ma}}, \bibinfo {author} {\bibfnamefont {W.}~\bibnamefont {Xia}}, \bibinfo {author} {\bibfnamefont {Z.}~\bibnamefont {Liu}}, \bibinfo {author} {\bibfnamefont {Q.}~\bibnamefont {Xiao}}, \bibinfo {author} {\bibfnamefont {Z.}~\bibnamefont {Liu}}, \bibinfo {author} {\bibfnamefont {Y.}~\bibnamefont {Yang}}, \bibinfo {author} {\bibfnamefont {J.}~\bibnamefont {Ding}}, \bibinfo {author} {\bibfnamefont {Z.}~\bibnamefont {Huang}}, \bibinfo {author} {\bibfnamefont {J.}~\bibnamefont {Liu}}, \bibinfo {author} {\bibfnamefont {Y.}~\bibnamefont {Qiao}}, \bibinfo {author} {\bibfnamefont {J.}~\bibnamefont {Liu}}, \bibinfo {author} {\bibfnamefont {Y.}~\bibnamefont {Peng}}, \bibinfo {author} {\bibfnamefont {S.}~\bibnamefont {Cho}}, \bibinfo {author} {\bibfnamefont {Y.}~\bibnamefont {Guo}}, \bibinfo {author} {\bibfnamefont {J.}~\bibnamefont {Liu}},\ and\ \bibinfo {author} {\bibfnamefont
  {D.}~\bibnamefont {Shen}},\ }\bibfield  {title} {\bibinfo {title} {Observation of electronic nematicity driven by the three-dimensional charge density wave in kagome lattice {KV$_{3}$Sb$_{5}$}},\ }\href {https://doi.org/10.1021/acs.nanolett.3c01151} {\bibfield  {journal} {\bibinfo  {journal} {Nano Lett.}\ }\textbf {\bibinfo {volume} {23}},\ \bibinfo {pages} {5625} (\bibinfo {year} {2023})}\BibitemShut {NoStop}%
\bibitem [{\citenamefont {Momma}\ and\ \citenamefont {Izumi}(2011)}]{momma.izumi.11}%
  \BibitemOpen
  \bibfield  {author} {\bibinfo {author} {\bibfnamefont {K.}~\bibnamefont {Momma}}\ and\ \bibinfo {author} {\bibfnamefont {F.}~\bibnamefont {Izumi}},\ }\bibfield  {title} {\bibinfo {title} {{{\sc vesta3} for three-dimensional visualization of crystal, volumetric and morphology data}},\ }\href {https://doi.org/10.1107/S0021889811038970} {\bibfield  {journal} {\bibinfo  {journal} {J. Appl. Crystallogr.}\ }\textbf {\bibinfo {volume} {44}},\ \bibinfo {pages} {1272} (\bibinfo {year} {2011})}\BibitemShut {NoStop}%
\bibitem [{\citenamefont {Kokalj}(1999)}]{kokalj.99}%
  \BibitemOpen
  \bibfield  {author} {\bibinfo {author} {\bibfnamefont {A.}~\bibnamefont {Kokalj}},\ }\bibfield  {title} {\bibinfo {title} {Xcrysden--a new program for displaying crystalline structures and electron densities},\ }\href {https://doi.org/10.1016/S1093-3263(99)00028-5} {\bibfield  {journal} {\bibinfo  {journal} {J. Mol. Graph. Model.}\ }\textbf {\bibinfo {volume} {17}},\ \bibinfo {pages} {176} (\bibinfo {year} {1999})}\BibitemShut {NoStop}%
\end{thebibliography}%

%%%%%%%%%%%%%%%%%%%%%%%%%%%%%%%%%%%
%%%%%%%%%%%%%%%%%%%%%%%%%%%%%%%%%%%
%%%%%%%%%%%%%%%%%%%%%%%%%%%%%%%%%%%

\clearpage
\newpage

\onecolumngrid

\begin{center}
  \textbf{\Large Supplemental Material}\\[.3cm]
  \textbf{\large Lattice dynamics study of electron-correlation-induced charge density wave \\[.1cm] in antiferromagnetic kagome metal FeGe}\\[.3cm]
  %%%%%%
  Andrzej Ptok {\it et al.}\\[.2cm]
  %%%%%%
  %{\itshape
%	$^{1}$Institute of Nuclear Physics, Polish Academy of Sciences, W. E. Radzikowskiego 152, PL-31342 Kraków, Poland
  %}
  (Dated: \today)
\\[0.3cm]
\end{center}

\setcounter{equation}{0}
\renewcommand{\theequation}{S\arabic{equation}}
\setcounter{figure}{0}
\renewcommand{\thefigure}{S\arabic{figure}}
\setcounter{section}{0}
\renewcommand{\thesection}{S\arabic{section}}
\setcounter{table}{0}
\renewcommand{\thetable}{S\arabic{table}}
\setcounter{page}{1}

%%%%%%%%%%%%%%%%%%%%%%%%%%%%%%%%%%%
%%%%%%%%%%%%%%%%%%%%%%%%%%%%%%%%%%%
%%%%%%%%%%%%%%%%%%%%%%%%%%%%%%%%%%%

\begin{figure}[!h]
\centering
\includegraphics[width=\columnwidth]{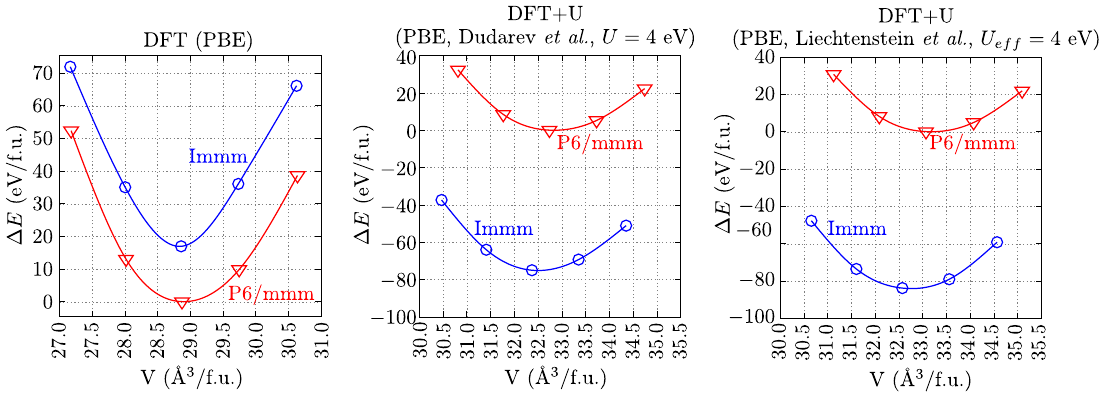}
\caption{
Role of the correlation effect on the energies of FeGe with the P6/mmm and Immm symmetries (red and blue line, respectively).
In the absence of correlations (standard DFT) the P6/mmm has always smaller energy than the Immm structure.
Introduction of the correlation leads to the dramatic decrease of energy for the Imma structure. 
Above some critical $U$, the Immm has smaller energy than P6/mmm regardless of the used DFT+U variant.
The ``zero'' energy is set at the energy of P6/mmm structure in equilibrium volume.
\label{fig.compare}}
\end{figure}

\begin{figure}[!h]
\centering
\includegraphics[width=\columnwidth]{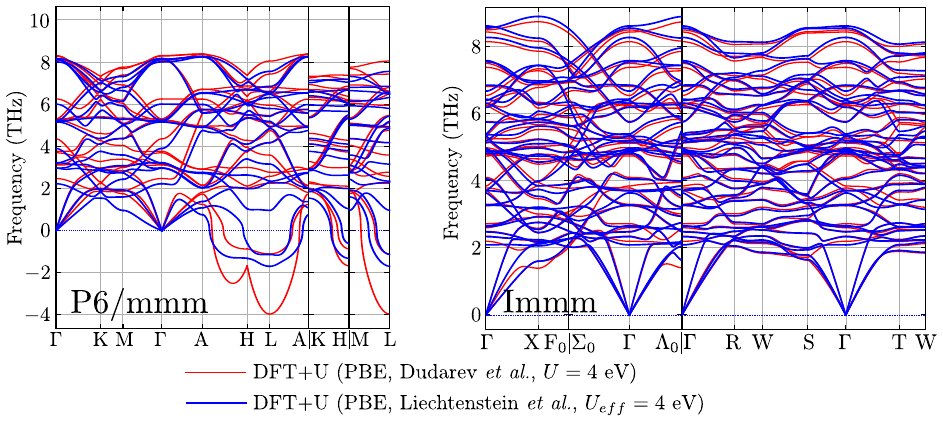}
\caption{
Comparison of the phonon dispersions of FeGe with P6/mmm and Immm symmetries depending on the applied DFT+U method: 
the simplified (rotationally invariant) approach to the DFT+U, introduced by Dudarev {\it et al.}~\cite{dudarev.botton.98} (red lines)
versus 
the rotationally invariant DFT+U introduced by Liechtenstein {\it et al.}~\cite{liechtenstein.anisimov.95} (blue lines).
Results for the P6/mmm and Immm structures are presented on left and right panel, respectively.
To sum up, the results are changed quantitatively, while qualitatively they are the same.
\label{fig.compare_ph}}
\end{figure}

\begin{figure}[!h]
\centering
\includegraphics[width=0.8\textwidth]{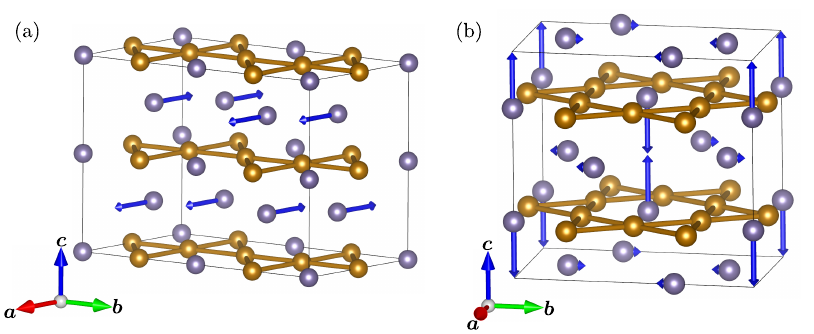}
\caption{
Displacement of atoms in (a) P6/mmm and (b) Ibam structures induced by the discussed soft modes. 
In the case of the P6/mmm structure, the soft mode is associated with alternating displacement of two Ge atoms within the $ab$ plane.
For the Ibma symmetry, the soft mode leads to simultaneous displacement of Ge(1) along the $c$ direction and Ge(2) within the $ab$ plane.
The final structure discussed in the main text possesses the Immm symmetry.
\label{fig.disp}
}
\end{figure}

\begin{figure}[!h]
\centering
\includegraphics[width=\textwidth]{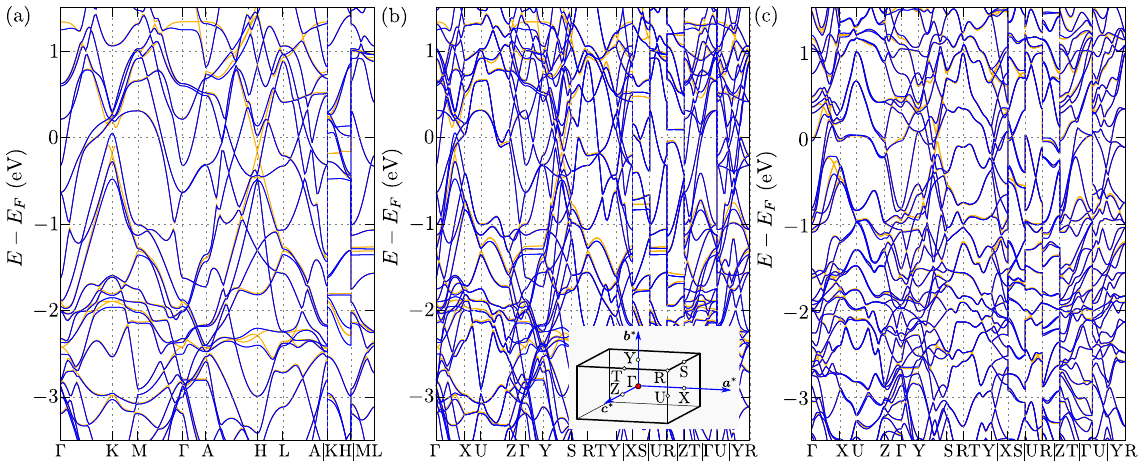}
\caption{
The comparison of electronic band structures for the P6/mmm and Immm phases.
(a) Electronic band structure for the P6/mmm magnetic unit cell.
(b) Folded band structure for P6/mmm calculated using the unit cell in the shape of the magnetic unit cell for the Immm phase.
(c) Electronic band structure for the Immm magnetic unit cell.
The orange and blue lines correspond to the band structure in the absence and presence of the spin--orbit coupling, respectively.
The inset in panel (b) shows the Brillouin zone for the Immm magnetic unit cell.
Results for $U = 4$~eV within the DFT+U Dudarev {\it et al.} approach.
\label{fig.71el}
}
\end{figure}

In this Supplemental Material, we present additional results:
\begin{itemize}
\item Fig.~\ref{fig.compare} -- 
Comparison of the energies of the FeGe system with P6/mmm and Immm symmetries depending on the DFT calculation scheme: without U, with U in the Dudarev variant ~\cite{dudarev.botton.98} and with U in the Liechtenstein variant~\cite{liechtenstein.anisimov.95}. 
%%%%%%%%%%%%%%%%%%%%%%
\item Fig.~\ref{fig.compare_ph} -- Dependence of phonon dispersions in FeGe with P6/mmm and Immm symmetries calculated using different DFT+U variants: 
the simplified (rotationally invariant) approach to the DFT+U, introduced by Dudarev {\it et al.}~\cite{dudarev.botton.98} 
versus 
the rotationally invariant DFT+U introduced by Liechtenstein {\it et al.}~\cite{liechtenstein.anisimov.95}.
%%%%%%%%%%%%%%%%%%%%%%
\item Fig.~\ref{fig.disp} -- Figure presenting the displacement of atoms induced by the discussed soft modes.
%%%%%%%%%%%%%%%%%%%%%%
\item Fig.~\ref{fig.71el} -- Comparison of the electronic band structure of the system with P6/mmm and Immm symmetries.
\end{itemize}

\vspace{3cm}

\vspace{4cm}

\vspace{3cm}

\newpage

The electronic band structures for the P6/mmm and Immm magnetic unit cells are presented in Figs.~\ref{fig.71el}(a) and~\ref{fig.71el}(c), respectively.
To facilitate comparison, we also present the folded band structure for the P6/mmm phase [Fig.~\ref{fig.71el}(b)], calculated using a cell with a shape similar to the magnetic unit cell of the Immm phase [presented in Fig.~\ref{fig.71}(b)].
As we can see, inclusion of the SOC opens band gaps in several points of the Brillouin zone (see orange and blue lines in Fig.~\ref{fig.71el}, to compare the electronic band structure in the absence and presence of the SOC).
We can simply find similarities between the electronic band structure for Immm [Fig.~\ref{fig.71el} (c)] and the folded band structure for P6/mmm [see Fig.~\ref{fig.71el}(b)].

\end{document}